\documentstyle[12pt]{article}
\makeatletter
\@addtoreset{equation}{section}
\renewcommand{\theequation}{\thesection.\arabic{equation}}
\makeatother
\makeatletter
\@addtoreset{equation}{subsection}

\def\R{\displaystyle \mathop{\bf R}}
\def\RR{\displaystyle \mathop{R}}

\def\V{\displaystyle \mathop{V}}
\def\G1{\displaystyle \mathop{G}}
\def\e1{\displaystyle \mathop{e}}
\def\U1{\displaystyle \mathop{U}}

\def\n1{\displaystyle \mathop{\nu}}
\def\ps1{\displaystyle \mathop{\Psi}}

\def\p1{\displaystyle \mathop{p}}
\def\J1{\displaystyle \mathop{J}}

\def\O1{\displaystyle \mathop{O}}

\def\hG{\displaystyle \mathop{\hat{G}}}
\def\gam{\displaystyle \mathop{\gamma}}

\def\pr{\displaystyle \mathop{\partial}}
\def\lpr{\displaystyle \mathop{\overleftarrow{\partial}}}
\def\lD1{\displaystyle \mathop{\overleftarrow{D}}}
\def\lhp{\displaystyle \overleftarrow{\hat{p}}}
\def\hgam{\displaystyle \mathop{\hat{\gamma}}}
\def\hpr{\displaystyle \mathop{\hat{\partial}}}
\def\Dlt{\displaystyle \mathop{\Delta}}
\def\hN{\displaystyle \mathop{\hat{N}}}
\def\he{\displaystyle \mathop{\hat{e}}}
\def\hp1{\displaystyle \mathop{\hat{p}}}
\def\hps{\displaystyle \mathop{\hat{\Psi}}}

\def\bp{\displaystyle \mathop{\bar{\Psi}}}

\def\S{\displaystyle \sum}
\def\IIn{\displaystyle \int}
\def\FFr{\displaystyle \frac}
\def\Lm{\displaystyle \lim}
\def\pd{\displaystyle \prod}
\def\hb{\displaystyle \mathop{\hat{b}}}
\def\F{\displaystyle \mathop{F}}
\def\sig1{\displaystyle \mathop{\sigma}}

\def\h1{\displaystyle \mathop{h}}
\def\H1{\displaystyle \mathop{H}}
\def\A1{\displaystyle \mathop{A}}

\def\D1{\displaystyle \mathop{D}}
\def\B1{\displaystyle \mathop{B}}
\def\L1{\displaystyle \mathop{L}}
\def\J1{\displaystyle \mathop{J}}
\def\A1{\displaystyle \mathop{A}}
\def\M1{\displaystyle \mathop{M}}
\def\g1{\displaystyle \mathop{g}}
\def\q1{\displaystyle \mathop{q}}
\def\x1{\displaystyle \mathop{x}}
\def\s1{\displaystyle \mathop{s}}
\def\X1{\displaystyle \mathop{X}}
\def\m1{\displaystyle \mathop{m}}
\def\Y1{\displaystyle \mathop{Y}}
\def\Z1{\displaystyle \mathop{Z}}
\def\W1{\displaystyle \mathop{W}}
\def\Bx{\displaystyle \mathop{\Box}}
\begin{document}

\begin{flushright} 
BAO/2000-3
\end{flushright}
\vskip 0.5truecm
\begin{center}
{\Large {\bf Microscopic Theory of the Standard Model}}
\vskip 1truecm
{\bf G.T.Ter-Kazarian}
\vskip 0.5truecm
{\em Byurakan Astrophysical 
Observatory, Armenia 378433 \\
E-mail:gago@bao.sci.am}
\end{center}
\vskip 1truecm
\begin{abstract}
The present article is the revised and updated 
version of the microscopic approach (MSM) [1,2] to the Standard Model 
(SM) of high energy physics. 
The operator manifold formalism (part I) enables the unification of the 
geometry and the field theory. It 
yields the quantization of geometry drastically different from earlier 
suggested schemes. We generalize this formalism via the concept of operator
multimanifold leading to the multiworld geometry.
The physical conditions are clarified at which the geometry,
as well the relativity (special and general) principle, together with the 
fermion fields serving as the basis for the 
constituent subquarks come into being. 
The subquarks emerge in the geometry only in certain permissible 
combinations utilizing the idea of subquark confinement principle, and 
have undergone the transformations yielding the internal
symmetries and gauge principle.
In part II we attempt to develop, further, the MSM, which enables an insight 
to the key problems of particle phenomenology. 
Particularly, we derive the Gell-Mann-Nishijima relation and flavour group,
infer the only possible low energy SM particle spectrum, and
conclude that the leptons are particles with integer 
electric and leptonic charges, and free of confinement, while quarks carry
fractional electric and baryonic charges, and imply the 
confinement. We suggest the microscopic theory of the unified electroweak
interactions with a small number of free parameters, wherein 
we exploit the background of the local expanded symmetry
$SU(2)\otimes U(1)$ and P-violation.
The Weinberg mixing angle is shown to have fixed value 
at $30^{o}$.
Due to the Bose-condensation of relativistic fermion pairs
the Higgs bosons have arisen
on an analogy of the Cooper pairs in superconductivity.
The resulting theory makes plausible following testable implications for the current 
experiments at LEP2, at the Tevatron and at LHC:

$\bullet$ {\em The Higgs bosons never could emerge in spacetime continuum, thus, they 
cannot be discovered in these experiments nor at any energy range}.

$\bullet$ {\em For each of the three SM families of quarks and 
leptons there are corresponding heavy family partners with the same 
quantum numbers 
\footnote{This prediction 
was directly  ensued from the MSM [1,2], as well the similar one was 
made in phenomenological consideration by S.L.Adler [3]).}
and common mass-shift coefficients $(1+k)$
given for the low-energy poles at $k_{1}>\sqrt{2}, \quad k_{2}=\sqrt{{8/3}}$ 
and $k_{3}= 2$, lying far above the electroweak scale, respectively, at the 
energy threshold values:} 
$
E_{1}>(419.6 \pm 12.0)GeV,\quad E_{2}= (457.6 \pm 13.2)GeV$ and 
$
E_{3}=(521.4 \pm 15.0)GeV.
$
We predict the Kobayashi-Maskawa 
quark flavour mixing patterns; the appearance of the CP-violation phase;
derive the mass-spectrum of leptons and quarks.
\end{abstract} 

\newpage
{\large {\bf Contents}}
\vskip 0.5truecm
\begin{tabbing}
\=aaaaaaaaaaaaaaaaaaaaaaaaaaaaaaaaaaaaaaaaaaaaaaaaaaaaaaaaaaaaaaaaaaaaaaaaaaa
\= \kill 
\>1. {\bf Introduction} . . . . . . . . . . . . . . . . . . . . . . . . . . . 
. . . . . . . . . . . . . .   
\> 4 \\\\
\>{\bf Part I} 
\hspace{0.5cm} 
{\bf Introduction to the Operator Manifold Formalism}       
\> 7\\\\
\>2. {\bf Preliminaries} . . . . . . . . . . . . . . . . . . . . . . . . . . 
. . . . . . . . . . . . . .  
\>7\\
\>2.1. Quantization of geometry . . . . . . . . . . . . . . . . . . . . . . .
. . . . . . . . . . 
\>8\\
\>2.2. Realization of the flat manifold $G$ . . . . . . . . . . . . . . . . .
. . . . . . . . . . .  
\>11\\
\>3. {\bf Mathematical background} . . . . . . . . . . . . . . . . . . . . . .
. . . . . . . . . 
\>13\\
\>4. {\bf Beyond the geometry and fields} . . . . . . . . . . . . . . . . . .
. . . . . . . . .   
\>14\\
\>4.1. The regular primordial structures . . . . . . . . . . . . . . . . . . . . .
. . . . . . . . . 
\>15\\
\>4.2. The distorted primordial structures . . . . . . . . . . . . . . . . . . . .
. . . . . . . . .  
\>16\\
\>4.3. ``Quark'' and ``colour'' confinement . . . . . . . . . . . . . . . . . . . .
. . . . . . . 
\>18\\
\>4.4. Gauge principle; internal symmetries . . . . . . . . . . . . . . . . . .
. . . . . . . . 
\>19\\
\>5. {\bf Operator multimanifold $\hat{G}_{N}$} . . . . . . . . . . . . . . . .
. . . . . . . . . . . . . . 
\>21\\
\>5.1. Operator vector and covector fields . . . . . . . . . . . . . . . . . . 
. . . . . . . . . 
\>21\\
\>5.2. Realization of the multimanifold $G_{N}$ . . . . . . . . . . . . . . . .
. . . . . . . . . .  
\>23\\
\>5.3. Subquarks and subcolour confinement . . . . . . . . . . . . . . . . . . 
. . . . . . . 
\>24\\\\
\>{\bf Part II}
\hspace{0.5cm} 
{\bf Realization of the Particle Physics} 
\>25\\\\
\>6. {\bf The MW-structure of the particles} . . . . . . . . . . . . . . . . .
. . . . . . . . 
\>25\\
\>6.1. Realization of the $Q-$world and Gell-Mann-Nisijima relation . . . . 
. . . . . . .   
\>27\\
\>6.2. The symmetries of the $W-$ and $B-$worlds . . . . . . . . . . . . . 
. . . . . . . . . 
\>28\\
\>6.3. The microscopic structure of leptons: lepton generations . . . . . .
. . . . . . . . 
\>29\\
\>6.4. The microscopic structure of quarks: quark generations . . . . . . 
. . . . . . . . . 
\>30\\
\>6.5. The flavour group $SU_{f}(6)$ . . . . . . . . . . . . . . . . . . .
. . . . . . . . . . . . . 
\>31\\
\>6.6. The primary field . . . . . . . . . . . . . . . . . . . . . . . . .
. . . . . . . . . . . .  
\>31\\
\>6.7. A generation of the fermion mass in the $Q-$world . . . . . . . . .
. . . . . . . . . 
\>33\\
\>6.8. The electroweak interactions: the $P-$violation . . . . . . . . . .
. . . . . . . . . .
\>35\\
\>6.9. The reduction coefficient and the Weinberg mixing angle . . . . . . . . . . . . .   
\>36\\
\>6.10. Emergence of composite isospinor-scalar bosons . . . . . . . . . . . . . . . . . .
\>37\\
\>7. {\bf The Higgs boson} . . . . . . . . . . . . . . . . . . . . . . . . . . . . . . . . . . . . .
\>39\\
\>7.1. The Bose-condensate of iso-pairs . . . . . . . . . . . . . . . . . . . 
. . . . . . . . . 
\>39\\
\>7.2. The non-relativistic approximation . . . . . . . . . . . . . . . . . . 
. . . . . . . . .
\>42\\
\>7.3. The relativistic treatment . . . . . . . . . . . . . . . . . . . . . . 
. . . . . . . . . . 
\>43\\
\>7.4. Self-interacting potential of Bose-condensate . . . . . . . . . . . . . . . . . . . . . 
\>45\\
\>7.5. The four-component Bose-condensate in magnetic field . . . . . . . . . 
. . . . . . 
\>47\\
\>7.6. Extension to lower temperatures . . . . . . . . . . . . . . . . . . . . . . . . . . . .
\>50\\
\>8. {\bf Lagrangian of electroweak interactions; the transmission of the} 
\> \\
\>\hspace{0.7cm}{\bf electroweak symmetry breaking from the $W-$world to 
spacetime} 
\> \\
\>\hspace{0.7cm}{\bf continuum} . . . . . . . . . . . . . . . . . . . . . . . . . . . . . . . . . . . . . . . . . 
\>52 \\
\>9. {\bf The two solid phenomenological implications of the MSM} . . . . . . 
. . . 
\>54 \\
\>10. {\bf Quark flavour mixing and the Cabibbo angles} . . . . . . . . . . . . . . . . . 
\>55 \\
\>11. {\bf The appearance of the CP-violating phase} . . . . . . . . . . . . . 
. . . . . . 
\>58\\
\>12. {\bf The mass-spectrum of leptons and quarks} . . . . . . . . . . . . . . . . . . . . 
\>59\\
\>13. {\bf The physical outlook and concluding remarks} . . . . . . . . . . . . . . . . . 
\>60\\ 
\>14. {\bf Acknowledgements} . . . . . . . . . . . . . . . . . . . . . . . . . 
. . . . . . . . . . 
\>63\\
\>15. {\bf Appendix A}: Mathematical background (a continuation). . .  . . . . . . . . . . . 
\>63\\
\>16. {\bf Appendix B}: Reflection of the Fermi fields . . . . . . . . . . . . 
. . . . . . . . . .  
\>72\\
\>17. {\bf Appendix C}: The solution of wave equation of distorted 
\> \\
\>\hspace{0.7cm}structure . . . . . . . . . . . . . . . . . . . . . . . . . . . . . . . . . . . . . . . . . . .     
\>74\\
\>18. {\bf Appendix D}: Field aspect of the OMM . . . . . . . . . . . . . . . 
. . . . . . . . . 
\>76\\
\>19. {\bf References} . . . . . . . . . . . . . . . . . . . . . . . . . . . . 
. . . . . . . . . . . . . 
\>76\\
\end{tabbing}
\newpage
\section {Introduction}
\label {int}
A definite pattern for the theoretical description of particle physics has 
emerged based on the framework of the Standard Model (SM) of high energy 
physics [4-22], which is built up from observation for prediction and 
correlation of the new data. 
Although the SM has proven to be in spectacular agreement with 
experimental measurements and quite successful in a predicting a wide range 
of phenomena, however, it is not exception to the rule that as the 
phenomenological approach it suffers from own difficulties. There were still 
many open key questions arisen inevitably that we have no understanding 
why the SM is as it is? Why is the gauge symmetry? 
Why is this the particle spectrum? 
The mechanism of the electroweak symmetry breaking is a complete mystery. The 
most problematic ingredient of such a breaking is the Higgs boson (in 
simplest version), which has not yet been discovered experimentally.
The untested aspects of SM are the mass spectrum of 
the particles, the mixing patterns and the CP violation. 
The latter is introduced through complex Yukawa couplings 
of fermions to Higgs bosons, resulting in complex parameters in the CKM 
matrix. The SM contains a large number of a {\em priori} free parameters, 
while a consistent complete theory would not have so many free parameters.
It is important then to develop the other schemes that attempt to reduce 
the number of free parameters. All the approaches proposed towards 
the unified gauge field theory, e.g. [11-42], have own advantages and 
difficulties. But, it is for one thing, the basic concepts and 
right symmetries,  as a rule, were inserted in all of them in {\em ad hoc} 
manner. However, their physical essence and origin still remain unknown 
and, therefore, the resulting theory is beset by various difficulties. 
The key of solution rested entirely in the searches for the new mathematical 
structures enabling an insight, in microscopic sense, to the conceptual 
foundation of particle physics. \\
To address to some of the above mentioned nagging questions of the SM 
recently the MSM is developed in [1,2], wherein the proliferation of lepton 
and quark flavours prompts us to consider the fields as composites. 
Certainly, it may seem foolhardy to set up such a picture in the spacetime 
continuum. The difficulties arisen here are well-known. The first problem is 
closely related to the expected mass differences 
of particles, which in this case would be too large ($\geq 1TeV$). 
Another problem concerns the transformations of particles. 
Our idea is to remove these difficulties by employing the multiworld (MW) 
geometry ensued from the operator manifold (OM) formalism [1].
The revised, expanded and completed version of [1,2] is suggested in the 
present article, which is organized as follows: 
The OM formalism is given in the part I, wherein to facilitate writing 
the main new features of the new 
version are firstly represented in the restructuring of the old one. We 
forbear to write out in the main text the pieces including only the 
technical issues, which have now replaced into corresponding appendices as 
they are a somewhat lengthy and so standard.
Both the quantum field and differential geometric aspects of the OM 
formalism are studied. The OM enables to develop an 
approach to the unification of the geometry and the field theory.
The former is equivalent to the configuration space 
wave mechanics incorporated with geometric properties leading to the 
quantization of geometry strongly different from earlier suggested 
schemes (subsec.2.1, App.A). Within the framework of algebraic approach we 
have reached to rigorous definition of the OM. While,
we consider one parameter group of operator diffeomorphisms, 
the operator differential forms and their integration, 
also operator exterior differentiation.
We generalize this formalism via the  concept of operator multimanifold 
(OMM) (sec.5), which yields the MW geometry (subsec.5.2)
involving the spacetime continuum and internal worlds of the given number.
In an enlarged framework of the OMM we define and 
clarify the conceptual basis of subquarks and their
characteristics stemming from the various symmetries of the internal worlds
(subsec.5.3).
The OMM formalism has the following features: 

{\em
$\bullet$ It provides a natural unification of the geometry-
yielding the 1) special and 2) general relativity principles, and the 
fermion fields serving as the basis for the constituent subquarks 
(subsec.5.3). 

$\bullet$ It has cleared up the physical conditions in which the geometry 
and particles come into being (subsec.2.2, 5.2).

$\bullet$ The subquarks
emerge in the geometry only in certain permissible combinations
utilizing the idea of the 3) subcolour (subquark) confinement 
principle, and have undergone the transformations yielding the internal
symmetries and 4) gauge principle}.

These results are used in the part II to develop 
the updated version of the MSM [2]. The main new features of the new 
version are as follows: 1) In dealing with a 
new round of the experiments, in this paper, we supplement this previous 
discussion, first by including a detailed description of the remarkable
specific mechanism of electroweak symmetry breaking arisen in the MSM 
(sec.8); 2) The two solid phenomenological testable implications of the 
MSM are given in the sec.9; 3) Finally, we derive a physically more 
realistic mass spectrum of the leptons and quarks (sec.12) instead of the 
former one inferred within the simplified scheme [2].\\
All the fields including the 
leptons and quarks along with the spacetime component have also the 
nontrivial composite internal MW-structure (sec.6). While, the various
subquarks then are defined on the corresponding internal worlds.
The possible elementary particles are thought to be 
composite dynamical systems in analogy to quantum mechanical stationary 
states of compound atom, but, now a dynamical treatment built up on the 
MW geometry is quite different and more amenable to qualitative 
understanding. The microscopic structure of leptons, quarks and other 
particles will be governed by the only possible conjunctions of constituent
subquarks implying concrete symmetries.
Although within considered schemes the subquarks are defined on the internal 
worlds, however the resulting spacetime components of particles, which we are 
going to deal with to describe the leptons and quarks defined on 
the spacetime continuum, are affected by them (subsec.6.4) in such a way 
that they carry exactly all the quantum numbers of the various constituent 
subquarks of the given composition.
The MSM enables an insight to the key concepts of particle physics 
(subsec.6.1-6.6), and to conclude that the leptons are particles with 
integer electric and leptonic charges and free of confinement, 
while the quarks 
carry fractional electric and baryonic charges and imply the confinement. 
The theoretical significance of the MSM resides in the microscopic 
interpretation of all physical parameters.
We derive the Gell-Mann-Nishijima relation and the flavour group.
The MW structure of primary field (subsec.6.6, 6.7) 
is described by the gauge invariant Lagrangian involving nonlinear 
fermion interactions of the internal field components somewhat similar to
the theory by Heisenberg and
his co-workers
[51,52], but still it will be defined on the MW geometry. 
This Lagrangian is the whole story since all the complexity of the leptons, 
quarks and their interactions arises from it.
The number of free parameters in this 
Lagrangian is reduced to primary coupling constant of the nonlinear
interaction and gauge coupling. Based on it, we consider the microscopic theory of the 
unified electroweak interactions (subsec.6.8-sub.12).
It follows that contemporary phenomenological
SM is an approximation to the suggested microscopic approach.
We exploit the background of the local symmetry
$SU^{loc}(2)\otimes U^{loc}(1)$ (subsec.6.8), the
weak hypercharge and P (mirror symmetry)-violation. 
The Weinberg mixing angle determining the symmetry reduction coefficient 
is shown to have a value fixed at $30^{o}$ (subsec.6.9).
We develop the microscopic approach to the isospinor Higgs boson with 
self-interaction and Yukawa couplings (subsec.6.10, sec.7). It involves 
Higgs boson as the collective excitations of bound quasi-particle pair.   
Tracing a resemblance with the Cooper pairs [53-55],
within the framework of local gauge invariance of the theory 
incorporated with the phenomenon of P-violation in weak interactions 
we suggest a mechanism providing the Bose-condensation of 
relativistic fermion pairs. In contrast to the SM, it predicts both the 
electroweak symmetry breaking in the $W$-world by the vacuum expectation 
value (VEV) of spin zero Higgs bosons and its transmission from the $W-$world 
to the spacetime continuum (sec.8). 
This is the most remarkable feature of our approach 
especially in the view of existing great belief 
of the conventional theories for a discovery of the Higgs boson 
with other new particles at next round of experiments at 
LEP2, at the Tevatron, at LHC and other colliders, which will explore the TeV energy 
range (e.g. see [50]). The LEP2 data (is currently running at 189 GeV) provide a lower 
limit $m_{H}>89.3 GeV$ on its mass in simplest version. 
Furthermore, there is a tight upper limit 
$(m_{h^{0}}< 150 GeV)$ on the mass of the lightest Higgs boson $h^{0}$ among the 5 
physical Higgs bosons predicted by the models of Minimal Supersymmetric Exstention of 
the SM (MSSM). The current direct search limits from LEP2 give $m_{h^{0}}> 75 GeV.$ 
Therefore, the future searches for this boson (if the mass is below 150 GeV or so)
would be a crucial point in testing the efforts made in the conventional models 
building as well in the present MSM based on a quite different approaches.
Actually, reflecting upon the results far obtained in the sec.9, in strong contrast 
to conventional theories, the MSM rejects drastically any expectation of discovery 
of the Higgs boson, but in the same time it expects to include a rich spectrum of new 
particles at higher energies. Thus, if the MSM proves viable it becomes 
an crucial issue to hold in experiments the following two solid tests: 

{\em
$\bullet$   
The Higgs bosons never could emerge in spacetime continuum since they have 
arisen only on the internal $W$-world, i.e., thus, the unobserved effects 
produced by such bosons cannot be discovered in experiments nor at any 
energy range. 

$\bullet$  For each of the three Standard Model families of quarks and 
leptons, there are corresponding heavy family partners with the same quantum 
numbers lying far above the electroweak scale}.

Regarding to the last phenomenological implication of the MSM, it is 
remarkable that the similar in many respects prediction is made in somewhat 
different context by S.L.Adler [3] within a phenomenological scheme of a 
compositeness of the quarks and leptons. It based on the generic group 
theoretical framework of rishon type models exploring the preon constituents. 
But, therein the most important specification of this scale is absent. 
Although one admits that such a scale could be much higher than 
electroweak scale, however, it is also necessary special argumentations   
in support of validity of this prediction in the case if this scale has 
turned out to be low enough, namely, if these heavy partners lie not too far 
above the electroweak scale. Even thus, as it is notified in [3], one must not 
worry for the existence of 6 heavy flavors, which is then marginally 
compatible with the current LEP data [18]. A complete analysis of this 
question, naturally, is possible now in suggested microscopic approach. 
The MSM enables oneself to study in detail the phenomenology 
associated with such extra heavy families and to estimate the value of 
energy threshold of their creation. While, the low energy scale could not 
be realized since it lies far below the energy threshold of the next pole necessary 
for appearing of the heavy partners. The estimate gives the common 
mass-shift coefficients $(1+k)$, where $k$ reads for the next few low 
energy poles with respect to the lowest one: 
$k_{0}=0,\quad k_{1}>\sqrt{2}, \quad k_{2}=\sqrt{{8/3}}$ and $k_{3}= 2$.
The first one obviously does not produce the extra families, but the energy 
thresholds corresponded to the next non-trivial poles can be respectively 
written:
$
E_{1}> (419.6 \pm 12.0)GeV,\quad E_{2}= (457.6 \pm 13.2)GeV$ and 
$
E_{3}=(521.4 \pm 15.0)GeV.
$
Thus, which of these schemes above, if any, is realized either exactly or at least 
approximately in nature remains to be seen in the years to come.\\
Finally we attempt to predict the mixing angles in the six-quark KM model 
(sec.10), the appearance of the CP-violation phase (sec.11) and  derive the 
mass spectrum of the leptons and quarks (sec.12). The physical outlook on 
suggested approach and concluding remarks are given in the sec.13, which 
have involved in order once again to resume a whole physical picture and to 
provide a sufficient background for its understanding without undue 
hardship. The appendices will complete the mathematical framework.\\
Our approach still should be considered as a preliminary one, wherein we have 
contended ourselves with a rather modest task and do not profess to have any 
clear-cut answers to all the problems of particle physics, the complete 
picture of which is largely beyond the scope of the present paper.  
The only argument that prompts us to consider present approach seriously is 
the remarkable feature that the most important properties of particle 
phenomenology can be derived naturally within its framework. Therefore we 
hope that it will be an attractive basis for the future theory.
Although many key problems are elucidated within outlined
approach, nevertheless some issues still remain to be solved.
To go any further in exploring the significance of obtained 
results it is entirely feasible, for instance, to promote the 
MSM into the supersymmetric framework in order to solve its technical 
aspects of the vacuum zero point energy and hierarchy problems, and attempt 
to develop the realistic viable microscopic theory (VMSM), which will be 
carried out in subsequent paper. Given therein the VMSM will make a new 
predictions about observing of the supersymmetric partners drastically 
different from those of conventional MSSM-models.

\begin{center}
{\Large {\bf Part I}}
\end{center}
\vskip 0.1truecm
\begin{center}
{\Large {\bf Introduction to the Operator Manifold Formalism}}
\end{center}
The mathematical framework of the OM formalism reveals primordial deeper 
structures underlying the fundamental concepts of the particle physics.
Here we explore the query how did the geometry and fields, as they are, come
into being. In the first our major purpose is to prove the idea that the 
geometry and fields, with the internal symmetries and all interactions, as 
well the four major principles of relativity (special and general), quantum, 
gauge and colour confinement are derivative, and they come into being 
simultaneously. The substance out of which the geometry and fields are made 
is the ``primordial structures'' involved into reciprocal ``linkage'' 
establishing processes (sec.4)
\renewcommand{\theequation}{\thesubsection.\arabic{equation}}
\section {Preliminaries}
\label{Prin}
This section contains some of the necessary preliminaries on generic of the 
OM formalism [1], which one to know in order to understand our approach.
We adopt then all the ideas and conventions of [1] and
to be brief we often suppress the indices without notice.
\\
We start by tracing at elementary level the relevant steps of motivation of 
the OM formalism:

$\bullet$ First step is an extension of the Minkowski space 
${\M1_{x}}{}_{4}\rightarrow M_{8}={\M1_{x}}{}_{4}\oplus {\M1_{u}}{}_{4}$ in 
order to introduce the particle mass operator defined on the internal world 
${\M1_{u}}{}_{4}$. 
For example, in the case of Dirac's particle one proceeds at once:
$$(\underbrace{\gamma p}_{x}-m)\ps1_{x}=0
\quad\rightarrow \quad
\gamma p\psi=0,$$ 
provided by
$\psi=\ps1_{x}\ps1_{u}, \quad
\gamma p=\underbrace{\gamma p}_{x} - \underbrace{\gamma p}_{u},
\quad
m\ps1_{u}\equiv\underbrace{\gamma p}_{u} \ps1_{u}
$
and
$$d\,x^{2}=inv \quad\rightarrow \quad
d\,x^{2}_{8}=d\,x^{2}-d\,u^{2}=0, \quad x_{8}\in M_{8}. 
$$
The same holds for the other fields of arbitrary spin.

$\bullet$ Next, a two-steps passage 
$M_{4}\rightarrow M_{6}  \stackrel{45^{0}}{\rightarrow} 
G_{6}$ will be performed for each sample of the $M_{4}$.
\\
a) A passage $M_{4}\rightarrow M_{6}$ restores the complete equivalence 
between the three spacial and three time components:
$$
\begin{array}{l}
e_{4}=(\vec{e},e_{0})\quad\rightarrow\quad 
\vec{e}_{6}=(\vec{e},\vec{e}_{0})\in M_{6},\quad
x_{4}=(\vec{x},x_{0})\quad\rightarrow \quad
x_{6}=(\vec{x},\vec{x}_{0})\in M_{6}.
\end{array}
$$
b) A rotation $M_{6} \stackrel{45^{0}}{\rightarrow} G_{6}$ of the basis 
vectors  on the angle $45^{0}$ provides an adequate algebra for 
quantization of the geometry (subsec.2.1):
$$
\begin{array}{l}
\vec{e}_{6} \stackrel{45^{0}}{\rightarrow} e_{(\lambda\alpha)},
\quad \lambda = \pm, \quad \alpha=1,2,3, \\ \\
e_{\pm\alpha}=\FFr{1}{\sqrt{2}}(e_{0\alpha}\pm e_{\alpha})=
O_{\pm}\otimes\sigma_{\alpha}, \quad 
<O_{\lambda},O_{\tau}>=1-\delta_{\lambda\tau}, 
\quad <\sigma_{\alpha},\sigma_{\beta}>=\delta_{\alpha\beta}.
\end{array}
$$ 
Accordingly one gets $M_{8}\rightarrow G_{12}$. Thus, within a simplified 
scheme (one $u$- channel) of the following it is convenient to 
deal in terms of smooth differentiable manifold 
$$G=\G1_{\eta}\oplus \G1_{u},$$  
$ Dim \,G=12,\quad Dim\,\G1_{i}=6 \quad (i=\eta, u)$. 

$\bullet$ Finally, in suggested approach we will be dealing in terms of 
first degree of the line element, which entails an additional phase 
multiplier $\Phi(\zeta$) for the vector defined on $G$:\\
$$d\,\zeta^{2}\quad\rightarrow \quad d\,\vec{\zeta}\,\,e^{iS}, 
\quad \vec{\zeta}\quad\rightarrow \quad\vec{\Phi}(\zeta)=
\vec{\zeta}\,\Phi(\zeta), \quad
\Phi(\zeta)\equiv e^{iS},$$ 
where 
$\vec{\zeta}=\vec{e}\, \,\zeta, \quad \vec{e}=
(\vec{\e1_{\eta}}, \,\vec{\e1_{u}}),\quad 
S(\zeta)$ is the invariant action defined on $G$.

\subsection{Quantization of Geometry}
\label{Vec}
The 
$\{e_{(\lambda,\mu,\alpha)}=O_{\lambda,\mu}\otimes
\sigma_{\alpha}\} \subset G$
$(\lambda,\mu=1,2; \quad \alpha=1,2,3)$
are linear independent $12$ unit vectors at the point $p$ of
the 12 dimensional smooth differentiable manifold $G$,
provided by the linear unit bipseudovectors $O_{\lambda,\mu}$
and the ordinary unit vectors $ \sigma_{\alpha}$ implying
$$<O_{\lambda,\mu},O_{\tau,\nu}>={}^{*}\delta_{\lambda,\tau}
{}^{*}\delta_{\mu,\nu}\quad
<\sigma_{\alpha}, \sigma_{\beta}>= \delta_{\alpha\beta},\quad
{}^{*}\delta=1-\delta,
$$
where $\delta$ is Kronecker symbol,  $\{ O_{\lambda,\mu}= 
O_{\lambda}\otimes  O_{\mu} \}$ is
the basis for tangent vectors of $2 \times 2$ dimensional linear 
pseudospace ${}^{*}{\bf R}^{4}={}^{*}{\bf R}^{2}\otimes
{}^{*}{\bf R}^{2}$, the $ \sigma_{\alpha}$ refer to three dimensional 
ordinary space ${\bf R}^{3}$. 
Henceforth, we always let the first two subscripts in the parentheses
to denote the pseudovector components, while the third refers to the
ordinary vector components.
The metric on $G$ is 
$\hat{\bf g}:{\bf T}_{p}\otimes {\bf T}_{p}\rightarrow C^{\infty}(G)$ 
a section of conjugate vector bundle $S^{2}{\bf T}$.
Any vector ${\bf A}_{p}\in{\bf T}_{p}$ reads ${\bf A}=
e A$, provided with
components $A$ in the basis
$\{e\}$.
In holonomic coordinate basis $\left(\partial/\partial\,
\zeta\right)_{p}$ one gets 
$A=\left.\FFr{d\,
\zeta}{d\,t}\right|_{p}$ and
$\hat{g}=g d\zeta\otimes d\zeta $.
The manifold $G$ decomposes as follows:
$$G={}^{*}{\bf R}^{2}
\otimes {}^{*}{\bf R}^{2} \otimes {\bf R}^{3}=\G1_{\eta}\oplus
\G1_{u}=\displaystyle\sum^{2}_{\lambda,\mu=1} 
\oplus {\bf R}^{3}_{\lambda \mu}=
{\R_{x}}^{3}\oplus 
{\R_{x_{0}}}^{3}\oplus
{\R_{u}}^{3}\oplus 
{\R_{u_{0}}}^{3}$$ with corresponding 
basis vectors  
${\e1_{i}}{}_{(\lambda\alpha)}={\O1_{i}}{}_{\lambda}\otimes 
\sigma_{\alpha}
\subset \G1_{i}$ $(\lambda =\pm,\quad 
i=\eta, u)$ of tangent sections, where 
$${\O1_{i}}{}_{+}=
\displaystyle \frac{1}{\sqrt{2}}(O_{1,1} +\varepsilon_{i} O_{2,1}),\quad
{\O1_{i}}{}_{-}=
\displaystyle \frac{1}{\sqrt{2}}(O_{1,2} +\varepsilon_{i} O_{2,2}),\quad
\varepsilon_{\eta}=1,\quad\varepsilon_{u}=-1.$$ Hence
$<{\O1_{i}}{}_{\lambda},{\O1_{j}}{}_{\tau}>=
\varepsilon_{i}\delta_{ij}{}^{*}\delta_{\lambda \tau}$.
The $\G1_{\eta}$ decomposes into three dimensional
ordinary and time  flat 
spaces $$\G1_{\eta}=
{\R_{x}}^{3}\oplus {\R_{x_{0}}}^{3}$$ with the signatures 
$sgn({\R_{x}}^{3})=(+++)$ 
and $sgn({\R_{x_{0}}}^{3})=(---)$. 
The same holds for the $\G1_{u}$ with opposite 
signatures $sgn({\R_{u}}^{3})=(---)$ 
and $sgn({\R_{u_{0}}}^{3})=(+++)$. \\
The positive metric forms are defined
on manifolds $\G1_{i}:\quad $   
$\eta^{2}\in \G1_{\eta}, \quad
u^{2}\in \G1_{u}.$
The passage to the four-dimensional Minkowski space is a further step as 
follows: since all directions in ${\R_{x_{0}}}^{3}$ are
equivalent, then by notion {\em time} one implies the projection of
time-coordinate on fixed arbitrary universal direction in ${\R_{x_{0}}}^{3}$.
This clearly respects the physical ground.
By such a reduction ${\R_{x_{0}}}^{3}\rightarrow {\R_{x_{0}}}^{1}$ the 
passage $$\G1_{\eta}\rightarrow M_{4}={\R_{x}}^{3}\oplus {\R_{x_{0}}}^{1}$$ 
may be performed whenever it will be necessary.\\
In the other case of the six dimensional curved manifold $\widetilde{G}$ 
(subsec.4.2), the passage to four dimensional Riemannian geometry $R_{4}$ is 
straightforward by making use of reduction of three time components 
$e_{0\alpha}=
\displaystyle \frac{1}{\sqrt{2}}(e_{(+\alpha)}+e_{(-\alpha)})$ of basis 
sixvector $e_{(\lambda \alpha)}$ to the single one $e_{0}$ in the given
universal direction, which merely has fixed the time 
coordinate. Actually, since Lagrangian of the fields defined on
$\widetilde{G}$ is a function of scalars such as
$A_{(\lambda \alpha)}B^{(\lambda \alpha)}=
A_{0 \alpha}B^{0 \alpha}+A_{\alpha}B^{\alpha}$, thus, taking into account 
that $A_{0 \alpha}B^{0 \alpha}=A_{0 \alpha}<e^{0\alpha},e^{0\beta}>B_{0 \beta}
=A_{0}<e^{0},e^{0}>B_{0}=A_{0}B^{0}$, one readily may perform the
required passage.
Hence
$$
d\,\zeta^{2}=d\,\eta^{2}-d\,u^{2}=0,
\quad \left.d\,\eta^{2} \right|_{6\rightarrow 4} \equiv
d\,s^{2}=g_{\mu\nu}d\,x^{\mu}d\,x^{\nu}=d\,u^{2}=inv.
$$
For more discussion see App.B or [1].\\
Unifying the geometry and particles into one framework
the OM formalism is analogous to the method of secondary 
quantization with appropriate expansion over the geometric objects.
We proceed at once to the secondary quantization of geometry
by substituting the basis elements for the 
creation and annihilation operators acting 
in the configuration space of occupation numbers.
Instead of pseudo vectors $O_{\lambda}$ we introduce the  
operators supplied by additional index ($r$) referring to the quantum 
numbers of corresponding state
\begin{equation}
\label {R24}
\hat{O}^{r}_{1}=O^{r}_{1}\alpha_{1},\quad
\hat{O}^{r}_{2}=O^{r}_{2}{\alpha}_{2},\quad
\hat{O}_{r}^{\lambda}={}^{*}\delta^{\lambda\mu}\hat{O}^{r}_{\mu}=
{(\hat{O}^{r}_{\lambda})}^{+},\quad
\{ \hat{O}^{r}_{\lambda},\hat{O}^{r'}_{\tau} \}=
\delta_{rr'}{}^{*}\delta_{\lambda\tau}I_{2}.
\end{equation}
The matrices ${\alpha}_{\lambda}$ satisfy the condition
$
\{ {\alpha}_{\lambda},{\alpha}_{\tau} \}={}^{*}\delta_
{\lambda\tau}I_{2},
$ where 
$
{\alpha}^{\lambda}={}^{*}\delta^{\lambda\mu}
{\alpha}_{\mu}={({\alpha}_{\lambda})}^{+},
$ and 
$I_{2}=\left( \matrix{
1 &0 \cr
0 &1\cr
}\right)$.
For example
${\alpha}_{1}=\left( \matrix{
0 &1 \cr
0 &0 \cr
}\right), \quad
{\alpha}_{2}=\left( \matrix{
0 &0 \cr
1 &0 \cr
}\right).$
The creation $\hat{O}^{r}_{1}$ and  annihilation $\hat{O}^{r}_{2}$ operators
are acting as follows:
$$\hat{O}^{r}_{1}\mid 0>={O}^{r}_{1}\mid 1>, 
\quad \hat{O}^{r}_{2}\mid 1>={O}^{r}_{2}\mid 0>,$$
where $\mid 0>\equiv  \mid 0,0,\ldots>$ and
$\mid 1>\equiv\mid 0,\ldots,1,\ldots>$ 
are respectively the nonoccupied vacuum state and the one occupied state.
Thus, 
$\hat{O}^{r}_{1}\mid 1>=0,\quad 
\hat{O}^{r}_{2}\mid 0>=$0.
A matrix realization of such states, for instance, can be:
$\mid 0>\equiv\chi_{1}=\left( \matrix{
0 \cr
1\cr
}\right), \\
\mid 1>\equiv\chi_{2}=\left( \matrix{
1 \cr
0 \cr
}\right).$
Hence
$\chi_{0}\equiv\mid 0>=\displaystyle\prod_{r=1}^{N}(\chi_{1})_{r}$ and
$\chi_{r'}\equiv\mid 1>=(\chi_{2})_{r'}\displaystyle\prod_{r\neq r'}
(\chi_{1})_{r}$.\\
Also, instead of ordinary basis vectors we introduce
the operators
$\hat{\sigma}^{r}_{\alpha}\equiv\delta_{\alpha\beta\gamma}
\sigma^{r}_{\beta}\widetilde{\sigma}_{\gamma}$,
where $\widetilde{\sigma}_{\gamma}$ are Pauli's matrices such that
\begin{equation}
\label{R27}
<\sigma_{\alpha}^{r},\sigma_{\beta}^{r'}>=\delta_{rr'}\delta_{\alpha\beta},
\quad
\hat{\sigma}^{\alpha}_{r}=\delta^{\alpha\beta}\hat{\sigma}^{r}_{\beta}=
{(\hat{\sigma}_{\alpha}^{r})}^{+}=\hat{\sigma}_{\alpha}^{r},
\quad
\{\hat{\sigma}_{\alpha}^{r},\hat{\sigma}_{\beta}^{r'}\}=2
\delta_{rr'}\delta_{\alpha\beta}I_{2}.
\end{equation}
The vacuum state $\mid 0>\equiv{\varphi}_{1(\alpha)}$
and the one occupied state $\mid 1_{(\alpha)}>
\equiv{\varphi}_{2(\alpha)}$ read:\\
${\varphi}_{1(\alpha)}\equiv\chi_{1}, \quad
{\varphi}_{2(1)}=\left( \matrix{
1 \cr
0 \cr
}
\right), \quad
{\varphi}_{2(2)}=\left( \matrix{
-i \cr
0\cr
}
\right), \quad
{\varphi}_{2(3)}=\left( \matrix{
0 \cr
-1\cr
}
\right),$
thus,
$$
{\hat{\sigma}}_{\alpha}^{r}\varphi_{1(\alpha)}=\sigma_
{\alpha}^{r}\varphi_{2(\alpha)}=(\sigma_{\alpha}^{r}\widetilde{\sigma}_
{\alpha})\varphi_{1(\alpha)},
\quad
{\hat{\sigma}}_{\alpha}^{r}\varphi_{2(\alpha)}=\sigma_
{\alpha}^{r}\varphi_{1(\alpha)}=(\sigma_{\alpha}^{r}\widetilde{\sigma}_
{\alpha})\varphi_{2(\alpha)}.
$$
Whence, the single eigenvalue
$(\sigma_{\alpha}^{r}\widetilde{\sigma}_{\alpha})$
associates with different $\varphi_{\lambda(\alpha)},$
namely it is degenerate with 
degeneracy degree equal 2. Thus, among quantum numbers $r$
there is also the quantum number of the half integer spin 
$\vec{\sigma}$ 
$(\sigma_{3}=\FFr{1}{2}s,\quad s=\pm1).$
This consequently gives rise to the spins of
particles.
The one occupied state reads
$\varphi_{r'(\alpha)}={(\varphi_{2(\alpha)})}_{r'}\displaystyle
\prod_{r\neq r'}{(\chi_{1})}_{r}.$
Next, we introduce the operators
$$
{\hat{\gamma}}^{r}_{(\lambda,\mu,\alpha)}\equiv{\hat{O}}^{r_{1}}_{\lambda}
\otimes{\hat{O}}^{r_{2}}_{\mu}\otimes{\hat{\sigma}}^{r_{3}}_{\alpha}$$
and the state vectors
$$
\chi_{\lambda,\mu,\tau(\alpha)}\equiv\mid\lambda,\mu,\tau(\alpha)>=
\chi_{\lambda}\otimes\chi_{\mu}\otimes\varphi_{\tau(\alpha)}, 
$$
where $\lambda,\mu,\tau,\nu=
1,2;\quad \alpha,\beta=1,2,3$ and $r\equiv (r_{1},r_{2},r_{3})$.
Omitting two valuedness of state vectors we apply
$\mid\lambda,\tau,\delta(\beta)>\equiv\mid\lambda,\tau>$,
and remember that always the summation 
must be extended over the double degeneracy of the spin states $(s=\pm 1)$.
The explicit matrix elements of basis vectors read
\begin{equation}
\label{R211}
<\lambda,\mu\mid{\hat{\gamma}}{}^{r}_{(\tau,\nu,\alpha)}\mid \tau,\nu>=
{}^{*}\delta_{\lambda\tau}{}^{*}\delta_{\mu\nu}
e^{r}_{(\tau,\nu,\alpha)},\quad
<\tau,\nu\mid{\hat{\gamma}}{}_{r}^{(\tau,\nu,\alpha)}\mid\lambda,\mu >=
{}^{*}\delta_{\lambda\tau}{}^{*}\delta_{\mu\nu}
e_{r}^{(\tau,\nu,\alpha)}.
\end{equation}
for given $\lambda,\mu.$
The operators of occupation numbers 
\begin{equation}
\label{R212}
{\hN_{1}}{}^{rr'}_{\alpha\beta}=
{\hat{\gamma}}{}^{r}_{(1,1,\alpha)}{\hat{\gamma}}{}^{r'}_{(2,2,\beta)},
\quad
{\hN_{2}}{}^{rr'}_{\alpha\beta}=
{\hat{\gamma}}{}^{r}_{(2,1,\alpha)}{\hat{\gamma}}{}^{r'}_{(1,2,\beta)}
\end{equation}
have the expectation values implying Pauli's exclusion principle
\begin{equation}
\label{R214}
\begin{array}{ll}
<2,2\mid{\hN_{1}}{}^{rr'}_{\alpha\beta}\mid 2,2>=
<1,2\mid{\hN_{2}}{}^{rr'}_{\alpha\beta}\mid 1,2>=\delta_{rr'}\delta_
{\alpha\beta},\\ \\
<1,1\mid{\hN_{1}}{}^{rr'}_{\alpha\beta}\mid 1,1>=
<2,1\mid{\hN_{2}}{}^{rr'}_{\alpha\beta}\mid 2,1>=0.
\end{array}
\end{equation}
The operators $\{{\hat{\gamma}}{}^{r}\}$
are the basis for tangent operator vectors 
$\hat{\Phi}(\zeta)={\hat{\gamma}}{}^{r}
\Phi_{r}(\zeta)$ of the 12 dimensional flat OM: $\,\,\hat{G}$,
where we introduce the vector function
belonging to the ordinary class of functions of $C^{\infty}$ smoothness 
defined on the manifold $G$:
$\quad \Phi_{r}^{(\lambda,\mu,\alpha)}(\zeta)=
\zeta^{(\lambda,\mu,\alpha)} \Phi_{r}^{\lambda,\mu}(\zeta),\quad
\zeta \in G$.
But, the operators $\{{\hat{\gamma}}{}_{r}\}$
is a dual basis for
operator covectors
$\bar{\hat{\Phi}}(\zeta)={\hat{\gamma}}{}_{r}
\Phi^{r}(\zeta)$, where
$\Phi^{r}=
{\bar{\Phi}}_{r}$ (charge conjugated).
Hence
\begin{equation}
\label{R215}
<\lambda,\mu\mid\hat{\Phi}(\zeta)\bar{\hat{\Phi}}(\zeta)\mid \lambda,\mu>=
{}^{*}\delta_{\lambda\tau}{}^{*}\delta_{\mu\nu}
\Phi_{r}^{(\tau,\nu,\alpha)}(\zeta)\Phi^{r}_{(\tau,\nu,\alpha)}(\zeta),
\end{equation}
for given $\lambda,\mu.$ Considering the  state vectors 
$\mid\chi_{\pm}>$  eq.(A.1.16) we get the matrix elements
\begin{equation}
\label{R217}
\begin{array}{l}
<\chi_{+}\mid\hat{\Phi}(\zeta)\bar{\hat{\Phi}}(\zeta)\mid\chi_{+}>
\equiv\Phi^{2}_{+}(\zeta)=\quad
\Phi_{r}^{(\lambda,1,\alpha)}(\zeta)\Phi^{r}_{(\lambda,1,\alpha)}(\zeta),\\ \\
<\chi_{-}\mid\hat{\Phi}(\zeta)\bar{\hat{\Phi}}(\zeta)\mid\chi_{-}>
\equiv\Phi^{2}_{-}(\zeta)=
\Phi_{r}^{(\lambda,2,\alpha)}(\zeta)\Phi^{r}_{(\lambda,2,\alpha)}(\zeta).
\end{array}
\end{equation}
The basis $\{{\hat{\gamma}}{}^{r}\}$
decomposes into
$\{ {\hgam_{i}}{}^{r} \}\quad (i=\eta,u),$ where
$${\hgam_{i}}{}^{r}_{(+\alpha)}=\FFr{1}{\sqrt{2}}
(\gamma^{r}_{(1,1\alpha)}+\varepsilon_{i}
\gamma^{r}_{(2,1\alpha)}),
\quad
{\hgam_{i}}{}^{r}_{(-\alpha)}=\FFr{1}{\sqrt{2}}
(\gamma^{r}_{(1,2\alpha)}+\varepsilon_{i}
\gamma^{r}_{(2,2\alpha)}).$$ 
The expansion of operator vectors $\hps_{i}\in\hG_{i}$ and 
operator covectors  $\bar{\hps_{i}}$ are written 
$\hps_{i}={\hgam_{i}}{}^{r}{\ps1_{i}}{}_{r},
\quad\bar{\hps_{i}}=
{\hgam_{i}}{}_{r}{\ps1_{i}}{}^{r},$
where the following vector functions of $C^{\infty}$ smoothness are 
defined on the manifolds $\G1_{i}:$ 
\begin{equation}
\label {R23}
{\ps1_{\eta} }{}_{r}^{(\pm\alpha)}(\eta,p_{\eta})=\eta^{(\pm\alpha)}
{\ps1_{\eta} }{}_{r}^{\pm}(\eta,p_{\eta}),\quad 
{\ps1_{u}}{}_{r}^{(\pm\alpha)}(u,p_{u})=u^{(\pm\alpha)}
{\ps1_{u}}{}_{r}^{\pm}(u,p_{u}).
\end{equation}
Namely, the probability of finding the vector function 
in the state $r$ with given sixvector of coordinate ($\eta$ or $u$) and 
momentum ($p_{\eta}$ or $p_{u}$) is determined by the square of its state 
wave function ${\ps1_{\eta} }{}_{r}^{\pm}(\eta,p_{\eta}),$ or 
${\ps1_{u}}{}_{r}^{\pm}(u,p_{u}).$ 
Due to the spin states, the ${\ps1_{i}}{}_{r}^{\pm}$ can be regarded as the 
Fermi field of the positive and negative frequencies 
${\ps1_{i}}{}^{\pm}_{r}\equiv{\ps1_{i}}{}^{r}_{\pm p}.$

\subsection {Realization of the Flat Manifold $G$} 
\label {quant}
The bispinor $\Psi(\zeta)$ defined on the manifold $G=\G1_{\eta}
\oplus\G1_{u}$ can be written  $\Psi(\zeta)=\ps1_{\eta}(\eta)\ps1_{u}(u)$,
where $\ps1_{i}$ is the bispinor defined on 
the manifold $\G1_{i}.$
The free state of $i$-type fermion with definite values of momentum
$p_{i}$ and spin projection $s$ is described by plane waves (App.A).
The relations of orthogonality and
completeness hold for the spinors.
Considering also the solutions of negative frequencies,
we make use of localized wave packets constructed by
means of superposition of plane wave solutions furnished by creation
and annihilation operators in agreement with Pauli's principle 
$$
{\hps_{i}}=\S_{\pm s}\IIn\frac{d^{3}p_{i}}{{(2\pi)}^{3/2}}
\left( {\hgam_{i}}_{(+\alpha)}{\ps1_{i}}^{(+\alpha)}+
{\hgam_{i}}_{(-\alpha)}{\ps1_{i}}^{(-\alpha)}\right), 
$$
etc, where the summation is extended over all dummy indices.
In such a manner we can treat as well the wave packets of operator vector
fields $\hat{\Phi}(\zeta)$. While the matrix element of the anticommutator 
of expansion coefficients reads
\begin{equation}   
\label{R35}
<\chi_{-}\mid \{ {\hgam_{i}}{}^{(+\alpha)}(p_{i},s),\,
{\hgam_{j}}{}_{(+\beta)}(p'_{j},s')\}\mid\chi_{-}>=
\varepsilon_{i}\delta_{ij}\delta_{ss'}\delta_{\alpha\beta}\delta^{(3)}
({\vec{p}}_{i}- {\vec{p'}}_{i}).
\end{equation}
In the aftermath, we get the most important relation
\begin{equation}
\label{R37}
\begin{array}{l}
\S_{\lambda=\pm}<\chi_{\lambda}\mid\hat{\Phi}(\zeta)
\bar{\hat{\Phi}}(\zeta)\mid
\chi_{\lambda}>= 
\S_{\lambda=\pm}<\chi_{\lambda}\mid
\bar{\hat{\Phi}}(\zeta)\hat{\Phi}(\zeta)\mid\chi_{\lambda}>= \\ \\
=-i\,\zeta^{2}\,\G1_{\zeta}(0)=-i\,\left(\eta^{2}\,\G1_{\eta}(0)-
u^{2}\,\G1_{u}(0)\right),
\end{array}
\end{equation}
where
$\G1_{i}(0)\equiv \Lm_{i\rightarrow i'}\G1_{i}(i-i'), \quad 
(i=\zeta,\eta, u,)$, etc., the Green's function 
$
\G1_{i}(i-i')=-(i\hpr_{i}+m)\Dlt_{i}(i-i')
$
is provided by the usual invariant singular functions 
$\Dlt_{i}(i-i').$ \\
Realization of the flat manifold $G$ ensued from the  
constraint imposed upon the matrix element eq.(2.2.2)
that, as the geometric object, it is required to be finite
\begin{equation}
\label{R919}
\S_{\lambda=\pm}<\chi_{\lambda}\mid\hat{\Phi}(\zeta)
\bar{\hat{\Phi}}(\zeta)\mid
\chi_{\lambda}> < \infty ,
\end{equation}
which gives rise to
\begin{equation}
\label{R919}
\zeta^{2}\,{\G1_{\zeta}}{}_{F}(0) < \infty ,
\end{equation}
and 
\begin{equation}
\label{R920}
\begin{array}{l}
{\G1_{\zeta}}{}_{F}(0)={\G1_{\eta}}{}_{F}(0)={\G1_{u}}{}_{F}(0)=\\
= \Lm_{u\rightarrow u'}\left[ -i\S_{{\vec{p}}_{u}}{\ps1_{u}}{}_{{p}_{u}}(u)
\,{\bp_{u}}{}_{{p}_{u}}(u')\,\theta (u_{0}-u'_{0})+
i\S_{{\vec{p}}_{u}}{\bp_{u}}{}_{{-p}_{u}}(u')\,{\ps1_{u}}{}_{{-p}_{u}}(u)
\,\theta (u'_{0}-u_{0}) \right],
\end{array}
\end{equation}
where the ${\G1_{\zeta}}{}_{F},{\G1_{\eta}}{}_{F}$ and ${\G1_{u}}{}_{F}$ are
causal Green's functions 
characterized by the boundary condition that only positive frequency
occur for $\eta_{0}>0\quad(u_{0}>0)$, only negative for
$\eta_{0}<0\quad(u_{0}<0)$. Here $\eta_{0}=\mid \vec{\eta}_{0}\mid $, 
$\eta_{0\alpha}=\FFr{1}{\sqrt{2}}
(\eta_{(+\alpha)}+\eta_{(-\alpha)})$ and the same holds for $u_{0}$.
Satisfying the condition eq.(2.2.4) the length of each vector
${\bf \zeta}=e\zeta\in G$
(see eq.(2.2.2)) compulsory must be equaled zero
\begin{equation}
\label{R921}
\zeta^{2}=\eta^{2}-u^{2}=0.
\end{equation}
Thus, the requirement eq.(2.2.3)
provided by eq.(2.2.5) yields the realization of the flat
manifold $G$, which subsequently leads to Minkowski flat
space $M_{4}$ (subsec.2.1) where, according to eq.(2.2.6),
the relativity principle holds
$$
\left. d\,\eta^{2}\right|_{6\rightarrow 4}\equiv d\,s^{2}=
d\,u^{2}=inv.
$$
\renewcommand {\theequation}{\thesection.\arabic {equation}} \section{Mathematical Background}
\label {appe}
$\bullet$ {\bf Field Aspect}\\\\
The quantum field theory of the OM
is equivalent to configuration space wave mechanics employing the 
antisymmetric state functions incorporated with geometric 
properties of corresponding objects (see App.A or [1]. 
Therein, by applying the algebraic approach we reach to rigorous definition 
of the OM: $\,\hat{G}$, construct the explicit forms of wave state 
functions and calculate the matrix elements of field operators.
While, the $\hat{G}$ reads
\begin{equation}
\label{R330}
\hat{G}=\S^{\infty}_{n=0}\hat{G}^{(n)}=
\S^{\infty}_{n=0}\left(\hat{\cal U}^{(n)}\otimes {\bar{\cal H}}^{(n)}\right),
\end{equation}
where 
$
\hat{\cal U}^{(n)}_{(r_{1},\ldots,r_{n})}=\hat{\cal U}^{(1)}_{r_{1}},
\otimes\cdots\otimes \hat{\cal U}^{(1)}_{r_{n}}
$ 
is the open neighbourhood of the n-points $\hat{\zeta}_{r_{i}}$ of the OM, 
$\,\,
{\bar{\cal H}}^{(n)}_{(r_{1},\ldots,r_{n})}=
{\cal H}^{(1)}_{r_{1}}\otimes\cdots\otimes
{\cal H}^{(1)}_{r_{n}}
$
is the Hilbert space for description of n particle system. 
Meanwhile, one has to modify the basis operators (the creation 
${\hat{\gamma}}{}_{r}$ and annihilation ${\hat{\gamma}}{}^{r}$ operators) 
in order to provide an anticommutation in arbitrary states. For example,
acting on free state $\mid 0>_{r_{i}}$ 
the creation operator ${\hat{\gamma}}_{r_{i}}$
now yields the one occupied state $\mid 1>_{r_{i}}$ with the phase 
$'+'$ or $'-'$ depending of parity of the number
of quanta in the states $ r < r_{i}$.
Modified operators satisfy the same anticommutation relations
of the basis operators (eq.(A.1.12)). 
Defining the secondary quantized form of one particle observable $A$ on 
the $\cal H$ we consider a set of identical samples $\hat{\cal H}_{i}$ of
one particle space ${\cal H}^{(1)}$ and operators $A_{i}$ acting on them.
The vacuum state given in eq.(A.1.16) satisfies the normalization condition. 
The state vectors eq.(A.1.17) are the eigenfunctions
of modified operators. They form a whole set of orthogonal vectors.
Considering an arbitrary superposition of state vectors
we get a whole set of explicit forms of the matrix elements of operator 
vector and covector fields (App.A).\\\\
$\bullet$ {\bf Differential Geometric Aspect}\\\\
For illustrative purposes here we consider a few examples from the 
differential geometric aspect of the OM by referring to the Appendix A for 
more details,.\\
The operators $\{ \hat{\gamma}{}^{r} \}$ are
the basis for all operator vectors of tangent section ${\hat{\bf T}}
_{\Phi_{p}}$
of principle bundle with the base $\hat{G}$ at the point 
${\bf \Phi}_{p}=\left. {\bf \Phi}({\bf \zeta}(t))\right|_{t=0} \in
\hat{G}$.
The smooth field of tangent operator vector 
$\hat{\bf A}({\bf \Phi}({\bf \zeta}))$ is a class of equivalence of the 
curves ${\bf f}({\bf \Phi}({\bf \zeta}))$,
${\bf f}({\bf \Phi}({\bf \zeta}(0)))={\bf \Phi}_{p}$.
While, the operator differential $\hat{d}\,A^{t}_{p}$ of the flux
$A^{t}_{p}:\hat{G} \rightarrow \hat{G}$ at the point
${\bf \Phi}_{p}$ with the velocity fields 
$\hat{\bf A}({\bf \Phi}({\bf \zeta}))$ is defined by one parameter
group of operator diffeomorphisms given for the curve 
${\bf \Phi}({\bf \zeta}(t)):R^{1} \rightarrow \hat{G}$. Provided one has
${\bf \Phi}({\bf \zeta}(0))={\bf \Phi}_{p}$ and
$\widehat{\dot{{\bf \Phi}}}({\bf \zeta}(0))=\hat{\bf A}_{p}$
\begin{equation}
\label{R41}
\hat{d}\,A^{t}_{p}({\bf A})=\left.\FFr{\hat{d}}{d\,t}\right|_{t=0}
A^{t}({\bf \Phi}({\bf \zeta}(t)))=\hat{\bf A}({\bf \Phi}({\bf \zeta}))=
\hat{\gamma}{}^{r}A_{p},
\end{equation}
where the $\{ A_{p}\}$ are the components of $\hat{\bf A}$ in the basis 
$\{ \hat{\gamma}{}^{r}\}$. According to eq.(3.2),
in holonomic coordinate basis 
$ \hat{\gamma}{}^{r} \rightarrow \left(
\hat{\partial}\left. \right/ \partial \Phi_{r}
({\bf \zeta}(t))\right)_{p}$ one gets
$
A_{p}=\left. \FFr{\partial \Phi_{r}}{\partial \zeta_{r}}
\FFr{d\,\zeta_{r}}{d\,t}\right|_{p}.
$
Hence, for any function $f:{\bf R}^{n}\rightarrow {\bf R}^{n}$ of the
ordinary class of functions of $C^{\infty}$ smoothness on $\hat{G}$ one 
may define an operator differential 
$$
<\hat{d}\,f,\hat{\bf A}>=\widehat{(A\,f)},
$$
by means of
smooth reflection 
$$\hat{d}\,f:\hat{\bf T}\left(\hat{G}\right)
\rightarrow \hat{R}\quad \left(\hat{\bf T}\left(\hat{G}\right)=
\displaystyle \bigcup_{\Phi_{p}}\hat{\bf T}_{\Phi_{p}}\right),
$$ 
where (see eq.(A.1.25))
\begin{equation}
\label{R416}
<\chi\|\hat{d}\,f,\hat{\bf A} \|\chi^{0}>=\S_{\lambda,\mu=1}^{2}
\S_{r^{\lambda\mu}=1}^{N_{\lambda\mu}} \hat{c}^{*}(r^{\lambda\mu})
<d\,f,{\bf A}>_{r^{\lambda\mu}}
=\S_{\lambda,\mu=1}^{2}
\S_{r^{\lambda\mu}=1}^{N_{\lambda\mu}} \hat{c}^{*}(r^{\lambda\mu})
\,({\bf A}\,f)_{r^{\lambda\mu}}.
\end{equation}
In coordinate basis
$
<d\, \Phi^{\widehat{\imath}}, \hat{\partial}\left.\right/ \partial 
\Phi^{j}>=\FFr{\partial \Phi^{\widehat{\imath}}}{\partial \Phi^{j}}=
\delta^{{\widehat{\imath}}}_{j},
$
provided by $d\, \Phi^{\widehat{\imath}}\equiv \hat{d}\Phi^{i}$ and\\
$
<\chi\|\hat{\delta}{}^{\imath}_{j} \|\chi^{0}>=\S_{\lambda,\mu=1}^{2}
\S_{r^{\lambda\mu}=1}^{N_{\lambda\mu}} \hat{c}^{*}(r^{\lambda\mu})
\delta^{i}_{j},$ 
where the $i$ and $j$ stand for a set of 
$(\lambda_{i},\mu_{i},\alpha_{i})$.
\\
The operator tensor $\hat{\bf T}$ of 
$\widehat{(n,0)}$-type at the point ${\bf \Phi}_{p}$ is a linear
function of the space
$
\hat{\bf T}^{n}_{0}=\underbrace{\hat{\bf T}_{\Phi_{p}}\otimes\cdots \otimes 
\hat{\bf T}_{\Phi_{p}}}_{n},
$
where the $\otimes $ denotes the tensor product.
It enables a correspondence between the element $(\hat{\bf A}_{1},\ldots,
\hat{\bf A}_{n})$ of $\hat{\bf T}^{n}_{0}$ and the number $T(\hat{\bf A}_{1},
\ldots,\hat{\bf A}_{n})$ furnished by linearity. 
Constructing matrix 
elements of operator tensors of $\hat{G}$ one produces the 
Cartan's exterior forms (A.1.28). Whence, the matrix elements of symmetric 
operator tensors equal zero. 
The differential operator
$n$ form $\left.{\bf \hat{\omega}}^{n}\right|_{\Phi_{p}}$ at the point
${\bf \Phi}_{p} \in \hat{G}$ can be defined as the exterior operator $n$ 
form on tangent operator space $\hat{\bf T}_{\Phi_{p}}$ of tangent
operator vectors $\hat{\bf A}_{1},\ldots,\hat{\bf A}_{n}$.
That is, if the $\wedge \hat{\bf T}^{*}_{\Phi_{p}}
\left(\hat{G}\right)$ means the exterior algebra on
$\hat{\bf T}^{*}_{\Phi_{p}}\left(\hat{G}\right)$, then
operator $n$ form $\left.{\bf \hat{\omega}}^{n}\right|_{\Phi_{p}}$
is an element of $n$-th degree out of $\wedge \hat{\bf T}^{*}_{\Phi_{p}}$
depending of the point ${\bf \Phi}_{p} \in \hat{G}$.
Hence ${\bf \hat{\omega}}^{n}=\displaystyle \bigcup_{\Phi_{p}}
\left.{\bf \hat{\omega}}^{n}\right|_{\Phi_{p}}$. Any differential operator
$n$ form of dual operator space
$\underbrace{\hat{\bf T}^{*}_{\Phi_{p}}\otimes\cdots \otimes 
\hat{\bf T}^{*}_{\Phi_{p}}}_{n}$ may be written 
$
{\bf \hat{\omega}}^{n}=\S_{i_{1}<\cdots<i_{n}}\alpha
_{i_{1} \cdots i_{n}} (\Phi)d\,\Phi^{\widehat{\imath}_{1}}\wedge\cdots\wedge
d\,\Phi^{\widehat{\imath}_{n}},
$
provided by the smooth differentiable functions $\alpha
_{i_{1} \cdots i_{n}}(\Phi)\in C^{\infty}$ and basis
$
d\,\Phi^{\widehat{\imath}_{1}}\wedge\cdots\wedge
d\,\Phi^{\widehat{\imath}_{n}}=
\S_{\sigma\in S_{n}}sgn(\sigma)
\gamma^{\sigma(\widehat{\imath}_{1}}\otimes\cdots\otimes
\gamma^{\widehat{\imath}_{n})}.
$
The matrix elements of some of the geometric objects of the $\hat{G}$ 
are given in the App.A.

\section{Beyond the Geometry and Fields}
\label{goyaks}
To facilitate the physical picture and provide sufficient background it
seems worth to bring few formal matters in concise form which one will 
have to know in order to understand the general structure of our approach
without undue hardship. Here we only outline briefly the relevant steps.
In the mean time we refer to [1] for more detailed justification of 
some of the procedures and complete exposition.
Before proceeding further, it is profitable to define 
the {\em pulsating gauge functions} 
and {\em fields} denoted by wiggles as follows:

$\bullet$ The function $\widetilde{W}(x)$ 
defined on the space $M$ ($x\in M$ ) and being   
an invariant with respect to the coordinate transformations 
is called the pulsating gauge function if it undergoes local gauge 
transformations
\begin{equation}
\label{R441}
\widetilde{W}'(x)=U(x)\widetilde{W}(x).
\end{equation}
Here $U(x)$ is the element of some simple Lie group $G$ the generators of 
which imply the algebra $[F^{a},F^{b}]=iC^{abc}F^{c}$, where $C^{abc}$
are wholly antisymmetric structure constants.

$\bullet$ A smooth function $\widetilde{\Phi}(\widetilde{W}(x))$ belonged to
some representation of the group $G$, where the generators are presented
by the matrices $T^{a}$, is called the 
pulsating field if under the transformation eq.(4.1) it transforms
\begin{equation}
\label{R441}
\widetilde{\Phi}'\equiv
\widetilde{\Phi}(\widetilde{W}'(x))=U(x)\widetilde{\Phi}(\widetilde{W}(x)).
\end{equation}
Let $L_{0}(\Phi,\partial\Phi)$ is the invariant Lagrangian of free field
$\Phi$ defined on $M$. Then, a simple gauge invariant Lagrangian of the 
pulsating field $\widetilde{\Phi}$ can be written
\begin{equation}
\label{R441}
L={\widetilde{W}}^{+}\widetilde{W}L_{0}(\Phi,\partial\Phi),
\end{equation}
which reduces to
\begin{equation}
\label{R441}
L\equiv L\left(\widetilde{\Phi},\widetilde{D\Phi}\right)=
L_{0}\left(\widetilde{\Phi}(\widetilde{W}(x)),
D\widetilde{\Phi}(\widetilde{W}(x))\right).
\end{equation}
Here we have noticed that due to eq.(4.1) and eq.(4.2)  
$\widetilde{\Phi}(\widetilde{W})=\widetilde{W}\Phi$, and introduced the 
covariant derivative 
$\widetilde{D\Phi}\equiv D\widetilde{\Phi}(\widetilde{W})=
\widetilde{W}\partial\Phi.$
Whence
\begin{equation}
\label{R441}
D=\partial - igT^{a}W^{a}, \quad T^{a}W^{a}= -\FFr{i}{g}\partial
\ln \widetilde{W},
\quad D\widetilde{W}=\left(D\widetilde{W}\right)^{+}=0,
\end{equation}
where $W^{a}$ is the gauge field, g is the coupling constant.
Hence, all the conventional matter fields interacting by gauge fields are
the pulsating fields. 

\renewcommand {\theequation}{\thesubsection.\arabic {equation}}

\subsection{The Regular Primordial Structures}
In [1] we have chosen a simple setting and considered the primordial 
structures designed to possess certain physical properties satisfying the 
stated general rules. These structures are the substance out of which the 
geometry and particles are made.
We distinguish the ``$\eta$- and $u$-types
primordial structures'' involved in the linkage establishing processes 
occurred between the structures of different types.
Let us recall that the $\eta$-type structure may accept the linkage
only from $u$-type structure, which is described by the link
function 
${\ps1_{\eta} }(s)$
belonged to the ordinary class of functions of $C^{\infty}$ smoothness, 
where $s\equiv\eta={\e1_{\eta}}{}_{(\lambda\alpha)}
\eta^{(\lambda\alpha)},\quad (\lambda = \pm; \alpha = 1,2,3,$ see subsec.2.1),
$\eta$ is the link coordinate.
Respectively the $u$-type structure may accept the linkage only
from $\eta$-type structure described by the link function
$
{\ps1_{u}}(s)$ (u-channel), where
$s\equiv u=\e1_{u} u $.
We assume that $s$ is the pulsating gauge function associated with the Abelian local gauge group $U(1)$ and  $\Psi(s)$ is the
pulsating field (the wiggles are left implicit). Thus, 
under local gauge transformations 
$$
s'=e^{-i\alpha}s,\quad \partial\alpha\neq 0,
$$
the link function 
$\Psi(s)$ transforms 
$$\Psi(s')=e^{-i\alpha}\Psi(s),$$
and the Lagrangian eq.(4.3) is invariant under gauge 
transformations.
It includes the covariant derivative 
$D(s)=\partial +ig\,b(s)$ and
gauge field $b(s)=\FFr{i}{g}\partial \ln s$ undergone gauge transformations
$
b(s')=b(s)+\FFr{1}{g}\,\partial\,\alpha.
$
Then,  ${\ps1_{i}}(s)=s{\ps1_{i}}=
{\e1_{i}}{}_{(\lambda\alpha)}{\ps1_{i}}^{(\lambda\alpha)}$ ($i=\eta, u$), 
where the eq.(2.1.8) holds
\begin{equation}
\label {eq: R2.2}
{\ps1_{\eta} }^{(\pm\alpha)}(\eta,p_{\eta})=
\eta^{(\pm\alpha)}
{\ps1_{\eta} }^{\pm}(\eta,p_{\eta}),\quad 
{\ps1_{u} }^{(\pm\alpha)}(u,p_{u})=
u^{(\pm\alpha)}
{\ps1_{u} }^{\pm}(u,p_{u}),
\end{equation}
a bispinor ${\ps1_{i} }^{\pm}$ is the invariant state wave function of 
positive or negative frequencies, $p_{i}$ is the corresponding link momentum.
Thus, a primordial structure can be considered as
a fermion found in external gauge field 
$b(s). $\\
The simplest system made of two structures of different types becomes
stable only due to the stable linkage
\begin{equation}
\label {R72}
\left|\p1_{\eta}\right|={({\p1_{\eta}}^{(\lambda\alpha)},
{\p1_{\eta}}{}_{(\lambda\alpha)})}^{1/2}=
\left|\p1_{u}\right|={({\p1_{u}}^{(\lambda\alpha)},
{\p1_{u}}{}_{(\lambda\alpha)})}^{1/2}.
\end{equation}
Otherwise they are unstable.
There is not any restriction on the number of primordial structures
of both types getting into the link establishing processes simultaneously.
In the stable system the link stability condition must be
held for each linkage separately.\\
The persistent processes of creation and annihilation of the primordial 
structures occur in different states $s, s',s'',...$ The "creation"
of structure in the given state $(s)$ is due to its transition to 
this state from other states $(s',s'',...)$, while the "annihilation"
means a vice versa.
Satisfying eq.(4.1.2) the primordial structures from the arbitrary states 
may establish a stable linkage.
Among the states $(s,s',s'',...)$ there is a lowest one ($s_{0}$),
in which all structures are regular. That is, they are in free (pure) state
and described by the plane wave functions
${\ps1_{\eta} }^{\pm }(\eta_{f}, p_{\eta})$ or 
${\ps1_{u} }^{\pm }(u_{f},p_{u})$
defined respectively on flat manifolds
$\G1_{\eta}$ and $\G1_{u}$. The index (f) specifies the points of 
corresponding flat manifolds $\eta_{f}\in\G1_{\eta}$, $u_{f}\in
\G1_{u}$. For example, in accordance with subsec.2.2, the equation of 
regular structure $\Psi(s_{+})\quad (s=s_{+}+s_{-})$ reads
$$
\left[i\gamma_{f}(\partial +igb(s_{+})) -m\right]\Psi(s_{+})=0,
$$
the matrices $\gamma_{f}$ are given in eq.(B.1.4).
Whence the equation of plane wave function  
$\Psi^{+}_{p}$ of positive frequencies stems
$$
(i\gamma_{f} \partial - m)\Psi^{+}_{p}=0.
$$
The processes of creation and annihilation of regular structures in lowest 
state are described by the OM formalism given in the previous sections.

\subsection{The Distorted  Primordial Structures}
\label{Vec}
In all the higher states the  primordial structures are distorted ones 
(interaction states) and described by distorted link functions defined on 
distorted manifolds $\widetilde{\G1_{\eta}}$ and $\widetilde{\G1_{u}}$.
A distortion 
$G\rightarrow \widetilde{G}$ 
with hidden Abelian local group
$G=U^{loc}(1)=SO^{loc}(2)$ and one dimensional trivial algebra
$\hat{g}=R^{1}$ has studied in [43].
It involves a drastic revision of a role of 
local internal symmetries in the concept of curved geometry.
Under the reflection of fields and their dynamics from Minkowski space 
to Riemannian a standard gauge principle of local internal symmetries is 
generalized. The gravitation gauge group is proposed, which is generated 
by hidden local internal symmetry.
This suggests an opportunity for the unification of all 
interactions on an equal footing.
Our scheme is implemented as follows:
Considering the principle bundle $p:E\rightarrow G$
the basis $e^{f}$ is transformed 
$
e=D\,e^{f},
$
under massless gauge distortion field $a_{f}$ associated with $U^{loc}(1)$.  
The matrix $D$ is in the form $D=C\otimes R$, where the distortion
transformations $O_{(\lambda\alpha)}=C^{\tau}_{(\lambda\alpha)}
O_{\tau}$ and
$\sigma_{(\lambda\alpha)}=R^{\beta}_{(\lambda\alpha)}
\sigma_{\beta}$ are defined. Here $C^{\tau}_{(\lambda\alpha)}=
\delta^{\tau}_{\lambda} + \kappa a_{(\lambda\alpha)}{}^{*}
\delta^{\tau}_{\lambda}$,
but $R$ is a matrix of the group $SO(3)$ of ordinary rotations of 
the planes involving two arbitrary basis vectors 
of the spaces $R^{3}_{\pm}$ around  the orthogonal third axes. 
The rotation angles are determined from the constraint imposed
upon distortion transformations that a sum of distorted parts
of corresponding basis vectors $O_{\lambda}$ and 
$\sigma_{\beta}$ should be zero for given $\lambda$ (App.B).
Whence $\tan\theta_{(\lambda\alpha)}=-\kappa a_{(\lambda\alpha)}$,
where $\theta_{(\lambda\alpha)}$ is the particular rotation around the axis
$\sigma_{\alpha}$. Next we construct
the diffeomorphism $G\rightarrow \widetilde{G}$ and introduce the invariant 
action of the fields.
The passage from six dimensional curved manifold $\widetilde{G}$
to four dimensional Riemannian geometry $R^{4}$ is straightforward 
(subsec.2.1).
Given a diffeomorphism
$u(u_{f}):\G1_{u}\rightarrow \widetilde{\G1_{u}}$ 
we consider the reflection
of the Fermi fields and their dynamics from the flat manifold $\G1_{u}$ 
to distorted one $\widetilde{\G1_{u}}$, and vice versa (App.B). 
In the aftermath, the relation between the wave functions of distorted 
and regular structures reads 
\begin{equation}
\label{eq: R2.4}
{\ps1_{u}}^{\lambda}(\theta_{+k})=
f_{(+)}(\theta_{+k})\,{\ps1_{u}}^{\lambda}, 
\quad
{\ps1_{u}}{}_{\lambda}(\theta_{-k})=
{\ps1_{u}}{}_{\lambda}\,f_{(-)}(\theta_{-k}).
\end{equation}
The ${\ps1_{u}}^{\lambda}({\ps1_{u}}{}_{\lambda})$ is the plane wave function 
of regular ordinary structure (antistructure) and
\begin{equation}
\label{eq: R2.5}
f_{(+)}(\theta_{+k})=e^{\chi_{R}(\theta_{+k})-
i\chi_{J}(\theta_{+k})}, \quad
f_{(-)}(\theta_{-k})=\left.f_{(+)}^{*}(\theta_{+k}) 
\right|_{\theta_{+k}=\theta_{-k}},
\end{equation}
where the $\chi_{R}$ and $\chi_{J}$ are given in Appendix C.\\
Next, we supplement the previous assumptions made in sec. 4 by a new one that 
now the $\eta$-type (fundamental) regular structure can not directly form
a stable system with the regular $u$-type (ordinary) structures.
Instead of it the $\eta$-type regular structure 
forms a stable system with the infinite number of distorted ordinary 
structures, where the link stability 
condition held for each linkage separately. Such structures take part 
in realization of the flat manifold $G$ (subsec.2.2). 
The laws regarding to this change apply in use of 
functions of distorted ordinary structures 
\begin{equation}
\label{eq:R2.6}
{\ps1_{u}}^{(\lambda\alpha)}(\theta_{+})=u^{(\lambda\alpha)}
{\ps1_{u}}^{\lambda}(\theta_{+}),\quad
{{\ps1_{u}}}{}_{(\lambda\alpha)}(\theta_{-})=u_{(\lambda\alpha)}
{\ps1_{u}}{}_{\lambda}(\theta_{-}),
\end{equation}
where $u \in \widetilde{\G1_{u}}$.
For our immediate purposes we employ the wave packets constructed by 
superposition of these functions furnished by generalized operators of 
creation and annihilation as the expansion coefficients
\begin{equation}
\label{eq: R2.11}
\begin{array}{l}
{\hps_{u}}(\theta_{+})=\S_{\pm s}\IIn\frac{d^{3}p_{u}}{{(2\pi)}^{3/2}}
\left( {\hgam_{u}}{}_{(+\alpha)}^{k}\,{\ps1_{u}}{}^{(+\alpha)}(\theta_{+k})+
{\hgam_{u}}{}_{(-\alpha)}^{k}\,{\ps1_{u}}{}^{(-\alpha)}(\theta_{+k})\right), 
\\ \\
{\bar{\hps_{u}}}(\theta_{-})=\S_{\pm s}\IIn\frac{d^{3}
p_{u}}{{(2\pi)}^{3/2}}
\left( {\hgam_{u}}{}^{(+\alpha)}_{k}\,{\ps1_{u}}{}_{(+\alpha)}(\theta_{-k})+
{\hgam_{u}}{}^{(-\alpha)}_{k}\,{\ps1_{u}}{}_{(-\alpha)}(\theta_{-k})\right),
\end{array}
\end{equation}
where as usual the summation is extended over all dummy indices.
The matrix element of anticommutator of generalized 
expansion coefficients reads 
\begin{equation}
\label{eq: R2.12}
<\chi_{-}\mid \{ {\hgam_{u}}{}^{(+\alpha)}_{k}(p,s),
\,{\hgam_{u}}{}_{(+\beta)}^{k'}(p',s') \}\mid\chi_{-}>= 
-\delta_{ss'}\delta_{\alpha\beta}\delta_{kk'}\delta^{3}(\vec{p}-\vec{p'}).
\end{equation}
The wave packets eq.(4.2.4) yield the causal Green's function
${\G1_{u}}^{\theta}_{F}(\theta_{+} - \theta_{-})$ of distorted 
ordinary structure.
Geometry realization requirement (eq.(2.2.5)) now should be satisfied for 
each ordinary structure in terms of
\begin{equation}
\label{eq: R2.13}
{\G1_{u}}{}^{\theta}_{F}(0)=\Lm_{\theta_{+} \rightarrow \theta_{-}}
{\G1_{u}}{}^{\theta}_{F}(\theta_{+} - \theta_{-})
={\G1_{\eta}}{}_{F}(0)= \Lm_{\eta'_{f} \rightarrow \eta_{f}}
{\G1_{\eta}}{}_{F}(\eta'_{f}-\eta_{f}).
\end{equation}
They are valid if following relations hold for
each distorted ordinary structure:
\begin{equation}
\label{eq: R2.14}
\S_{k}{\ps1_{u}}(\theta_{+k})
\,{\bar{\ps1_{u}}}(\theta_{-k})
=\S_{k}{\ps1_{u}}'(\theta'_{+k})
\,{\bar{\ps1_{u}}}'(\theta'_{-k})=
\cdots = inv. 
\end{equation}
Namely, the distorted ordinary structures have met in the permissible
combinations to realize the geometry in the stable system.
Below, in simplified schematic way we exploit the background of the
known colour confinement and gauge principles.
This scheme still should be considered as a preliminary one, which will be 
further elaborated in the sec.5.

\subsection{``Quarks'' and ``Colour'' Confinement}
\label{quark}
At the very first to avoid irrelevant complications, here, for illustrative 
purposes, we will attempt to 
introduce temporarily skeletonized ``quark'' and 
``antiquark'' fields emerged in confined phase in the simplified  
geometry with the one-u channel given in the previous subsections. 
The complete picture of such a dynamics is beyond the scope of this 
subsection, but some relevant discussions on this subject will also be 
presented in the subsec.5.3.
We may think 
of the function ${\ps1_{u}}{}^{\lambda}(\theta_{+k})$ at fixed $(k)$ as 
being the $u$-component of bispinor field of ``quark'' ${q}_{k}$, and of
${\bar{\ps1_{u}}}{}_{\lambda}(\theta_{-k})$ - the $u$-component of 
conjugated bispinor field of ``antiquark'' ${\bar{q}}_{k}$.
The index $(k)$ refers to colour degree of freedom in the 
case of rotations through the angles  $\theta_{+k}$ and anticolour 
degree of freedom in the case of $\theta_{-k}$.
The $\eta$-components of quark fields are plane waves.
In both cases of local and global rotations we respectively distinguish 
two types of quarks: local ${q}_{k}$ and global ${q}_{k}^{c}$.
Hence, the quark is a fermion with the half integer spin and certain 
colour degree of freedom. There are exactly three colours.
The rotation through the angle $\theta_{+k}$ yields
a total quark field defined on the flat manifold $G=\G1_{\eta}\oplus\G1_{u}$
\begin{equation}
\label{eq: R2.15}
{q}_{k}(\theta)=\Psi(\theta_{+k})={\ps1_{\eta}}^{0}
\ps1_{u}(\theta_{+k})
\end{equation}
where ${\ps1_{\eta}}^{0}$ is a plane wave defined on $\G1_{\eta}$.
According to eq.(4.2.1), one gets
\begin{equation}
\label{eq: R2.16}
{q}_{k}(\theta)={\ps1_{\eta}}^{0}{\q1_{u}}_{k}(\theta)=
{\q1_{\eta}}_{k}(\theta){\ps1_{u}}^{0},
\quad {\q1_{\eta}}_{k}(\theta)\equiv f_{(+)}(\theta_{+k}){\ps1_{\eta}}^{0}, 
\end{equation}
where ${\ps1_{u}}^{0}$ is a plane wave, ${\q1_{u}}{}_{k}(\theta)$ and
${\q1_{\eta}}{}_{k}(\theta)$ may be considered as the quark fields with the 
same quantum numbers 
defined respectively on flat manifolds $\G1_{u}$ and  $\G1_{\eta}$. 
By making use of the rules stated in subsec.2.1
one may readily return to Minkowski space
$\G1_{\eta}\rightarrow M_{4}$. In the sequel, the quark 
field defined on $M_{4}$ will be ensued
${\q1_{\eta}}{}_{k}(\theta) \rightarrow q_{k}(x)$, $x \in M_{4}$.
Due to eq.(4.2.7) and eq.(4.3.1) they imply
\begin{equation}
\label{eq: R2.17}
\S_{k}{q}_{k}\,{\bar{q}}_{k} =
\S_{k}{q'}_{k}\,{\bar{q'}}_{k}=
\cdots = inv,
\end{equation}
namely
\begin{equation}
\label{eq:R2.18}
\S_{k}f_{(+)}(\theta_{+k})f_{(-)}(\theta_{-k} )
=\S_{k}f'_{(+)}(\theta'_{+k})f'_{(-)}(\theta'_{-k} )=
\cdots =inv.
\end{equation}
The eq.(4.3.3) utilizes the idea of colour (quark) confinement 
principle: the quarks emerge in the geometry only in special combinations
of colour singlets.
Only two colour singlets are available (see below)
\begin{equation}
\label{eq: R2.20}
(q\bar{q})=\FFr{1}{\sqrt{3}}\delta_{kk'}{\hat{q}}_{k}{\bar{\hat{q}}}_{k'}=
inv, 
\quad
(qqq)=\FFr{1}{\sqrt{6}}\varepsilon_{klm}{\hat{q}}_{k}{\hat{q}}_{l}
{\hat{q}}_{m}=inv.
\end{equation}
These results will be generalized in the next section where the physically 
more realistic scheme of the MW geometry with the multi u-channel should be 
subject for discussion. 

\subsection {Gauge Principle; Internal Symmetries}
\label {gauge}
Following [44,45], the principle of identity holds for ordinary
regular structures, namely each regular structure in the lowest state
can be regarded as a result of transition from an
arbitrary state, in which they assumed to be distorted. 
This is stated in terms of link-functions below
\begin{equation}
\label{eq: R2.22}{\ps1_{u}}{}_{\lambda}=f^{-1}_{(+)}(\theta_{+k})
{\ps1_{u}}{}_{\lambda}(\theta_{+k})=
f^{-1}_{(+)}(\theta'_{+l})
{\ps1_{u}}{}'_{\lambda}(\theta'_{+l})=
\cdots. 
\end{equation}
Hence, the following transformations may be implemented upon
distorted ordinary structures occurred in the stable system:
\begin{equation}
\label{eq: R2.21}
\begin{array}{l}
{\ps1_{u}}'^{\lambda}(\theta'_{+l})=
f^{(+)}_{lk}{\ps1_{u}}^{\lambda}(\theta_{+k})=
f(\theta'_{+l},\theta_{+k}){\ps1_{u}}^{\lambda}(\theta_{+k}),\\\\ 
{\ps1_{u}}{}'_{\lambda}(\theta'_{-l})=
{\ps1_{u}}{}_{\lambda}(\theta_{-k})f^{(-)}_{kl}=
\left.{\ps1_{u}}{}_{\lambda}(\theta_{-k})f^{*}(\theta'_{-l},\theta_{-k})
\right|_{\begin{array}{l}
\theta'_{-l}=\theta'_{+l}\\
\theta_{-k}=\theta_{+k},
\end{array}}
\end{array}
\end{equation}
provided by
\begin{equation}
\label{eq: R2.22}
f^{(+)}_{lk}=\exp \{ \chi^{R}_{lk}-i\chi^{J}_{lk} \}, 
\quad \left. f^{(-)}_{kl}={(f^{(+)}_{lk})}^{*}
\right|_{\begin{array}{l}
\theta'_{-l}=\theta'_{+l}\\
\theta_{-k}=\theta_{+k},
\end{array}},
\end{equation} 
\begin{equation}
\label{eq: R2.22}
\chi^{R}_{lk}=\chi_{R}(\theta'_{+l})-\chi_{R}(\theta_{+k}),
\quad
\chi^{J}_{lk}=\chi_{J}(\theta'_{+l})-\chi_{J}(\theta_{+k}).
\end{equation}
The transformation functions are the operators in the space 
of internal degrees of freedom labeled by $(\pm k)$ corresponding to 
distortion rotations around the axes $(\pm k)$ by the
angles  $\theta_{\pm k}$. 
We make proposition that the distortion rotations are incompatible, namely the
transformation operators $f^{(\pm)}_{lk}$ obey the incompatibility relations
 \begin{equation}
\label{eq: R2.25}
\begin{array}{l}
f^{(+)}_{lk}f^{(+)}_{cd}-f^{(+)}_{ld}f^{(+)}_{ck}=\|f^{(+)}\|
\varepsilon_{lcm}\varepsilon_{kdn}f^{(-)}_{nm},\\ \\
f^{(-)}_{kl}f^{(-)}_{dc}-f^{(-)}_{dl}f^{(-)}_{kc}=\|f^{(-)}\|
\varepsilon_{lcm}\varepsilon_{kdn}f^{(+)}_{mn},
\end{array}
\end{equation}
where $l,k,c,d,m,n=1,2,3$.
The relations eq.(4.4.5) hold in general for both the local and global 
rotations. In the following we shall often be concerned with these most 
important relations.
Making use of eq.(4.3.1), eq.(4.3.2) and eq.(4.4.2), one gets the 
transformations 
implemented upon the quark field, which in matrix notation take the form
$
q'(\zeta)=U(\theta(\zeta))\,q(\zeta), 
\quad
\bar{q'}(\zeta)=\bar{q}(\zeta)\,U^{+}(\theta(\zeta)),
$
where $q = \{ {q}_{k} \}, \quad U(\theta)=\{ f^{(+)}_{lk}  \}$. 
Under the incompatibility commutation relations
(4.4.5), the transformation matrices $\{ U \}$  generate
the unitary group of internal symmetries $U(1), SU(2),$ 
$SU(3)$. 
Since the distorted ordinary structures have made contribution
in the realization of geometry $G$ instead of
regular ones, then, stated somewhat differently
the principle of identity of regular structures 
directly leads to the equivalent principle (the gauge principle):
an action integral of any dynamical physical system must be invariant under
arbitrary transformations eq.(4.4.2).
Below we discuss different possible models.\\
1. In the simple case of one dimensional local
transformations through the local angles $\theta_{+1}(\zeta)$ and
$\theta_{-1}(\zeta) one has $
$\quad
f^{(+)}=\left( \matrix{
f^{(+)}_{11}  & 0  & 0 \cr 
0             & 1  & 0 \cr
0             & 0  & 1\cr
}\right), \quad
f^{(-)}={(f^{(+)})}^{+}.
$
The incompatibility relations eq.(4.4.5) 
reduce to identity $f^{(+)}_{11}=\|f^{(+)}\|.$
At
$\chi_{R}(\theta_{+1})=\chi_{R}(\theta_{-1})$
the transformations
\begin{equation}
\label{eq: R2.28}
f^{(+)}_{11}=U(\theta)=f(\theta_{+1}(\zeta),\theta_{-1}(\zeta))=
\exp \{ -i\chi^{(+)}_{J}(\theta_{+1})+i\chi^{(-)}_{J}(\theta_{-1}) \}.
\end{equation}
generate a commutative
Abelian unitary local group of electromagnetic interactions  
realized as the Lie group $U^{loc}(1)= SO^{loc}(2)$ with
one dimensional trivial algebra $\hat{g}_{1}=R^{1}$:
$U(\theta)=e^{-i\theta}$,
where
$\theta \equiv \chi^{(+)}_{J}(\theta_{+1})-\chi^{(-)}_{J}(\theta_{-1})$.
The strength of interaction is specified by a single coupling $Q$ of 
electric charge. 
The invariance under
the local group $U^{loc}(1)$ leads to electromagnetic field, the massless
quanta of which - the photons are {\em electrically neutral}, because of the
condition eq.(4.3.4):
\begin{equation}
\label{eq: R2.29}
f(\theta_{+1},\theta_{-1})=f(\theta'_{+1},\theta'_{-1})=
\cdots = inv.
\end{equation}
2. Next, we consider the case of two dimensional local transformations
through the angles $\theta_{\pm m}(\zeta)$ around two axes $(m=1,2)$.
The matrix function of transformation is written 
$
f^{(+)}=\left( \matrix{
f^{(+)}_{11}  & f^{(+)}_{12}  & 0 \cr 
f^{(+)}_{21}    & f^{(+)}_{22}  & 0 \cr
0               & 0             & 1\cr
}\right), \quad
f^{(-)}={(f^{(+)})}^{+}.
$
The incompatibility relations eq.(4.4.5) give rise to nontrivial conditions
\begin{equation}
\label{eq: R2.31}
\begin{array}{ll}
f^{(+)}_{11}=\left\| f^{(+)}\right\| {(f^{(+)}_{22})}^{*}, \qquad
f^{(+)}_{21}=-\left\| f^{(+)}\right\| {(f^{(+)}_{12})}^{*},\\ \\
f^{(+)}_{12}=-\left\| f^{(+)}\right\| {(f^{(+)}_{21})}^{*},\quad
f^{(+)}_{22}=\left\| f^{(+)}\right\| {(f^{(+)}_{11})}^{*},
\end{array} 
\end{equation}
Hence $\left\| f^{(+)}\right\|=1.$ One readily derives the matrix 
$U(\theta)$ of gauge transformations of collection of fundamental fields
$
U=e^{-i\vec{T}\vec{\theta}}=
\left( \matrix{
f^{(+)}_{11}  & f^{(+)}_{12}\cr
f^{(+)}_{21}  & f^{(+)}_{22}\cr 
} \right),
$
where $T_{i} \quad(i=1,2,3)$ are the generators
of the group $SU(2)$.
The fields will come in multiplets forming a 
basis for representations of the isospin group $SU(2)$. Meanwhile
\begin{equation}
\label{eq: R2.33}
\begin{array}{l}
\FFr{\theta_{1}}{\theta}=\FFr{e^{\chi^{R}_{12}}\sin\chi^{J}_{12}}
{\sqrt{1-e^{2\chi^{R}_{11}}{\cos}^{2}\chi^{J}_{11}}}, 
\quad
\FFr{\theta_{2}}{\theta}=-\FFr{e^{\chi^{R}_{12}}\cos\chi^{J}_{12}}
{\sqrt{1-e^{2\chi^{R}_{11}}{\cos}^{2}\chi^{J}_{11}}},\quad
\FFr{\theta_{3}}{\theta}=\FFr{e^{\chi^{R}_{11}}\sin\chi^{J}_{11}}
{\sqrt{1-e^{2\chi^{R}_{11}}{\cos}^{2}\chi^{J}_{11}}}, \\ \\
\theta = \mid\vec{\theta}\mid 
= 2\arccos \left(e^{\chi^{R}_{11}}\cos\chi^{J}_{11}\right),
\quad e^{\chi^{R}_{11}}\leq 1,
\end{array}
\end{equation}
provided by
\begin{equation}
\label{eq:R2.34}
\chi^{R}_{11}=\chi^{R}_{22}, \quad\chi^{R}_{12}=\chi^{R}_{21}
\quad
\chi^{J}_{11}+\chi^{J}_{22}=0, \quad \chi^{J}_{21}+\chi^{J}_{12}=\pi,
\quad
\chi^{R}_{12}=\FFr{1}{2}\ln \left(1-e^{2\chi^{R}_{11}}\right).
\end{equation}
That is, three functions $\chi^{R}_{11}, \chi^{J}_{11}$ and $\chi^{J}_{12}$
or the angles $\theta'_{+1},\theta_{+1}$ and $\theta_{+2}$ are parameters
of the group $SU^{loc}(2)$ 
\begin{equation}
\label{eq; R2.35}
\chi^{R}_{11}=\chi_{R}(\theta'_{+1})-\chi_{R}(\theta_{+1}),
\quad
\chi^{J}_{11}=\chi_{J}(\theta'_{+1})-\chi_{J}(\theta_{+1}),
\quad
\chi^{J}_{12}=\chi_{J}(\theta'_{+1})-\chi_{J}(\theta_{+2}).
\end{equation}
3. In the case of gauge transformations occurred around all three
axes $(l,k=1,2,3)$:
$
f^{(+)}=\left( \matrix{
f^{(+)}_{11}  & f^{(+)}_{12}  & f^{(+)}_{13} \cr 
f^{(+)}_{21}  & f^{(+)}_{22}  & f^{(+)}_{23} \cr
f^{(+)}_{31}  & f^{(+)}_{32}  & f^{(+)}_{33}\cr
} \right), \quad
f^{(-)}={(f^{(+)})}^{+},
$
the incompatibility relations eq.(4.4.5) yield the unitary condition
$U^{-1}=U^{+}, \,\,
f^{(+)}\equiv U$, and also $\left\| U \right\| =1$. Then
$U(\theta)=e^{-\frac{i}{2}\vec{\lambda}\vec{\theta}}$,
where $\FFr{\lambda_{i}}{2} \quad(i=1,\ldots,8)$ are the matrix
representation of generators of the group $SU(3)$.
Right through  differentiation one derives 
$\vec{\lambda}\vec{d\theta}=2iU^{+}dU,$
or
$
\vec{\theta}=-\IIn Im\left(tr \left(\vec{\lambda}
\left( f^{(-)}df^{(+)} \right) 
\right)\right),
$
provided by
$Re\left(tr \left(\vec{\lambda}\left( f^{(-)}df^{(+)}\right) \right)
\right)\equiv 0.$
For the infinitesimal transformations $\theta_{i} \ll 1$ we get
\begin{equation}
\label{eq: R2.40}
\begin{array}{ll}
\theta_{1}\approx 2e^{\chi^{R}_{12}}\sin\chi^{J}_{12},\quad   
\theta_{3}\approx \sin\chi^{J}_{33}+2\sin\chi^{J}_{11},\quad       
\theta_{5}\approx 2(1- e^{\chi^{R}_{13}}\cos\chi^{J}_{13}),\\
\theta_{2}\approx 2(1- e^{\chi^{R}_{12}}\cos\chi^{J}_{12}),\quad  
\theta_{4}\approx 2e^{\chi^{R}_{13}}\sin\chi^{J}_{13},\quad       
\theta_{6}\approx 2e^{\chi^{R}_{23}}\sin\chi^{J}_{23},\\ 
\theta_{7}\approx 2(1- e^{\chi^{R}_{23}}\cos\chi^{J}_{23}),\quad
\theta_{8}\approx -\sqrt{3}\sin\chi^{J}_{33},
\end{array}
\end{equation}
provided by
\begin{equation}
\label{eq:R2.41}
\chi^{R}_{ll}\approx 0,
\quad
\chi^{R}_{lk}\approx \chi^{R}_{kl}, \quad 
\chi^{J}_{lk}\approx \chi^{J}_{kl}, \quad (l \neq k)\quad
\sin\chi^{J}_{11}+\sin\chi^{J}_{22}+\sin\chi^{J}_{33}\approx 0.
\end{equation}
The internal symmetry group
$SU^{loc}_{C}(3)$ enables to introduce the gauge theory in colour space,
with the colour charges as exactly conserved quantities. While, the local 
colour transformations are implemented on the coloured quarks right through 
the $SU^{loc}_{C}(3)$ rotation matrix U in the fundamental 
representation. 

\section {Operator Multimanifold $\hat{G}_{N}$}
\label {Oper}
\subsection{Operator Vector and Covector Fields}
\label{ov}
The OM formalism of $\hat{G}=
\hG_{\eta}\oplus\hG_{u}$ is built up by assuming 
an existence only of ordinary primordial structures of one sort 
(one u-channel).
Being confronted by our major
goal to develop the microscopic approach to field theory
based on multiworld geometry, henceforth we generalize the OM formalism via 
the concept of the OMM. Then, instead of one sort of ordinary structures we 
are going to deal with different species of ordinary structures. 
But before proceeding further and to enlarge the previous model it is 
profitable to assume an existence of infinite number of 
${}^{i}u$-type ordinary
structures of different species $i=1,2,\ldots ,N $ (multi-u channel).
These structures will be specified by the superscript $i$
to the left. 
This hypothesis,
as it will be seen in the subsequent part II, leads to 
the substantial progress of understanding of the properties of particles.
At the very outset we consider the processes
of creation and annihilation of regular structures of $\eta$- and 
${}^{i}u$-types in the lowest state ($s_{0}$).
The general rules stated in subsec 2.1 regarding to this change apply 
a substitution of operator basis pseudo vectors and covectors by a new
ones $(i=1,2,\ldots, N)$
\begin{equation}
\label {eq: R3.7}
{}^{i}\hat{O}^{r_{1}r_{2}}_{\lambda,\mu}=
{}^{i}\hat{O}^{r_{1}}_{\lambda}\otimes
{}^{i}\hat{O}^{r_{2}}_{\mu}\equiv
{}^{i}\hat{O}^{r}_{\lambda,\mu}={}^{i}O^{r}_{\lambda,\mu}
(\alpha_{\lambda}\otimes \alpha_{\mu}),
\end{equation}
provided by $ r\equiv(r_{1},r_{2})$ and
$$
\begin{array}{l}
{}^{i}O^{r}_{1,1}=\FFr{1}{\sqrt{2}}(\nu_{i}\,{\O1_{\eta}}{}_{+}^{r}+
{}^{i}{\O1_{u}}{}_{+}^{r}), \quad
{}^{i}O^{r}_{2,1}=\FFr{1}{\sqrt{2}}(\nu_{i}\,{\O1_{\eta}}{}_{+}^{r}-
{}^{i}{\O1_{u}}{}_{+}^{r}), \\\\ 
{}^{i}O^{r}_{1,2}=\FFr{1}{\sqrt{2}}(\nu_{i}\,{\O1_{\eta}}{}_{-}^{r}+
{}^{i}{\O1_{u}}{}_{-}^{r}), \quad
{}^{i}O^{r}_{2,2}=\FFr{1}{\sqrt{2}}(\nu_{i}\,{\O1_{\eta}}{}_{-}^{r}-
{}^{i}{\O1_{u}}{}_{-}^{r}),
\end{array}
$$
where 
\begin{equation}
\label {eq: R3.9}
<\nu_{i},\nu_{j}>=\delta_{ij},\quad
<{}^{i}{\O1_{u}}{}_{\lambda}^{r},\,{}^{j}{\O1_{u}}{}_{\tau}^{r'}>=
-\delta_{ij}\delta_{rr'}{}^{*}\delta_{\lambda\tau},\quad
<{\O1_{\eta}}{}_{\lambda}^{r},\,{}^{i}{\O1_{u}}{}_{\tau}^{r'}>=0.
\end{equation}
We consider then the operators
$
{}^{i}{\hat{\gamma}}{}^{r}_{(\lambda,\mu,\alpha)}=
{}^{i}{\hat{O}}^{r_{1}r_{2}}_{\lambda, \mu}
\otimes{\hat{\sigma}}^{r_{3}}_{\alpha}.$ 
and calculate nonzero matrix elements
\begin{equation}
\label{eq: R3.11}
<\lambda,\mu\mid {}^{i}{\hat{\gamma}}{}^{r}_{(\tau,\nu,\alpha)}\mid \tau,\nu>
=
{}^{*}\delta_{\lambda\tau}{}^{*}\delta_{\mu,\nu}
\,{}^{i}e^{r}_{(\tau,\nu,\alpha)},
\end{equation}
where ${}^{i}e^{r}_{(\lambda,\mu,\alpha)}=
{}^{i}O^{r}_{\lambda,\mu} \otimes\sigma_{\alpha}$.
The operators $\left\{{}^{i}{\hat{\gamma}}{}^{r} \right\}$
are the basis for all the operator vectors 
$\hat{\Phi}(\zeta)={}^{i}{\hat{\gamma}}{}^{r}\,\,
{}^{i}\Phi_{r}(\zeta)$ of tangent section of 
principle bundle with the base of operator multimanifold
$\hat{G}_{N}=\left( \S_{i}^{N}\oplus {}^{*}\hat{\RR_{i}}{}^{4} \right)
\otimes\hat{R}{}^{3}$.
Here $^{*}\hat{\RR_{i}}{}^{4}$ is the $2\times2$ dimensional linear 
pseudo operator space, with the set of the linear unit operator 
pseudo vectors eq.(5.1.1) as the basis of tangent vector section,
and $\hat{R}^{3}$ is the three dimensional real linear operator space with
the basis consisted of the ordinary unit operator vectors  
$\{ {\hat{\sigma}}^{r}_{\alpha}\}$. 
The 
$\hat{G}_{N}$ decomposes as follows:
\begin{equation}
\label{eq: R3.15}
\hat{G}_{N}=\hG_{\eta}\oplus\hG_{u_{1}}
\oplus \cdots \oplus\hG_{u_{N}},
\end{equation}
where $\hG_{u_{i}}$ is the six dimensional operator
manifold of the given species $(i)$ with the basis \\
$\left\{ {}^{i}{\hgam_{u}}{}^{r}_{(\lambda\alpha)} =
{}^{i}{\hat{\O1_{u}}}{}_{\lambda}^{r}\otimes {\hat{\sigma}}^{r}_{\alpha}
\right\}$.
The expansions of operator vectors and 
covectors are written 
$\hps_{\eta}={\hgam_{\eta}}{}^{r}
{\ps1_{\eta}}{}_{r}, \quad
\hps_{u}= {}^{i}{\hgam_{u}}{}^{r}\,
{}^{i}{\ps1_{u}}{}_{r}, 
\quad
\bar{\hps_{\eta}}=
{\hgam_{\eta}}{}_{r}
{\ps1_{\eta}}{}^{r},\quad
\bar{\hps_{u}}=
{}^{i}{\hgam_{u}}{}_{r}\,
{}^{i}{\ps1_{u}}{}^{r},$
where the components ${\ps1_{\eta}}{}_{r}(\eta)$ and
${}^{i}{\ps1_{u}}{}_{r}(u)$ are respectively the 
link functions of $\eta$-type and ${}^{i}u$-type structures.

\subsection {Realization of the Multimanifold $G_{N}$}
\label {manif}
Now, we consider the special system of the regular structures, which is made 
of one fundamental structure of $\eta$-type and 
infinite number of ${}^{i}u$-type ordinary structures of different species 
$(i=1,\ldots, N)$. 
The primordial structures establish the stable linkage to form the stable 
system
\begin{equation}
\label{eq: R4.1}
p^{2}=p^{2}_{\eta}- \S^{N}_{i=1}p^{2}_{u_{i}}=0.
\end{equation}
The free field defined on the
multimanifold $G_{N}=\G1_{\eta}\oplus
\G1_{u_{1}}\oplus \cdots\oplus\G1_{u_{N}}$ is written
$$
\Psi =\ps1_{\eta}(\eta)\ps1_{u}(u),\quad 
\ps1_{u}(u)=\ps1_{u_{1}}(u_{1})\cdots\ps1_{u_{N}}(u_{N}), 
$$
where $\ps1_{u_{i}}$ is  the bispinor
defined on the internal manifold $\G1_{u_{i}}$.
On analogy of subsec.2.2  
we make use of localized wave packets by
means of superposition of plane wave solutions furnished 
by creation and annihilation operators in agreement with Pauli's 
principle.
Straightforward calculations now give the generalization of the relation 
eq.(2.2.2)
\begin{equation}
\label{eq: R4.7}
\begin{array}{l}
\S_{\lambda=\pm}<\chi_{\lambda}\mid\hat{\Phi}(\zeta)
\bar{\hat{\Phi}}(\zeta)\mid
\chi_{\lambda}>= 
\S_{\lambda=\pm}<\chi_{\lambda}\mid
\bar{\hat{\Phi}}(\zeta)\hat{\Phi}(\zeta)\mid\chi_{\lambda}>= \\ \\
-i\,\zeta^{2}{\G1_{\zeta}}(0) = -i\,\left(\eta^{2}{\G1_{\eta}}(0)-
\S^{N}_{i=1}{u_{i}}^{2}{\G1_{u_{i}}}(0)\right).
\end{array}
\end{equation}
Along the same line the realization of the multimanifold stems from the  
condition eq.(2.2.3), which is now imposed  upon the matrix element eq.(5.2.2).
Let denote
$
u^{2}\G1_{u}(0)\equiv
\Lm_{u_{i}\rightarrow u'_{i}}\S_{i=1}^{N}(u_{i}u'_{i})
\G1_{u_{i}}(u_{i}-u'_{i})
$
and consider a stable system eq.(5.2.1). Hence
\begin{equation}
\label{eq: R4.12}
{\G1_{u}}{}_{F}(0)=
{\G1_{\eta}}{}_{F}(0)=
{\G1_{\zeta}}{}_{F}(0),
\end{equation}
where ${\G1_{\eta}}{}_{F},
{\G1_{u}}{}_{F}$ and ${\G1_{\zeta}}{}_{F}$ are the causal Green's functions
of the $\eta-,u-$ and $\zeta$-type structures, and
$m\equiv \left| p_{u}\right|=\left( \S_{i=1}^{N}
{p_{u_{i}}}^{2} \right)^{1/2}=\left| p_{\eta}\right|.$
In the aftermath, the length of each vector
${\bf \zeta}={}^{i}e\,\,
{}^{i}\zeta\in G_{N}$
should be equaled zero (subsec.2.2)
$\zeta^{2}=\eta^{2}-u^{2}=\eta^{2}-\S_{i=1}^{N}({u^{G}_{i}})^{2}=0,$
where use is made of
$$
\left(u^{G}_{i}\right)^{2}\equiv u_{i}^{2}\,
\Lm_{
\begin{array}{l}
u_{i}\rightarrow u'_{i}\\
\eta\rightarrow\eta'
\end{array}
}
{\G1_{u_{i}}}{}_{F}(u_{i}-u'_{i})
\left. \right/
{\G1_{\eta}}{}_{F}(\eta-\eta')
$$
and 
$u^{G}_{i}={}^{i}{\he_{u}}{}_{(\lambda,\alpha)}
\,u_{i}^{G\,(\lambda,\alpha)}$.
Thus, the multimanifold $G_{N}$ comes into being, which 
decomposes as follows:
\begin{equation}
\label{eq: R4.18}
G_{N}=\G1_{\eta}\oplus\G1_{u_{1}}\oplus\cdots
\oplus \G1_{u_{N}}.
\end{equation}
It brings us to the conclusion: the major requirement eq.(2.2.3)
provided by stability condition eq.(5.2.1) or eq.(5.2.3) yields 
the flat multimanifold $G_{N}$.
Meanwhile, the  Minkowski flat space $M_{4}$ stems from the
flat submanifold $\G1_{\eta}$ (subsec. 2.1), in which the line 
element turned out to be invariant. That is, the principle of relativity 
comes into being with the $M_{4}$ ensued from
the MW geometry $G_{N}$.
In the following we shall use a notion of the $i$-th 
internal world for the submanifold $\G1_{u_{i}}$.

\subsection {Subquarks and Subcolour Confinement}
\label {sub}
Since our discussion within this section in many respects is similar to 
that of sec.4, here we will be brief.
We assume that the distortion rotations
(${\G1_{u_{i}}}\stackrel{\theta}{\rightarrow} \widetilde{\G1_{u_{i}}}$, for 
given $i$) through the angles
${}^{i}\theta_{+k}$ and ${}^{i}\theta_{-k}\quad k=1,2,3$ occur 
separately in the three dimensional internal spaces 
${\RR_{u_{i}}}{}_{+}^{3}$ and ${\RR_{u_{i}}}{}_{-}^{3}$ composing six 
dimensional distorted submanifold $\widetilde{\G1_{u_{i}}}=
{\RR_{u_{i}}}{}_{+}^{3}\oplus{\RR_{u_{i}}}{}_{-}^{3}$ (see eq.(5.2.4)). 
As it is exemplified in previous section, the laws apply 
in use the wave packets constructed by superposition of the link functions
of distorted ordinary structures furnished by generalized operators of
creation and annihilation as the expansion coefficients
\begin{equation}
\label{eq: R5.3}
{\hps_{u}}(\theta_{+})=
\S_{\pm s}\IIn\frac{d^{3}p_{u_{i}}}{{(2\pi)}^{3/2}}
\left( {}^{i}{\hgam_{u}}{}_{(+\alpha)}^{k}
{}^{i}{\ps1_{u}}^{(+\alpha)}({}^{i}\theta_{+k})+
{}^{i}{\hgam_{u}}{}_{(-\alpha)}^{k}
{}^{i}{\ps1_{u}}^{(-\alpha)}({}^{i}\theta_{+k})\right), 
\end{equation}
etc. The fields ${}^{i}{\ps1_{u}}(\theta_{+k})$
and ${}^{i}{\ps1_{u}}(\theta_{-k})$
are defined on the distorted internal spaces
${\RR_{u_{i}}}{}_{+}^{3}$ and ${\RR_{u_{i}}}{}_{-}^{3}$.
The generalized expansion coefficients in eq.(5.3.1) imply
\begin{equation}   
\label{eq: R5.4}
<\chi_{-}\mid \{ {}^{i}{\hgam_{u}}{}^{(+\alpha)}_{k}(p_{u_{i}},s_{i}),\,\,
{}^{j}{\hgam_{u}}{}_{(+\beta)}^{k'}(p'_{u_{j}},s'_{j})\}\mid\chi_{-}>=
-\delta_{ij}\delta_{kk'}\delta_{ss'}\delta_{\alpha\beta}\delta^{3}
({\vec{p}}_{u_{i}}- {\vec{p'}}_{u_{i}}).
\end{equation}
The condition of the MW geometry realization eq.(5.2.3) now
reduced to be
\begin{equation}
\label{eq: R5.5}
\S_{i=1}^{N}\omega_{i}
\left[ \Lm_{{}^{i}\theta_{+}\rightarrow {}^{i}\theta_{-}}
{\G1_{u_{i}}}^{\theta}{}_{F}({}^{i}\theta_{+}-{}^{i}\theta_{-})\right]=
\Lm_{\eta_{f}\rightarrow\eta'_{f}}
\G1_{\eta}{}_{F}(\eta_{f}-\eta'_{f}),
\end{equation}
provided by
$\omega_{i}=\FFr{u_{i}^{2}}{u^{2}}.$
Taking into account the expression of causal Green's function
at given $(i)$, in the case if 
$$
\Lm_{
u_{i_{1}}\rightarrow u_{i_{2}}
}
{\G1_{u_{i_{1}}}}{}_{F}(u_{i_{1}}-u_{i_{2}})
= 
\Lm_{u'_{i_{1}}\rightarrow u'_{i_{2}}}
{\G1_{u_{i'_{1}}}}{}_{F}(u'_{i_{1}}-u'_{i_{2}})
=\cdots=inv,
$$
one gets
\begin{equation}
\label{eq: R5.16}
\S_{k}\,
{}^{i}{\ps1_{u}}({}^{i}\theta_{+k})\,\,
{}^{i}{\bar{\ps1_{u}}}({}^{i}\theta_{-k})=
\S_{k}\,
{}^{i}{\ps1_{u}}'({}^{i}\theta'_{+k})\,\,
{}^{i}{\bar{\ps1_{u}}}'({}^{i}\theta'_{-k})=\cdots = inv.
\end{equation}
Thus, in the context of the physically more realistic MW geometry it is 
legitimate now to substitute 
the concept of quark $(q_{k})$ schematically introduced in sec.4 by the
subquark $({}^{i}q_{k})$.
Everything said will then remain valid,
provided we make a simple change of quarks into subquarks, the 
colours into 
subcolours.
Hence, we may think of the function 
${}^{i}{\ps1_{u}}({}^{i}\theta_{+k})$ as the $u$-component of
bispinor field of subquark $({}^{i}q_{k})$ of species $(i)$
with subcolour $k$, and respectively 
${}^{i}{\bar{\ps1_{u}}}({}^{i}\theta_{-k})$ -the conjugated bispinor field 
of antisubcolour $(k)$.
The subquarks and antisubquarks may be local $({}^{i}q_{k})$ or
global $({}^{i}q_{k}^{c})$.
Then, the subquark $({}^{i}q_{k})$ is the fermion with the 
half integer 
spin and subcolour degree of freedom, and,
according to eq.(5.3.4), could emerge on the mass shell only in confined 
phase 
\begin{equation}
\label{eq: R5.19}
\S_{k}\,
{}^{i}{q}_{k}\,{}^{i}\,{\bar{q}}_{k} =
\S_{k}\,
{}^{i}{q'}_{k}\,{}^{i}{\bar{q'}}_{k}=
\cdots=inv.
\end{equation}
To trace a resemblance with the previous section, the internal 
symmetry group
${}^{i}G=U(1), SU(2), SU(3)$ enables to introduce the gauge theory
in internal world with the subcolour charges as exactly conserved 
quantities. Furthermore, the subcolour transformation have implemented on
subquark fields right through local and global rotation matrices of
group ${}^{i}G$ in fundamental representation. 
Due to the Noether procedure the 
conservation of global charges ensued from the global 
gauge invariance of physical system, meanwhile reinforced requirement
of local gauge invariance may be satisfied as well by introducing the 
gauge fields with the values in Lie algebra ${}^{i}\hat{g}$ of the group
${}^{i}G$.

\renewcommand {\theequation}{\thesection.\arabic {equation}}
\vskip 1truecm
\begin{center}
{\Large {\bf Part II}}
\end{center}
\vskip 0.1truecm
\begin{center}
{\Large {\bf Realization of the Particle Physics}}
\end{center}
Continuing our program based on the OMM formalism, in this part we attempt to 
develop, further, the microscopic approach to the SM, which enables an 
insight to the key problems of particle phenomenology. Particularly, we 
suggest the microscopic theory of the unified electroweak interactions with a 
small number of free parameters. Besides the microscopic interpretation of 
all physical parameters the resulting theory has two testable 
solid implications, which are drastically different from those of conventional 
models.
\section {The MW-Structure of the Particles}
\label {Struct}
For our immediate purpose to describe the particles 
as the composite dynamical systems defined on the MW geometry, 
we shall consider the collection of matter 
fields $\Psi(\zeta)$ with nontrivial internal structure
$
\Psi(\zeta)=\ps1_{\eta}(\eta)\,\,{}^{1}\ps1_{u}(\theta_{1})
\cdots\,\,{}^{N}\ps1_{u}(\theta_{N}).
$
We suppose that
the component ${}^{i}\ps1_{u}(\theta_{i})$ is made of product of 
some constituent subquarks and antisubquarks, which
form the multiplets transformed by fundamental 
${}^{i}D(j)$ and contragradient ${}^{i}\bar{D}(j)$ irreducible
representations of group ${}^{i}G$. 
Hereinafter we suppose that the MW index $(i)$ will be running
only through $i=Q,W,B,$ $s,c,b,t$ specifying the internal worlds formally
taken to denote in following nomenclature: Q-world of electric 
charge; W-world of weak interactions;
B-baryonic world of strong interactions; the s,c,b,t  are
the worlds of strangeness, charm, bottom and top.
We admit also that the distortion rotations 
in the worlds Q,W and B are local ${}^{i}\theta_{\pm k}(\eta)$, while
they are global  in the worlds s,c,b,t.  
Below we introduce the fields of leptons $(l)$ and 
quarks $(q_{f})$ with different flavours f=u,d,s,c,b,t. 
To develop some feeling for this problem and to avoid irrelevant 
complications, here we may temporarily skeletonize it by taking
the leptons to have following MW- structure:
\begin{equation}
\label{eq: R7.1}
l \equiv \Psi_{l}(\zeta)=\ps1_{\eta}(\eta)\ps1_{Q}(u_{Q})\ps1_{W}(u_{W}),
\end{equation}
while the quarks are in the form
\begin{equation}
\label{eq: R7.2}
q_{f} \equiv \Psi_{f}(\zeta)=\ps1_{\eta}(\eta)
\ps1_{Q}(u_{Q})\ps1_{W}(u_{W})\ps1_{B}(u_{B})q_{f}^{c},
\end{equation}
where the superscript $(c)$ specified the worlds in which rotations are 
global
\begin{equation}
\label{eq: R7.3}
\begin{array}{l}
q_{u}^{c}=q_{d}^{c}=1, \quad q_{s}^{c}={\ps1_{s}}^{c}(u_{s}),
\quad q_{c}^{c}\equiv{\ps1_{c}}(u_{c}),\quad
q_{b}^{c}\equiv{\ps1_{b}}(u_{b}),
\quad q_{t}^{c}\equiv{\ps1_{t}}(u_{t}).
\end{array}
\end{equation}
We can take this scheme as a starting point for our considerations, which 
will become clearer in the next sections where the explicit forms of the 
$\ps1_{i}(u_{i})$ will be subject for discussion. We assign a scale $1/3$ to  
each distortion rotational mode in the three dimensional spaces
${\RR_{u_{i}}}{}_{+}^{3}$ and ${\RR_{u_{i}}}{}_{-}^{3},$ namely
the subquark arisen after the rotation around the given axis of the given 
world carries the  $1/3$ charge of corresponding species; while, the 
antisubquark carries respectively the $(-1/3)$ charge.
In the case of the worlds C=s,c,b,t, where distortion rotations are
global and diagonal with respect to axes 1,2,3, the physical system of
corresponding subquarks is invariant under the global transformations
$f^{(3)}_{C}(\theta^{c})$ of the global unitary group $SU^{c}_{3}$:
$$
f^{(3)}_{C}=\left(\matrix{
f^{c}_{11}  & 0  & 0 \cr 
0             & f^{c}_{22}  & 0 \cr
0             & 0  & f^{c}_{33}\cr
} \right)=
\exp \left\{ -\FFr{i}{3}
\left( \matrix{
\theta^{c}_{1}  & 0  & 0 \cr
0             & \theta^{c}_{2}  & 0 \cr
0             & 0  & \theta^{c}_{3}\cr
} \right)
\right\},
$$
where 
$
f^{(3)}_{C}\left( f^{(3)}_{C}\right)^{+}=1, \quad 
\|f^{(3)}_{C}\| = 1
$.
That is 
$
\theta^{c}_{1}+\theta^{c}_{2}+\theta^{c}_{3}=0.
$
The simplest possibility gives
$\theta^{c}\equiv \theta^{c}_{1}=\theta^{c}_{2}$. Hence, one gets
\begin{equation}
\label{eq: R7.6}
f^{(3)}_{C}=\exp \left\{ -\FFr{i}{3}
\left( \matrix{
1  & 0  & 0 \cr
0             & 1  & 0 \cr
0             & 0  & -2\cr
} \right)\theta^{c}
\right\}=
\exp\left( 
-i\FFr{\lambda_{8}}{\sqrt{3}}\theta^{c}
\right)=
e^{-iY^{c}\theta^{c}},
\end{equation}
provided by the operator of hypercharge $Y^{c}$ of diagonal group 
$SU^{c}_{i}$. 
If one has all the worlds involved, then
$
Y^{c} = s + c + b + t.
$
The Q-world, which has an essential role for realization of the 
condition of the MW connections, will be discussed in detail in the
next section. We only note that conservation of each rotation mode in 
Q- and B- worlds, where the distortion rotations are local, means that
corresponding subquarks carry respectively the conserved charges
Q and B in the scale $1/3$, and antisubquarks - $(-1/3)$ charges.
It can be provided by including the matrix $\lambda_{8}$ as
the generator  with the others in the symmetries of corresponding worlds
(Q, B), and expressed in the invariance of the system of corresponding 
subquarks under the transformations of these symmetries.
The incompatibility relations eq.(4.4.5) for global 
distortion rotations in the worlds C=s,c,b,t reduced to
$$
f^{c}_{11}f^{c}_{22}=\bar{f}^{c}_{33},\quad
f^{c}_{22}f^{c}_{33}=\bar{f}^{c}_{11},\quad
f^{c}_{33}f^{c}_{11}=\bar{f}^{c}_{22},
$$
where 
$
\|f^{(3)}_{C}\| = f^{c}_{11} f^{c}_{22} f^{c}_{33}=1, \quad
f^{c}_{ii}\bar{f}^{c}_{ii}=1 \quad \mbox{for} \quad i=1,2,3.
$
It means that the two subcolour singlets are available:
$
\left( q\bar{q}\right)^{c}_{i}=inv, 
\quad
\left(q_{1}q_{2}q_{3} \right)^{c}=inv,
$
carried the charges
$
C_{\left( q\bar{q}\right)^{c}_{i}}=0 \quad 
C_{\left(q_{1}q_{2}q_{3} \right)^{c}}=1,
$
respectively, where we denote
$
\left( q\bar{q}\right)^{c}_{i}\equiv {}^{c}q_{i}{}^{c}\bar{q}_{i}, 
\quad \left(q_{1}q_{2}q_{3} \right)^{c}\equiv
{}^{c}q_{1}{}^{c}q_{2}{}^{c}q_{3} .
$
Including the baryonic charge into {strong hypercharge} 
\begin{equation}
\label{eq: R8.11}
Y = B + s + c + b + t,
\end{equation}
we conclude that the hypercharge Y is the sum of all the conserved 
quantum numbers associated with the corresponding rotational modes of the 
internal worlds B,s,c,b,t involved in the MW geometry realization 
condition eq.(5.2.3).

\renewcommand{\theequation}{\thesubsection.\arabic{equation}}
\subsection {Realization of Q-World and Gell-Mann-Nishijima Relation}
\label {Rel}
The symmetry of the Q-world is assumed to be a 
local unitary symmetry $diag\left( SU^{loc}(3)\right)$ is diagonal with 
respect to axes 1,2,3. The unitary unimodular matrix 
$f^{(3)}_{Q}$ of local distortion rotations takes the form
$$
f^{(3)}_{Q}=\left( \matrix{
f^{Q}_{11}  & 0  & 0 \cr
0             & f^{Q}_{22}  & 0 \cr
0             & 0  & f^{Q}_{33}\cr
} \right)=
Q_{1}e^{-i\theta_{1}}+Q_{2}e^{-i\theta_{2}}+
Q_{3}e^{-i\theta_{3}}=
e^{-i\vec{Q}\vec{\theta}}=e^{-i\lambda_{Q}\theta_{Q}},
$$
where $f^{(3)}_{Q}\left( f^{(3)}_{Q}\right)^{+}=1, \quad 
\|f^{(3)}_{Q}\| = 1$,
provided
$$
Q_{1}=\left( \matrix{
1  & 0  & 0 \cr
0  & 0  & 0 \cr
0  & 0  & 0\cr
} \right), \quad
Q_{2}=\left( \matrix{
0  & 0  & 0 \cr
0  & 1  & 0 \cr
0  & 0  & 0\cr
} \right), \quad
Q_{3}=\left( \matrix{
0  & 0  & 0 \cr
0  & 0  & 0 \cr
0  & 0  & 1\cr
} \right),
$$
and
$
\theta_{1}+\theta_{2}+\theta_{3}=0.
$
Taking into account the scale of rotation mode,
in the other simple case than those of eq.(6.4):
$
\theta_{2}=\theta_{3}=-\FFr{1}{3}\theta_{Q}, 
$
it follows that
$
\theta_{1}=\FFr{2}{3}\theta_{Q} .
$
Among the generators of the group $SU(3)$ only the matrices
$\lambda_{3}$ and $\lambda_{8}$ are diagonal. Therefore, the matrix 
$\lambda_{Q}$ may be written 
$\lambda_{Q}=\FFr{1}{2}\lambda_{3}+\FFr{1}{2\sqrt{3}}\lambda_{8}.
$
Making use of the corresponding operators of the group $SU(3)$
we arrive at Gell-Mann-Nishijima relation
\begin{equation}
\label{eq: R9.7}
Q= T_{3} + \FFr{1}{2}Y,
\end{equation}
where $Q=\lambda_{Q}$ is the generator of electric charge,
$T_{3}=\FFr{1}{2}\lambda_{3}$ is the third component of isospin 
$\vec{T}$, and $Y=\FFr{1}{\sqrt{3}}\lambda_{8}$ is the hypercharge.
The eigenvalues of 
these operators will be defined in due course considering the concrete
symmetries and microscopic structures of fundamental fields. We 
think of operators $T_{3}$ and  $Y$ as the MW connection
charges and of relation eq.(6.1.1) as the condition of
the realization of the MW connections. Thus, during realization of 
the MW- structure the symmetries of corresponding internal
worlds must be unified into more higher symmetry including also the 
$\lambda_{3}$ and $\lambda_{8}$. The realization conditions of
the MW- structure are embodied in eq.(5.2.3) and eq.(6.1.1), 
provided by the conservation law of each rotational mode
eq.(6.4) of the given internal worlds involved into the MW geometry 
realization condition. For example, in the case of quarks eq.(6.2), one has
\begin{equation}
\label{eq: R9.8}
\S_{i=B,s,c,b,t}\omega_{i}\,{\G1_{i}}^{\theta}{}_{F}(0)=
{\G1_{\eta}}{}_{F}(0),
\end{equation}
and the Gell-Mann-Nishijima relation is written down
\begin{equation}
\label{eq: R9.9}
Q= T_{3} + \FFr{1}{2}(B+s+c+b+t).
\end{equation}
The other case of leptons eq.(6.1) related closely to
the realization of W-world of weak interactions will be discussed in detail
in subsec.6.8, the realization 
condition for which reduces to following:
\begin{equation}
\label{eq: R9.10}
{\G1_{Q}}^{\theta}{}_{F}(0)={\G1_{\eta}}{}_{F}(0), \quad u\equiv u_{Q},
\quad (i\equiv Q)
\end{equation}
and
\begin{equation}
\label{eq: R9.11}
Q= T_{3}^{w} + \FFr{1}{2}Y^{w},
\end{equation}
where $T_{3}^{w}$ and $Y^{w}$ are respectively the operators of third 
component of weak isospin $\vec{T}^{w}$ and weak hypercharge
(subsec.6.2, 6.8).
The incompatibility relations eq.(4.4.5) 
for the local distortion transformations in Q-world lead to
$$
f^{Q}_{11}f^{Q}_{22}=\bar{f}^{Q}_{33},\quad
f^{Q}_{22}f^{Q}_{33}=\bar{f}^{Q}_{11},\quad
f^{Q}_{33}f^{Q}_{11}=\bar{f}^{Q}_{22},
$$
where 
$
f^{Q}_{ii}\bar{f}^{Q}_{ii}=1, \quad \mbox{for} \quad i=1,2,3,\quad
\|f^{(3)}_{Q}\| = f^{Q}_{11} f^{Q}_{22} f^{Q}_{33}=1.
$
This in turn suggests two subcolour singlets
$
\left( q\bar{q}\right)^{Q}_{i}=inv,
\quad
\left(q_{1}q_{2}q_{3} \right)^{Q}=inv,
$
with the electric charges 
$
Q_{\left( q\bar{q}\right)^{Q}_{i}}=0, \quad 
Q_{\left(q_{1}q_{2}q_{3} \right)^{Q}}=1,
$
respectively.
The singlets $\left( q\bar{q}\right)^{Q}_{i}$ for given $i$ allow us to think 
of the $\left( q\bar{q}\right)^{Q}$ system as the mixed ensemble, such that a 
fraction of the members with relative population $L_{1}$ are characterized by 
the $(q_{1})^{Q}$, some other fraction with relative population  $L_{2}$, by 
$(q_{2})^{Q}$, and so on. Namely, the $\left( q\bar{q}\right)^{Q}$ ensemble 
can be regarded as a mixture of pure ensembles. The fractional populations are 
constrained to satisfy the normalization condition
\begin{equation}
\label{eq: R4.6}
\S_{i} L_{i}=1, \quad L_{i}\equiv \left( q\bar{q}\right)^{Q}_{i}\left.\right/
\left( q\bar{q}\right)^{Q}.
\end{equation}
The $ L_{i}$ also imply the orthogonality condition ensued from the symmetry 
of the $Q$-world
\begin{equation}
\label{eq: R4.7}
<L_{i}, L_{j}>=0 \quad \mbox{if}\quad i\neq j.
\end{equation}
This prompts us to define the usual quantum mechanical density operator
\begin{equation}
\label{eq: R4.8}
\rho_{1}^{Q}=\S_{i}L_{i}\,(q_{i})^{Q})\,(q_{i})^{Q\,+}, \quad 
tr(\rho_{1}^{Q})=1.
\end{equation}
The eq.(6.1.6) suggests another singlets as well
\begin{equation}
\label{eq: R4.9}
\left(q_{1}q_{2}q_{3} \right)^{Q}_{i}\equiv
L_{i}\,\left(q_{1}q_{2}q_{3} \right)^{Q},
\end{equation}
which will be used to build up the MW-structures of the leptons.

\subsection{The Symmetries of the W- and B-Worlds}
$\bullet$ The W-World\\
Invoking local group of weak hypercharge $Y^{w}$ ($U^{loc}(1)$),
it will be seen in subsec. 6.8 that the symmetry of 
W-world of weak interactions rather is $SU^{loc}(2)\otimes U^{loc}(1)$.
However, for the present it is worthwhile to restrict oneself 
by admitting that the symmetry of W-world is simply expressed by the group
of weak isospin  $SU^{loc}(2)$. Namely, from the very first we consider a 
case of two dimensional distortion transformations through the angles 
$\theta_{\pm}$ around two arbitrary axes in the W-world. 
In accordance with the results of sec.4, the fields of subquarks and 
antisubquarks will come in doublets, which form the basis for fundamental 
representation of weak isospin group $SU^{loc}(2)$ often called a 
``custodial'' symmetry [5,20]. The doublet states
are complex linear combinations of up and down states of weak isotopic 
spin. Three possible doublets of six subquark states are
$
\left( \begin{array}{c} q_{1}\\ q_{2}
\end{array} \right)^{w},\quad
\left( \begin{array}{c} q_{2}\\ q_{3}
\end{array} \right)^{w},\quad
\left( \begin{array}{c} q_{3}\\ q_{1}
\end{array} \right)^{w}.
$
\\
\\
$\bullet$ The B-World\\
The B-world is responsible for strong interactions. The internal 
symmetry group is $SU^{loc}_{c}(3)$ enabling to introduce
gauge theory in subcolour space with subcolour charges as exactly 
conserved quantities (sec.4).
The local distortion transformations are implemented
on the subquarks $(q_{i})^{B},\quad i=1,2,3$ through the
$SU^{loc}_{c}(3)$ rotation matrix $U$ in the fundamental 
representation. Taking into account a conservation of rotation 
mode, each subquark carries $(1/3)$ baryonic  
charge, while the antisubquark carries the  $(-1/3)$ baryonic charge.

\subsection {The Microscopic Structure of Leptons: \\
Lepton Generations}
\label {Lept}
After a quantitative discussion of the properties of symmetries of internal
worlds, below we will attempt to show how the known fermion fields of 
leptons and quarks fit into this scheme. 
In this section we start with the leptons.
Taking into account the eq.(6.1.1), eq(6.1.4), we may
consider six possible lepton fields forming three doublets of 
lepton generations
$
\left( \begin{array}{c} \nu_{e}\\ e
\end{array} \right),\quad
\left( \begin{array}{c} \nu_{\mu}\\ \mu
\end{array} \right),\quad
\left( \begin{array}{c} \nu_{\tau}\\ \tau
\end{array} \right),
$
where
\begin{equation}
\label{eq: R11.2}
\begin{array}{l}
\left\{ \begin{array}{l}
\nu_{e}\equiv {\ps1_{\eta}}{}_{\nu_{e}}(\eta)\, 
(q_{1}\bar{q_{1}})^{Q}(q_{1})^{w}=
L_{e}{\ps1_{\eta}}{}_{\nu_{e}}(\eta)\, (q\bar{q})^{Q}(q_{1})^{w},\\\\
e \equiv {\ps1_{\eta}}{}_{e}(\eta)\,
(\overline{q_{1}q_{2}q_{3}})^{Q}_{1}(q_{2})^{w}=
L_{e}{\ps1_{\eta}}{}_{e}(\eta)\, (\overline{q_{1}q_{2}q_{3}})^{Q}(q_{2})^{w},
\end{array} \right. 
\\\\
\left\{ \begin{array}{l}
\nu_{\mu}\equiv {\ps1_{\eta}}{}_{\nu_{\mu}}(\eta)\, 
(q_{2}\bar{q_{2}})^{Q}(q_{2})^{w}=
L_{\mu}{\ps1_{\eta}}{}_{\nu_{\mu}}(\eta)\, (q\bar{q})^{Q}(q_{2})^{w},\\\\
\mu \equiv {\ps1_{\eta}}{}_{\mu}(\eta)\,
(\overline{q_{1}q_{2}q_{3}})^{Q}_{2}(q_{3})^{w}=
L_{\mu}{\ps1_{\eta}}{}_{\mu}(\eta)\, (\overline{q_{1}q_{2}q_{3}})^{Q}(q_{3})^{w},
\end{array} \right. 
\\ \\
\left\{ \begin{array}{l}
\nu_{\tau}\equiv  {\ps1_{\eta}}{}_{\nu_{\tau}}(\eta)\,
(q_{3}\bar{q_{3}})^{Q}(q_{3})^{w}=
L_{\tau}{\ps1_{\eta}}{}_{\nu_{\tau}}(\eta)\, (q\bar{q})^{Q}(q_{3})^{w},\\\\
\tau \equiv  {\ps1_{\eta}}{}_{\tau}(\eta)\,
(\overline{q_{1}q_{2}q_{3}})^{Q}_{3}(q_{1})^{w}=
L_{\tau}{\ps1_{\eta}}{}_{\tau}(\eta)\, (\overline{q_{1}q_{2}q_{3}})^{Q}(q_{1})^{w}
\end{array} \right. .
\end{array} 
\end{equation}
Here $e,\mu, \tau$ are the electron, the muon and the tau meson, 
$\nu_{e},\nu_{\mu},\nu_{\tau}$ are corresponding neutrinos,
$L_{e}\equiv L_{1},L_{\mu}\equiv L_{2},L_{\tau}\equiv L_{3},$
are leptonic charges. The leptons carry leptonic charges 
as follows: $L_{e}: \,(e,\nu_{e}),\,\,$
$ L_{\mu}:\,(\mu,\nu_{\mu})$ and
$ L_{\tau}:\,(\tau,\nu_{\tau}),$ which are conserved in all interactions. 
The leptons carry also the weak isospins:
$T^{w}_{3}=\FFr{1}{2}$ for  $\nu_{e},\nu_{\mu},\nu_{\tau}$; and
$T^{w}_{3}=-\FFr{1}{2}$ for $e,\mu, \tau$, respectively,
and following electric charges:
$
Q_{\nu_{e}}=Q_{\nu_{\mu}}=Q_{\nu_{\tau}}=0, \quad
Q_{e}=Q_{\mu}=Q_{\tau}=-1.
$
The Q-components $\ps1_{Q}(u_{Q})$ of lepton fields eq.(6.3.1) are made
of singlet combinations of subquarks in Q-world. They imply subcolour
confinement eq.(6.1.4). Then, the MW geometry realization condition 
is already satisfied and leptons may emerge in free combinations without any
constraint. Thus, in suggested scheme there are only three possible
generations of six leptons with integer electric and leptonic charges
have being free of confinement.

\subsection {The Microscopic Structure of Quarks: \\
Quark Generations}
\label {Quark}
The only possible MW- structures of 18 quark 
fields read
\begin{equation}
\label{eq: R12.1}
\begin{array}{l}
\left\{ \begin{array}{l}
u_{i}\equiv {\ps1_{\eta}}{}_{u}(\eta)\,
(q_{2}q_{3})^{Q}(q_{1})^{w}(q_{i}^{B}),\\\\
d_{i} \equiv  {\ps1_{\eta}}{}_{d}(\eta)\,
(\bar{q}_{1})^{Q}(q_{2})^{w}(q_{i}^{B}),
\end{array} \right. 
\quad
\left\{ \begin{array}{l}
c_{i}\equiv {\ps1_{\eta}}{}_{c}(\eta)\,
(q_{3}q_{1})^{Q}(q_{2})^{w}(q_{i}^{B})(q_{c}^{c}),\\\\
s_{i} \equiv {\ps1_{\eta}}{}_{s}(\eta)\,
(\bar{q}_{2})^{Q}(q_{3})^{w}(q_{i}^{B})(\bar{q}_{s}^{c}),
\end{array} \right. 
\\\\
\left\{ \begin{array}{l}
t_{i}\equiv {\ps1_{\eta}}{}_{t}(\eta)\,
(q_{1}q_{2})^{Q}(q_{3})^{w}(q_{i}^{B})(q_{t}^{c}),\\\\
b_{i} \equiv {\ps1_{\eta}}{}_{b}(\eta)\,
(\bar{q}_{3})^{Q}(q_{1})^{w}(q_{i}^{B})(\bar{q}_{b}^{c}),
\end{array} \right. ,
\end{array} 
\end{equation}
where the subcolour index $(i)$ runs through $i=1,2,3$, the 
$(q_{f}^{c})$ are given in eq.(6.3).
Henceforth the subcolour index will be left implicit, but always a 
summation must be extended over all subcolours in B-world.
These fields form three possible doublets of weak isospin in the W-world
$
\left( \begin{array}{c} u \\ d
\end{array} \right),\quad
\left( \begin{array}{c} c \\ s
\end{array} \right),\quad
\left( \begin{array}{c} t \\ b
\end{array} \right).
$
The quark flavour mixing and similar issues are left for treatment 
in sec.6.10.
The corresponding electric charges of quarks
read
$
Q_{u}=Q_{c}=Q_{t}=\FFr{2}{3},\quad
Q_{d}=Q_{s}=Q_{b}=-\FFr{1}{3},
$
in agreement with the rules governing the MW connections 
eq.(6.1.1), where the electric charge difference of up and down quarks
implies
$
\Delta Q=\Delta T^{w}_{3}=1.
$
The explicit form of structure of 
$(q_{f}^{c})$ will be discussed in next section. Here we only note 
that all components of $(q_{f}^{c})$ are made of singlet combinations of 
global subquarks in corresponding internal worlds. They obey a condition
of subcolour confinement. According to eq.(6.1.2), the subcolour confinement
condition for B-world still remains to be satisfied such that the
total quark fields obey to confinement.
Then quarks would not be free particles and unwanted states (since not seen)
like quarks or diquarks etc. are eliminated by construction at the very
beginning. Thus, three quark generations of six possible quark 
fields exist. They carry fractional electric and baryonic charges
and imply a confinement.
Their other charges are left to be discussed below.
Although within considered schemes the subquarks are 
defined on the internal worlds, however the resulting 
$\eta$-components , which we are going to deal with to describe the leptons 
and quarks defined on the spacetime continuum, are affected by 
them. Actually, as it is seen in subsec.4.3 
the rotation through the angle $\theta_{+k}$ yields
a total subquark field 
$$
{q}_{k}(\theta)=\Psi(\theta_{+k})={\ps1_{\eta}}^{0}
\ps1_{u}(\theta_{+k})
$$
where ${\ps1_{\eta}}^{0}$ is the plane wave defined on $\G1_{\eta}$.
Hence, one gets
$$
{q}_{k}(\theta(\eta))={\ps1_{\eta}}^{0}\,{\q1_{u}}{}_{k}(\theta(\eta))=
{\q1_{\eta}}{}_{k}(\theta(\eta))\,{\ps1_{u}}^{0},
\quad {\q1_{\eta}}{}_{k}(\theta(\eta))\equiv f_{(+)}(\theta_{+k}(\eta))
\,{\ps1_{\eta}}^{0}, 
$$
where ${\ps1_{u}}^{0}$ is a plane wave defined on $\G1_{u}$. The
${\q1_{\eta}}{}_{k}(\theta(\eta))$ can be considered as the subquark field
defined on the flat manifold $\G1_{\eta}$ with the same quantum numbers of
${\q1_{u}}{}_{k}(\theta(\eta))$.
Thus, instead of the eq.(6.3.1) and eq.(6.4.1) we may consider on 
equal footing only the resulting $\eta$-components of leptons and quarks 
implying the given same structures. This
enables to pass back to the Minkowski spacetime continuum 
$\G1_{\eta}\rightarrow M_{4}$ (subsec.2.1).

\subsection {The Flavour Group $SU_{f}(6)$}
\label {flavour}
We adopt a simplified view-point on the field
component $(q_{f}^{c})$ (f=u,d,s,c,b,t) eq.(6.3) associated with the 
global distortion rotations in the given worlds s,c,b,t, such that they have 
following microscopic structure with corresponding global charges:
\begin{equation}
\label{eq: R13.1}
\begin{array}{l} 
q^{c}_{u}=q^{c}_{d}=1, \quad 
\bar{q}^{c}_{s}=\left( \overline{q_{1}^{c}q_{2}^{c}q_{3}^{c}} \right)^{s}, 
\quad s =-1;\quad
q^{c}_{c}=\left( q_{1}^{c}q_{2}^{c}q_{3}^{c} \right)^{c}, 
\quad c = 1;\\\\
\bar{q}_{b}^{c}=\left( \overline{q_{1}^{c}q_{2}^{c}q_{3}^{c}} \right)^{b}, 
\quad b = -1;\quad
q_{t}^{c}=\left( q_{1}^{c}q_{2}^{c}q_{3}^{c} \right)^{t}, 
\quad t = 1.
\end{array} 
\end{equation}
To realize the MW-structure the global symmetries
of internal worlds have unified into more higher symmetry including
the generators $\lambda_{3}$ and $\lambda_{8}$ (subsec.6.1).
This global group is the flavour group $SU_{f}(6)$ 
unifying all the symmetries $SU^{c}_{i}$ of the worlds Q,B,s,c,b,t:
$
SU_{f}(6)\supset
SU_{f}(2)\otimes SU^{c}_{B}\otimes SU^{c}_{s}\otimes 
SU^{c}_{c}\otimes SU^{c}_{b}\otimes SU^{c}_{t}.
$
The total symmetry reads
$
G_{tot}\equiv G^{loc}\otimes G^{glob}=G^{loc}\otimes SU_{f}(6),
$
provided
$
G^{loc}\equiv SU^{loc}(3) \otimes G^{loc}_{w},
$
where $G^{loc}_{w}$ is the local symmetry of the electroweak 
interactions (subsec.6.8). The other important aspects
of standard model are left for investigation in the next sections. 
However, below we proceed at once with further exposition of our approach
to consider a gauge invariant Lagrangian of primary field 
with the MW- structure and nonlinear fermion interactions of the 
components.

\subsection{The Primary Field}
\label{Fund}
All the fields including the leptons eq.(6.3.1) and
quarks eq.(6.4.1), along with the spacetime components have also the MW
internal components made of the various constituent subquarks defined on the 
given internal worlds, such that the internal components are consisted of 
distorted ordinary structures (sec.5)
\begin{equation}
\label{eq: R14.1}
\Psi(\theta)=\ps1_{\eta}(\eta)\ps1_{Q}(\theta_{Q})\ps1_{W}(\theta_{W})
\ps1_{B}(\theta_{B})\ps1_{C}(\theta^{c}).
\end{equation}
The components $\ps1_{Q}(\theta_{Q}), \ps1_{W}(\theta_{W}),
\ps1_{B}(\theta_{B})$ are primary massless bare Fermi fields.
We assume that this field has arisen from 
primary field in the lowest state $(s_{0})$ with the same
field components consisted of regular ordinary structures, 
which is motivated by the argument given in the sec.5 that
the  regular ordinary structures directly could not take part 
in link exchange processes with the $\eta$-type regular structure.
Therefore, the primary field  defined on $G_{N}$ 
\begin{equation}
\label{eq: R14.2}
\Psi(0)=\ps1_{\eta}(\eta)\ps1_{Q}(0)\ps1_{W}(0)
\ps1_{B}(0)\ps1_{C}(0)
\end{equation}
serves as the ready made frame into which the distorted
ordinary structures of the same species should be involved.
We apply the Lagrangian of this field possessed local gauge 
invariance written in the notations of App. D:
\begin{equation}
\label{eq: R14.3}
\widetilde{L}_{0}(D)=
\FFr{i}{2} \{ \bar{\Psi}_{e}(\zeta)\,
{}^{i}\gamma
{\D1_{i}}\Psi_{e}(\zeta)-
{\D1_{i}}\bar{\Psi}_{e}(\zeta)\,
{}^{i}\gamma
\Psi_{e}(\zeta) \},
\end{equation}
with the vector indices contracted to form scalars, where
$
{\D1_{i}}=
{\pr_{i}}-ig{\bf \B1_{i}}
(\zeta), $
${\bf \B1_{i}}$ are gauge fields.
Since the components $\ps1_{B}$ and $\ps1_{C}$ will be of no consequence for
a discussion, then we temporarily leave them implicit, 
namely $i=\eta,Q,W$. 
The equation of primary field of the MW- structure with nonlinear fermion 
interactions of the components may be derived from an invariant action 
in terms of local gauge invariant Lagrangian,
which looks like Heisenberg theory [51,52]
\begin{equation}
\label{eq: R14.5}
\widetilde{L}(D)=\widetilde{L}_{0}(D)+\widetilde{L}_{I}+
\widetilde{L}_{B},
\end{equation}
provided by the Lagrangians of nonlinear fermion interactions of the 
components 
$
\widetilde{L}_{I}=\sqrt{2}\widetilde{O}_{1}\otimes L_{I}, 
$
and gauge field
$
\widetilde{L}_{B}=\sqrt{2}\widetilde{O}_{1}\otimes L_{B}.
$
The binding interactions are in the form
\begin{equation}
\label{eq: R14.7}
\begin{array}{l}
L_{I} = {\L1_{Q}}{}_{I} + {\L1_{W}}{}_{I},\quad 
{\L1_{Q}}{}_{I}=\FFr{\lambda}{4}({\J1_{Q}}{}_{L}{\J1_{Q}}{}_{R}^{+}
+{\J1_{Q}}{}_{R}{\J1_{Q}}{}_{L}^{+}), \quad 
{\L1_{W}}{}_{I}=\FFr{\lambda}{2}S_{W}S_{W}^{+}, \\\\
L_{B}=-\FFr{1}{2}Tr(\bf B\bar{\bf B})=-\FFr{1}{2}Tr
\left( {\bf \B1_{i}}
{\bf \B1_{i}} \right),
\end{array}
\end{equation}
where
$$
\begin{array}{l}
{\J1_{Q}}{}_{L,R}=\V_{Q} \mp \A1_{Q},\quad
{\V_{Q}}{}_{(\lambda\alpha)}=\bp_{Q}\gamma_{(\lambda\alpha)}\ps1_{Q},
\quad 
{\V_{Q}}{}_{(\lambda\alpha)}^{+}=
{\V_{Q}}{}^{(\lambda\alpha)}=
\bp_{Q}\gamma^{(\lambda\alpha)}\ps1_{Q} \\\\
{\A1_{Q}}{}_{(\lambda\alpha)}=\bp_{Q}\gamma_{(\lambda\alpha)}
\gamma_{5}\ps1_{Q},
\quad 
{\A1_{Q}}{}_{(\lambda\alpha)}^{+}=
{\A1_{Q}}^{(\lambda\alpha)}=
\bp_{Q}\gamma^{5}\gamma^{(\lambda\alpha)}\ps1_{Q},\quad
S_{W}=\bp_{W}\ps1_{W}, 
\end{array}
$$
$\gamma_{\mu}$ and
$\gamma_{5}=i\gamma_{0}\gamma_{1}\gamma_{2}\gamma_{3}$
are Dirac matrices.
According to Fiertz theorem 
the interaction Lagrangian 
$
{\L1_{Q}}{}_{I}=\FFr{\lambda}{2}(VV^{+} - AA^{+})
$
may be written
$
{\L1_{Q}}{}_{I}=-\lambda (S_{Q}S_{Q}^{+} - P_{Q}P_{Q}^{+}),
$ provided by
$
S_{Q}=\bp_{Q}\ps1_{Q}, \quad
P_{Q}=\bp_{Q}\gamma_{5}\ps1_{Q}.
$
Hence
\begin{equation}
\label{eq: R14.9}
\widetilde{L}(D)=\sqrt{2}\widetilde{O}_{1}\otimes L(D),
\quad
L(D)=\L1_{\eta}(\D1_{\eta})-
\L1_{Q}(\D1_{Q})-
\L1_{W}(\D1_{W}),
\end{equation}
where
$$
\begin{array}{l}
\L1_{\eta}(\D1_{\eta})=
{\L1_{\eta}}'\,{}^{(0)}_{0}(\D1_{\eta})-
\FFr{1}{2}Tr ({\bf \B1_{\eta}}\bar{\bf \B1_{\eta}}),
\quad
\L1_{Q}(\D1_{Q})=
{\L1_{Q}}'\,{}^{(0)}_{0}(\D1_{Q})-
{\L1_{Q}}{}_{I}-
\FFr{1}{2}Tr ({\bf \B1_{Q}}\bar{\bf \B1_{Q}}),
\\\\
\L1_{W}(\D1_{W})=
{\L1_{W}}'\,{}^{(0)}_{0}(\D1_{W})-
{\L1_{W}}{}_{I}-
\FFr{1}{2}Tr ({\bf \B1_{W}}\bar{\bf \B1_{W}}).
\end{array}
$$
Here
$$
\begin{array}{l}
{\L1_{\eta}}'\,{}^{(0)}_{0}=
\FFr{i}{2} \{ 
\bar{\Psi}\stackrel{\frown}{\gamma\D1_{\eta}}\Psi-
\bar{\Psi}\stackrel{\frown}{\gamma\lD1_{\eta}}\Psi
\}= {\ps1_{u}}^{+}
{\L1_{\eta}}{}_{0}^{(0)}
\ps1_{u},\\\\
{\L1_{u}}'\,{}^{(0)}_{0}=
\FFr{i}{2} \{ 
\bar{\Psi}\stackrel{\frown}{\gamma\D1_{u}}\Psi-
\bar{\Psi}\stackrel{\frown}{\gamma\lD1_{u}}\Psi
\}= {\ps1_{\eta}}^{+}
{\L1_{u}}{}_{0}^{(0)}
\ps1_{\eta},
\end{array}
$$
and 
$$
{\L1_{\eta}}{}_{0}^{(0)}=
\FFr{i}{2} \{ 
\bp_{\eta}\stackrel{\frown}{\gamma\D1_{\eta}}\ps1_{\eta}-
\bp_{\eta}\stackrel{\frown}{\gamma\lD1_{\eta}}\ps1_{\eta}
\},\quad
{\L1_{u}}{}_{0}^{(0)}=
\FFr{i}{2} \{ 
\bp_{u}\stackrel{\frown}{\gamma\D1_{u}}\ps1_{u}-
\bp_{u}\stackrel{\frown}{\gamma\lD1_{u}}\ps1_{u}
\}.
$$
The Lagrangian eq.(6.6.6) has the global 
$\gamma_{5}$ and local gauge symmetries. We consider 
only $\gamma_{5}$ symmetry in Q-world, namely  
${\bf \B1_{Q}}\equiv 0$.\\
According to the OMM formalism, it is important to fix the mass shell of 
the stable MW- structure (eq.(5.2.1). It means that
we must take at first the variation of the Lagrangian eq.(6.6.3) with 
respect to primary field eq.(6.6.2), then have
switched on nonlinear fermion interactions of the components.
In other words we take the variation of the Lagrangian eq.(14.6) with
respect to the components on the fixed mass shell.
The equations of free field (${\bf B}=0$) of the MW-structure follow 
at once
\begin{equation}
\label{eq: R14.20}
\hat{p}\Psi_{e}(\zeta)=i\stackrel{\frown}{\gamma \partial}\Psi_{e}(\zeta)=
i\,{}^{i}\gamma
{\pr_{i}}\Psi_{e}(\zeta)=0,\quad
\bar{\Psi}_{e}\lhp=-i{\pr_{i}}\bar{\Psi}_{e}
\,{}^{i}\gamma=0,
\end{equation}
which lead to separate equations for the massless components
$\ps1_{\eta}$, $\ps1_{Q}$ and $\ps1_{W}$:
\begin{equation}
\label{eq: R14.21}
\gamma^{(\lambda\alpha)}
{\p1_{\eta}}{}_{(\lambda\alpha)}\ps1_{\eta}=
i\gamma
{\pr_{\eta}}\ps1_{\eta}=0, \quad
\gamma
{\p1_{Q}}\ps1_{Q}=
i\gamma
{\pr_{Q}}\ps1_{Q}=0, \quad
\gamma
{\p1_{W}}\ps1_{W}=
i\gamma{\pr_{W}}\ps1_{W}=0. 
\end{equation}
The important feature is that the field equations (6.6.7) remain invariant
under the substitution
${\ps1_{Q}}^{(0)}\rightarrow {\ps1_{Q}}^{(m)}$,
where
${\ps1_{Q}}^{(0)}$ and ${\ps1_{Q}}^{(m)}$
are respectively the massless and massive $Q$-component fields, to which
merely the substitution
${\ps1_{\eta}}^{(0)}\rightarrow {\ps1_{\eta}}^{(m)}$ is
corresponded.
In free state the massless field components
$\ps1_{\eta}$, $\ps1_{Q}$ and $\ps1_{W}$ are independent and
due to eq.(6.6.8), the Lagrangian 
\begin{equation}
\label{eq: R14.22}
L'\,{}^{(0)}_{0}= 
{\ps1_{u}}^{+}{\L1_{\eta}}{}_{0}^{(0)}{\ps1_{u}}-
{\ps1_{\eta}}{}^{+}{\L1_{u}}_{0}^{(0)}\ps1_{\eta}=
{\ps1_{u}}^{+}{\L1_{\eta}}{}_{0}^{(0)}{\ps1_{u}}-
{\ps1_{\eta}}^{+}({\L1_{Q}}{}_{0}^{(0)}+{\L1_{W}}{}_{0}^{(0)})\ps1_{\eta}
\end{equation}
reduces to the following:
\begin{equation}
\label{eq: R14.21}
L'\,^{(0)}_{0}= 
{\L1_{\eta}}{}_{0}^{(0)}-{\L1_{u}}{}_{0}^{(0)}=
{\L1_{\eta}}{}_{0}^{(0)}-{\L1_{Q}}{}_{0}^{(0)}-{\L1_{W}}{}_{0}^{(0)}.
\end{equation}
Hence, we implement our scheme as follows:
starting with the reduced Lagrangian 
$L'\,^{(0)}_{0}$ of free field
we shall switch on nonlinear fermion interactions of the 
components. After a generation of nonzero mass of the $\ps1_{Q}$ component
in Q-world (next sec.) we shall look for the corresponding corrections via 
the eq.(6.6.9) to the reduced Lagrangian eq.(6.6.10) of free field.
These corrections mean the interaction between the components governed by the
eq.(6.6.7) and eq.(6.6.9), and do not imply at all the mass acquiring process
for the $\eta$-component (see eq.(6.7.6)).

\subsection{A Generation of the Fermion Mass in the Q-World}
\label{Qrear}
We apply now a well known Nambu-Jona-Lasinio model [56] to 
generate a fermion mass in the Q-world and
start from the chirality invariant total Lagrangian of the field $\ps1_{Q}:
$ 
$
\L1_{Q}={\L1_{Q}}_{0}^{(0)}-{\L1_{Q}}_{I},
$
where a primary field $\ps1_{Q}$ is the massless bare spinor implying 
$\gamma_{5}$ invariance. However,
due to interaction the rearrangement of vacuum state has caused a 
generation of 
nonzero mass of fermion such like to appearance of energy gap in 
superconductor [53-55]. Actually, one may say that particle physicists have 
always shown greater openness to adopt creatively concepts of condensed 
matter physics. In this case as well pursuing the analogy with the 
BCS-Bogoliubov theory of superconductivity, wherein the energy gap is 
created by the electron-electron interaction of Cooper pairs,
in the [56, 57] it was assumed that the mass of Dirac quasi-particle 
excitation is due to some interaction between massless bare fermions, which 
may be considered as a self-consistent (Hartree-Fock) representation of it. 
This approach based on the main 
idea that due to a dynamical instability the field theory in general 
may admit also nontrivial solutions with less symmetry than the
initial symmetry of Lagrangian. Hence, it is considered such possibility
that the field equations may possess higher symmetry, while their solutions 
may reflect some asymmetries arisen due to fact that nonperturbative 
solutions to nonlinear equations do not in general possess the symmetry
of the equations themselves. In [56, 57] the solution of massive fermion 
is obtained which lack the initial $\gamma_{5}$ symmetry of the Lagrangian.
On the analogy of Gor'kov's 
theory [58, 59] it is shown that if one takes into account only the
qualitative dynamical effects connected with rearrangement of vacuum state,
in addition to the trivial solution of equation of massless fermion
a real Dirac quasi-particle will satisfy the equation with non-zero 
self-energy operator $\Sigma(p,m,\widetilde{\lambda},\Lambda)$ 
depending on mass $(m)$,
coupling constant $(\widetilde{\lambda})$ and cut-off 
$(\Lambda)$. In the mean time
$\widetilde{\lambda}=\lambda\,\Gamma(m,\widetilde{\lambda},\Lambda)$, where $\lambda$
is a bare coupling, $\Gamma$ is the vertex function. This theory leads to the 
expression of self-energy operator $\Sigma_{Q}$ for the field $\ps1_{Q}$. 
In lowest order it is quadratically divergent, but with a cutoff can be made 
finite.
Making use of passage $\G1_{Q}\rightarrow {\M1_{Q}}_{4}$ (subsec.2.1), 
one shall proceed directly with the calculation.
In momentum space one gets [56, 57]
\begin{equation}
\label{eq: R15.2}
\begin{array}{l}
\Sigma_{Q}=m_{Q}=-\FFr{8\lambda i}{{(2\pi)}^{4}}\IIn\FFr{m_{Q}}
{p^{2}_{Q}+m^{2}_{Q}-i\varepsilon}F(p_{Q},\Lambda)d^{4}p_{Q},
\end{array}
\end{equation}
where $F(p_{Q},\Lambda)$ is a cutoff factor, $m_{Q}=\mid \Delta_{Q}\mid$,
$\Delta_{Q}=4 \lambda <{\ps1_{Q}}_{R},{\ps1_{Q}}_{L}^{+}>$,
${\ps1_{Q}}_{L,R}=\FFr{1\mp \gamma_{5}}{2}\ps1_{Q}$, $<\cdots>$ specifies
the physical vacuum averaging. 
Besides of trivial solution  $m_{Q}=0$, this equation has also nontrivial 
solution determining $m_{Q}$ in terms of $\lambda$ and $\Lambda$.
Straightforward calculations with invariant cutoff yield the relation 
$
\FFr{2\pi^{2}}{\lambda\Lambda^{2}}=1-\FFr{m^{2}_{Q}}{\Lambda^{2}}
\ln \left( \FFr{\Lambda^{2}}{m^{2}_{Q}}+ 1 \right).
$
The latter is valid only if 
$\FFr{\lambda\Lambda^{2}}{2\pi^{2}}\simeq 1$. After a vacuum 
rearrangement the total Lagrangian of initial massless bare field
${\ps1_{Q}}^{0}$ gives rise to corresponding Lagrangian ${\L1_{Q}}^{(m)}$ 
of massive field ${\ps1_{Q}}^{(m)}:\quad$
$
\L1_{Q}={\L1_{Q}}_{0}^{(0)}-{\L1_{Q}}_{I}={\L1_{Q}}^{(m)}
$
describing Dirac particle 
$
(\gamma p_{Q}-\Sigma_{Q}){\ps1_{Q}}^{(m)}=0.
$
In lowest order 
$
\Sigma_{Q}=m_{Q}\ll \lambda^{-1/2}.
$
Within the refined theory of superconductivity, the collective 
excitations of quasi-particle pairs arise in addition to the individual
quasi-particle excitations when a quasi-particle accelerated in the
medium [55, 60-64]. This leads to the conclusion given in [56, 57] that, in
general, the Dirac quasi-particle is only an approximate description
of an entire system with the collective excitations as the stable
or unstable bound quasi-particle pairs. In a simple approximation 
there arise CP-odd excitations of zero mass as well as CP-even
massive bound states of nucleon number zero and two.
Along the same line we must substitute 
in eq.(6.6.7) the massless field
$\Psi^{(0)}\equiv \ps1_{\eta}{\ps1_{Q}}^{(0)}\ps1_{W}$ by massive
field $\Psi^{(m)}\equiv \ps1_{\eta}{\ps1_{Q}}^{(m)}\ps1_{W}$. 
We obtain
\begin{equation}
\label{eq: R15.7}
\gamma p_{Q}{\ps1_{Q}}^{(m)}=
\Sigma_{Q}{\ps1_{Q}}^{(m)}, \quad
\gamma p_{W}{\Psi}^{(m)}=0,\quad
\gamma p_{\eta}{\Psi}^{(m)}=
(\gamma p_{Q}+\gamma p_{W}){\Psi}^{(m)}=
\Sigma_{Q}{\Psi}^{(m)}. 
\end{equation}
This applies following corrections to eq.(6.6.9):
\begin{equation}
\label{eq: R15.8}
\begin{array}{l}
{\L1_{\eta}}'\,^{(m)}_{0}=
{\ps1_{u}}{}^{+}{\L1_{\eta}}_{0}^{(m)}\ps1_{u}=
{\ps1_{u}}^{+}({\L1_{\eta}}{}_{0}^{(0)}-
\Sigma_{Q}\bp_{\eta}\ps1_{\eta})\ps1_{u} \rightarrow
{\L1_{\eta}}{}_{0}^{(0)}-
\Sigma_{Q}\bar{\Psi}\Psi, \\ \\
{\L1_{Q}}'\,^{(m)}_{0}=
({\ps1_{\eta}}{\ps1_{W}})^{+}
{\L1_{Q}}{}_{0}^{(m)}
({\ps1_{\eta}}{\ps1_{W}})=
({\ps1_{\eta}}{\ps1_{W}})^{+}({\L1_{Q}}{}_{0}^{(0)}-
\Sigma_{Q}\bp_{Q}\ps1_{Q})({\ps1_{\eta}}{\ps1_{W}}) \rightarrow
{\L1_{Q}}{}_{0}^{(0)}-
\Sigma_{Q}\bar{\Psi}\Psi,
\end{array}
\end{equation}
where suffix $(m)$ in $\Psi^{(m)}$ is left implicit. 
A redefinition 
${\ps1_{Q}}^{(0)}\rightarrow {\ps1_{Q}}^{(m)}$
leaves the structure of the piece of the Lagrangian eq.(12.10) involving
only the fields $\ps1_{\eta}$ and
$\ps1_{W}$ unchanged 
\begin{equation}
\label{eq: R15.9}
\begin{array}{l}
L_{0}={\L1_{\eta}}{}_{0}^{(0)}-{\L1_{W}}{}_{0}^{(0)}=
\left( {\L1_{\eta}}{}_{0}^{(0)}-
\Sigma_{Q}\bar{\Psi}\Psi \right)-
\left( 
{\L1_{W}}{}_{0}^{(0)}-
\Sigma_{Q}\bar{\Psi}\Psi
\right)=
{\L1_{\eta}}{}_{0}^{(m)}-{\L1_{W}}{}_{0}^{(m)},
\end{array}
\end{equation}
where the component $\ps1_{Q}$ is left implicit.
The gauge invariant Lagrangian eq.(6.6.6)
takes the form
\begin{equation}
\label{eq: R15.10}
L(D)=\L1_{\eta}(\D1_{\eta})-
\L1_{W}(\D1_{W}),
\end{equation}
where upon combining and rearranging relevant terms we separate the
Lagrangians
\begin{equation}
\label{eq: R15.11}
\L1_{\eta}(\D1_{\eta})=
\FFr{i}{2} \{ 
\bp_{\eta}\stackrel{\frown}{\gamma\D1_{\eta}}\ps1_{\eta}-
\bp_{\eta}\stackrel{\frown}{\gamma\lD1_{\eta}}\ps1_{\eta}
\}-f_{Q}\bar{\Psi}\Psi -\FFr{1}{2}Tr({\bf \B1_{\eta}}\bar{\bf \B1_{\eta}})
\end{equation}
\begin{equation}
\label{eq: R15.12}
\L1_{W}(\D1_{W})=
\FFr{i}{2} \{ 
\bp_{W}\stackrel{\frown}{\gamma\D1_{W}}\ps1_{W}-
\bp_{W}\stackrel{\frown}{\gamma\lD1_{W}}\ps1_{W}
\}-\Sigma_{Q}\bar{\Psi}\Psi-
\FFr{\lambda}{2}S_{W}{S_{W}}^{+}-
\FFr{1}{2}Tr({\bf \B1_{W}}\bar{\bf \B1_{W}}),
\end{equation}
provided by
$
f_{Q}\equiv \Sigma_{Q}(p_{Q},m_{Q},\lambda,\Lambda),\quad 
\Psi=\ps1_{\eta}\ps1_{W}. 
$
\\
The eq.(6.7.6) and eq.(6.7.7) are the Lagrangians that will be further 
evaluated and we shall be concerned within the following.

\subsection {The Electroweak Interactions: the P-Violation}
\label{Symm}
The microscopic approach creates a particular incentive for 
the pertinent concepts and ideas of the unified electroweak interactions. 
We admit that the local rotations in the W-world are occurred at very 
beginning around two arbitrary axes (subsec.6.2), 
namely
$
Dim W^{loc}_{(2)} = N_{(q^{w}_{1},q^{w}_{2})}=2,
$
where N is a subquark number, {\em Dim} is a dimension of local rotations.
The subquarks come up in doublets forming the basis of fundamental 
representation of weak isospin group $SU(2)$
$
q^{w}_{(2)}=
\left(\matrix{
q_{1}\cr
q_{2} \cr}\right)^{w}.
$
The transformations $U$ of local group $SU^{loc}(2)$ are implemented upon the
left- and right-handed fields
$
q_{L,R}({\vec{T}}^{w})=\FFr{1 \mp \gamma_{5}}{2}
q^{w}_{(2)}({\vec{T}}^{w}).
$
If P-symmetry holds, one has
$
q'_{L,R}=U \,q_{L,R}, \quad U\in SU^{loc}(2).
$
But, under such circumstances  
the weak interacting particles could not be realized, because of the
condition of the MW connections eq.(6.1.5), which is not satisfied yet, 
i.e. the Q- and W-worlds could not be realized 
separately.
A simple way of effecting a reconciliation is to assume that
during a realization of weak interacting charged fermions, under the 
action of the Q-world, instead the spanning of the initial world 
$W^{loc}_{(2)}$ into the world of unified electroweak interaction 
$W^{loc}_{(3)}$ took place, where the local rotations always occur around all 
the three axes: 
$W^{loc}_{(2)}\rightarrow W^{loc}_{(3)}$
provided by
$
Dim W^{loc}_{(3)} = 3 \neq N_{(q^{w}_{1},q^{w}_{2})}=2.
$
As far as at the very beginning all the subquark fields in W-world are 
massless, we cannot rule out the possibility that they are transformed independently.
On the other hand, when this situation prevails the spanning 
$
W^{loc}_{(2)}\rightarrow W^{loc}_{(3)}
$
must be occurred compulsory in order to provide a necessary background for 
the condition eq.(6.1.5)
to be satisfied.
The most likely attitude here is that doing away
this shortage the subquark fields $q_{L_{1}}\,\,,q_{L_{2}}\,\,,q_{R_{1}},$
and $q_{R_{2}}$ tend to give rise to triplet.
The three dimensional effective space 
$W^{loc}_{(3)}$ will then arise
\begin{equation}
\label{eq: R16.7}
W^{loc}_{(2)} \ni  q^{w}_{(2)}\,
({\vec{T}}^{w}=\FFr{1}{2})
\rightarrow q^{w}_{(3)}=
\left(\matrix{
q_{R}({\vec{T}}^{w}=0)\cr
\cr
q_{L}({\vec{T}}^{w}=\FFr{1}{2})\cr}
\right)= 
\left(\matrix{
q^{w}_{3}\cr
q^{w}_{1}\cr
q^{w}_{2}\cr}
\right) \equiv
\left(\matrix{
q_{R_{2}}\cr
q_{L_{1}}\cr
q_{L_{2}}\cr}
\right)\in W^{loc}_{(3)}.
\end{equation}
The latter holds if violating initial P-symmetry 
the components $q_{R_{1}},q_{R_{2}}$ have remained in the isosinglet 
states, i.e. the components $q_{L}$ form isodoublet while
$q_{R}$ is a isosinglet:
$
q_{L}\,({\vec{T}}^{w}=\FFr{1}{2}),\quad
q_{R}\,({\vec{T}}^{w}=0).
$
Hence, the mirror symmetry is broken.
Corresponding local transformations are implemented upon triplet
$
{q^{w}}'_{(3)}=f_{W}^{(3)}q^{w}_{(3)},
$
where the unitary matrix of three dimensional local rotations reads
$$
f^{(3)}_{W}=\left( \matrix{
f_{33}  & 0  & 0 \cr 
0       & f_{11}  & f_{12} \cr
0       & f_{21}  & f_{22} \cr
} \right)^{w}.
$$
Making use of incompatibility relations eq.(4.4.5) one gets
\begin{equation}
\label{eq: R16.11}
\|f^{(3)}_{W}\|=f_{33}(f_{11}f_{22}-f_{12}f_{21})=f_{33}
\varepsilon_{123}\varepsilon_{123}
\|f^{(3)}_{W}\| f_{33}^{*},
\end{equation}
or
$
f_{33}f_{33}^{*}=1.
$
That is
$
f_{33}=e^{-i\beta},
$
and
$$
\|f^{(2)}_{W}\|=f_{11}f_{22}-f_{12}f_{21}=
\|f^{(3)}_{W}\| f_{33}^{*}=
\|f^{(3)}_{W}\| e^{i\beta}.
$$
Due to condition $\|f^{(3)}_{W}\|=1$ it reads
$
\|f^{(2)}_{W}\|=e^{i\beta}\neq 1,
$
thus, the initial symmetry $SU^{loc}(2)$ is broken. Restoring it
the fields $q_{L}$ should be undergone to additional transformations
$$
f^{(2)}_{W}\rightarrow{f^{(2)}_{W}}'=\left( \matrix{
f_{11}e^{-i\frac{\beta}{2}}  & f_{12}e^{-i\frac{\beta}{2}} \cr
f_{21}e^{-i\frac{\beta}{2}}  & f_{22}e^{-i\frac{\beta}{2}} \cr
} \right)^{w}
$$
in order to satisfy the unimodularity condition of the matrix of the group
$SU^{loc}(2)$:
$
\|{f^{(2)}_{W}}'\|=\| f^{(2)}_{W}\| e^{-i\beta}=1, \quad
{f^{(2)}_{W}}'\in SU^{loc}(2).
$
While, the expanded group of local rotations in W-world has arisen
\begin{equation}
\label{eq: R16.18}
f^{(3)}_{exp}=\left( \matrix{
e^{-i\beta}  & 0  & 0 \cr
0  & f_{11}e^{-i\frac{\beta}{2}}  & f_{12}e^{-i\frac{\beta}{2}} \cr
0  & f_{21}e^{-i\frac{\beta}{2}}  & f_{22}e^{-i\frac{\beta}{2}} \cr
} \right)\in SU^{loc}(2)_{L}\otimes U^{loc}(1),
\end{equation}
where
$
U=e^{-i\vec{T}^{w}\vec{\theta^{w}}}\in SU^{loc}(2)_{L}, \quad
U_{1}=e^{-iY^{w}\theta_{1}}\in U^{loc}(1).
$
Here $U^{loc}(1)$ is the group of weak hypercharge $Y^{w}$ taking
the following values for left- and right-handed subquark fields:
$
q_{R}:Y^{w}=0,-2,\quad q_{L}:Y^{w}=-1.
$
Whence 
$
q'^{w}_{(3)}=f_{exp}^{(3)}q^{w}_{(3)},
$
and
$$
q'_{L}= {\displaystyle e^{-i\vec{T}^{w}\vec{\theta}^{w}-iY^{w}_{L}\theta_{1}}}
q_{L}, \quad
q'_{R}={\displaystyle e^{-iY^{w}_{R}\theta_{1}}}q_{R}.
$$
\subsection{The Reduction Coefficient and the Weinberg Mixing Angle}
\label{Reduc}
The realization of weak interacting 
particles has always incorporated with the spanning eq.(6.8.1).
This implies P-violation in W-world expressed in the
reduction of initial symmetry group of local transformations of
right-handed components $q_{R}$:
\begin{equation}
\label{eq: R17.1}
\left[ SU(2)\right]_{R}\rightarrow \left[ U(1)\right]_{R},
\end{equation}
The invariance of physical system of the fields
$q_{R}$ under initial group $\left[ SU(2)\right]_{R}$ may be
realized as well by introducing non-Abelian massless vector gauge fields
${\bf A}=\vec{T}^{w}\vec{A}$ with the values in Lie algebra 
of the group $\left[ SU(2)\right]_{R}$. 
Under the reduction eq.(6.9.1) the coupling 
constant $(g)$ changed into $(g')$ specifying the interaction strength
between $q_{R}$ and the Abelian gauge field $B$ associated 
with the local group $\left[ U(1)\right]_{R}$. While
$
g = g'\tan \theta_{w},
$
where $\theta_{w}$ is the Weinberg mixing angle, in terms of which the
reduction coefficient reads
$
r_{p}=\FFr{g-g'}{g+g'}=\FFr{1-\tan \theta_{w}}{1+\tan \theta_{w}}.
$
To define the $r_{p}$ we consider the interaction vertices corresponding
to the groups $\left[ SU(2)\right]_{R}:$
$
g {\bf A}\bar{q}_{R}\gamma\FFr{\bf \tau}{2}q_{R}
\quad$
and $\quad \left[ U(1)\right]_{R}:\quad $
$
g' B\bar{q}_{R}\gamma\FFr{Y^{w}}{2}q_{R}.
$
Notifying that the matrix $\FFr{\lambda_{8}}{2}$ is in the same 
normalization scale as each of the matrices $\FFr{\lambda_{i}}{2}\quad 
(i=1,2,3):$
$
Tr\left( \FFr{\lambda_{8}}{2}\right)^{2}=
Tr\left( \FFr{\lambda_{i}}{2}\right)^{2}=\FFr{1}{2}
$
the vertex scale reads
$
(\mbox{Scale})_{SU(2)}=g \FFr{\lambda_{3}}{2},
$
which is equivalent to $g \FFr{\lambda_{8}}{2}$.
It is obvious that per generator scale should not be changed at the
reduction eq.(6.9.1), i.e.
$
\FFr{(\mbox{Scale})_{SU(2)}}{N_{SU(2)}}=
\FFr{(\mbox{Scale})_{U(1)}}{N_{U(1)}},
$
where $N_{SU(2)}$ and $N_{U(1)}$ are the numbers of generators 
respectively in the groups $SU(2)$ and $U(1)$. 
Hence
$
(Scale)_{U(1)}=\FFr{1}{3}(Scale)_{SU(2)}.
$
Stated somewhat differently, the normalized vertex
for the group $\left[ U(1)\right]_{R}$  reads
$
\FFr{1}{3}g B\bar{q}_{R}\gamma\FFr{\lambda_{8}}{2}q_{R}.
$
In comparing the coefficients can then be equated 
$
\FFr{g'}{g} = \tan \theta_{w}=\FFr{1}{\sqrt{3}},
$
and
$
r_{p}\simeq 0.27.
$
We may draw a statement that during the realization of the MW-structure
the spanning eq.(6.8.1) compulsory occurred, which underlies the
P-violation in W-world incorporated with the reduction eq.(6.9.1). The
latter is characterized by the Weinberg mixing angle 
with the value fixed at $30^{0}$.

\subsection {Emergence of Composite Isospinor-Scalar 
Bosons}
\label{Mes}
The field $q^{w}_{(2)}$ is the W-component of total field
$
q_{(2)}=\q1_{\eta}{}_{(2)}\,{\q1_{W}}{}_{(2)}\,(\equiv
{\ps1_{\eta}}{}_{(2)}\,{\ps1_{W}}{}_{(2)}),
$
where the field component $\q1_{Q}\, (\equiv \ps1_{Q})$ is left implicit.
Instead of it, below we introduce the additional suffix 
$(Q=0,\pm)$ specifying electric charge of the field.
At the very beginning there is an absolute 
symmetry between the components
$
q_{1}={\q1_{\eta}}{}_{1}\,{\q1_{W}}{}_{1}$ and
$
q_{2}={\q1_{\eta}}{}_{2}\,{\q1_{W}}{}_{2}.$
Hence, left-
and right-handed components of fields may be written
\begin{equation}
\label{eq: R12.11.1}
q_{1L}={\q1_{\eta}}{}_{1L}^{(0)}\,{\q1_{W}}{}_{1L}^{(-)}, \quad
q_{2L}={\q1_{\eta}}{}_{2L}^{(-)}\,{\q1_{W}}{}_{2L}^{(0)},\quad
q_{1R}={\q1_{\eta}}{}_{1R}^{(0)}\,{\q1_{W}}{}_{1R}^{(-)}, \quad
q_{2R}={\q1_{\eta}}{}_{2R}^{(-)}\,{\q1_{W}}{}_{2R}^{(0)}.
\end{equation}
On the example of one lepton generation $e$ and $\nu$, without loss of 
generality, we shall exploit the properties of these fields. A
further implication of other fermion generations will be straightforward.
One has
$$
\begin{array}{l}
{\q1_{\eta}}{}_{L}=
\left(\matrix{
{\q1_{\eta}}{}_{1L}^{(0)}\cr
\cr
{\q1_{\eta}}{}_{2L}^{(-)}\cr}
\right)\equiv L =
\left(\matrix{
\nu_{L}\cr
e^{-}_{L}\cr}
\right), \quad
{\q1_{\eta}}{}_{R}=\left({\q1_{\eta}}{}_{1R}^{(0)},\,
{\q1_{\eta}}{}_{2R}^{(-)}
\right)\equiv
R =\left(\nu_{R},\,e^{-}_{R}\right),\\\\
{\q1_{W}}_{L}=
\left(\matrix{
{\q1_{W}}{}_{1L}^{(-)}\cr
\cr
{\q1_{W}}{}_{2L}^{(0)}\cr}
\right),\quad
{\q1_{W}}{}_{R}=\left({\q1_{W}}{}_{1R}^{(-)},\,{\q1_{W}}{}_{2R}^{(0)}\right).
\end{array}
$$
We evaluate the term $f_{Q}\bar{\Psi}\Psi$ in the Lagrangian eq.(6.7.6)
$$
\bar{\Psi}\Psi=\Psi^{+}_{L}\Psi_{R}+\Psi^{+}_{R}\Psi_{L}
\equiv q^{+}_{L}q_{R}+q^{+}_{R}q_{L},
$$
provided by
$$
\begin{array}{l}
q^{+}_{L}q_{R}={\q1_{\eta}}{}_{L}^{+}\,{\q1_{W}}{}_{L}^{+}\,
{\q1_{W}}{}_{R}\,{\q1_{\eta}}{}_{R}=
{\bar{\q1_{\eta}}}{}_{L}\left( 
\gamma^{0}\,{\q1_{W}}{}_{L}^{+}\,{\q1_{W}}{}_{R}
\right){\q1_{\eta}}{}_{R}=\bar{L}\, \varphi \,R,\\\\
q^{+}_{R}q_{L}={\q1_{\eta}}{}_{R}^{+}\,{\q1_{W}}{}_{R}^{+}\,
{\q1_{W}}{}_{L}\,{\q1_{\eta}}{}_{L}=
{\bar{\q1_{\eta}}}{}_{R}\left( 
\gamma^{0}\,{\q1_{W}}{}_{R}^{+}\,{\q1_{W}}{}_{L}
\right){\q1_{\eta}}{}_{L}=\bar{R}\,\varphi^{+} \,L.
\end{array}
$$
For appropriate values of the parameters this term causes
$$
\bar{\Psi}\Psi=\bar{q}_{(2)}\,q_{(2)}=\bar{L}\,\varphi \,R+
\bar{R}\, \varphi^{+}\, L,
$$
where the isospinor-scalar meson field $\varphi$ reads
$$
\varphi \equiv  
\gamma^{0}\,{\q1_{W}}{}_{L}^{+}\,{\q1_{W}}{}_{R},\quad
\varphi^{+}\equiv 
\gamma^{0}\,{\q1_{W}}{}_{R}^{+}\,{\q1_{W}}{}_{L}.
$$
A calculation gives
$$
\varphi=
\left(\matrix{
\varphi_{1}\cr
\varphi_{2}\cr}
\right), \quad
\varphi_{1}\equiv \left( {\q1_{W}}{}_{1L}^{(-)}\right)^{+}
{\q1_{W}}{}_{R},\quad
\varphi_{2}\equiv \left( {\q1_{W}}{}_{2L}^{(0)}\right)^{+}
{\q1_{W}}{}_{R},\quad
\varphi^{+}=\left( \varphi^{+}_{1}, \varphi^{+}_{2} \right).
$$
Hence, the possible two doublets of the composite isospinor-scalar bosons 
read
$$
\varphi_{u}=
\left(\matrix{
\varphi_{1\,u}^{(+)}\cr
\cr
\varphi_{2\,u}^{(0)}\cr}
\right), \quad
\varphi_{d}=
\left(\matrix{
\varphi_{1\,d}^{(0)}\cr
\cr
\varphi_{2\,d}^{(-)}\cr}
\right).
$$
where
$$
\varphi_{1\,u}^{+}\equiv \left( {\q1_{W}}{}_{1L}^{(-)}\right)^{+}
{\q1_{W}}{}_{2R}^{(0)},\quad
\varphi_{2\,u}^{0}\equiv \left( {\q1_{W}}{}_{2L}^{(0)}\right)^{+}
{\q1_{W}}{}_{2R}^{(0)},
$$
and
$$
\varphi_{1\,d}^{0}\equiv \left( {\q1_{W}}{}_{1L}^{(-)}\right)^{+}
{\q1_{W}}{}_{1R}^{(-)},\quad
\varphi_{2\,d}^{-}\equiv \left( {\q1_{W}}{}_{2L}^{(0)}\right)^{+}
{\q1_{W}}{}_{1R}^{(-)}.
$$
In accordance with eq.(6.1.5), the isospinor-scalar
meson carries following weak hypercharge
$
\varphi:Y^{w}=1.
$
Thus, the term $-f_{Q}\bar{\Psi}\Psi$ arisen in the total Lagrangian
of fundamental fermion field eq.(6.7.6) accommodates the Yukawa 
couplings between the fermions and corresponding isospinor-scalar bosons 
in fairy conventional form
\begin{equation}
\label{eq: R12.11.2}
-f_{Q}\bar{\Psi}\Psi=
-f_{e}\left( \bar{L}\,\varphi \,e_{R}+\bar{e_{R}}\,\varphi^{+} \,L\right)-
f_{\nu}\left( \bar{L}\,\varphi_{c} \,\nu_{R}+\bar{\nu_{R}}\,\varphi^{+}_{c} 
\,L\right),
\end{equation}
where the charge conjugated field $\varphi_{c}$ is defined 
$
\left( \varphi_{c}\right)_{i}=\varphi^{*\,k}\varepsilon_{ik}$.
To compute the coupling constants $f_{e}$ and $f_{\nu}$ for the leptons one 
must retrieve their implicit field-components $\ps1_{Q}$. Hence
\begin{equation}
\label{eq: R12.11.2}
f_{i}=tr(\rho_{i}^{Q}\,\Sigma_{Q}), \quad
f_{i}^{\nu}=tr(\rho^{Q\,\nu}_{i}\,\Sigma_{Q}), \quad
\end{equation}
where the density operators $\rho_{i}^{Q}$ and 
$\rho^{Q\,\nu}_{i}$ for given $i$ of the pure ensembles are used
\begin{equation}
\label{eq: R12.11.2}
\begin{array}{l}
\rho_{i}^{Q}=\left(q_{1}q_{2}q_{3}\right)^{Q\,+}_{i}\,
\left(q_{1}q_{2}q_{3}\right)^{Q}_{i}, \quad
\rho^{Q\,\nu}_{i}=\left(q_{i}\bar{q}_{i}\right)^{Q\,+}\,
\left(q_{i}\bar{q}_{i}\right)^{Q},\\\\
tr(\rho_{i}^{Q})^{2}=tr(\rho_{i}^{Q})=1, \quad
tr(\rho_{i}^{Q\,\nu})^{2}=tr(\rho_{i}^{Q\,\nu})=1.
\end{array}
\end{equation}
According to eq.(6.1.6), one gets
\begin{equation}
\label{eq: R12.11.2}
f_{i}=L_{i}^{2}\,\bar{\Sigma}_{Q},\quad 
f_{i}^{\nu}=L_{i}^{2}\,\bar{\Sigma}_{Q}^{\nu},\quad 
\bar{\Sigma}_{Q}\equiv\Sigma_{Q}(\lambda, L)\, \rho^{Q},
\quad
\bar{\Sigma}_{Q}{}^{\nu}\equiv\Sigma_{Q}(\lambda, L)\, \rho^{Q\,\nu},
\end{equation}
where
\begin{equation}
\label{eq: R12.11.2}
\rho^{Q}=\left(q_{1}q_{2}q_{3}\right)^{Q\,+}\,
\left(q_{1}q_{2}q_{3}\right)^{Q}, \quad
\rho^{Q\,\nu}=\left(q\,\bar{q}\right)^{Q\,+}\,
\left(q\,\bar{q}\right)^{Q}.
\end{equation}
An implication of the quarks into this scheme is straightforward if one 
retrieves their implicit field-components 
$\ps1_{Q}, \ps1_{B},\ps1_{C},\quad (C=s,c,b,t)$ (subsec.6.6). On the analogy of 
previous case the coupling constants read
\begin{equation}
\label{eq: R12.11.2}
f_{i}=tr(\rho_{i}\,\Sigma_{Q}),
\end{equation}
where $i=u,d,s,c,b,t.$ Taking into account the MW structure of the 
quarks eq.(6.4.1), we may write down the corresponding density operators 
\begin{equation}
\label{eq: R12.11.2}
\rho_{i}=\rho_{i}^{Q}\rho_{i}^{B}\rho_{i}^{C}
\end{equation}
given in a convention
\begin{equation}
\label{eq: R12.11.2}
\rho_{i}^{A}=\ps1_{A}{}_{i}^{+}\ps1_{A}{}_{i},
\end{equation}
where $\rho_{u}^{C}=\rho_{d}^{C}=1.$

\renewcommand{\theequation}{\thesubsection.\arabic{equation}}
\section{The Higgs Boson}
\label{Higgs}
To break the gauge symmetry down and leading to masses of the
fields , one needs in general, several kinds of spinless Higgs bosons , with 
conventional Yukawa couplings to fermion currents and transforming by an
irreducible representation of gauge group. The Higgs theory like [76-79]
involves these bosons as the ready made fundamental elementary fields, which
entails various difficulties. 
Within outlined here microscopic approach the 
self-interacting isospinor-scalar Higgs bosons arise in the $W$-world
as the collective modes of excitations of the bound quasi-particle iso-pairs. 

\subsection{The Bose Condensate of Iso-Pairs}
\label{Cond}
The ferromagnetism [80],
Bose superfluid [81] and BCS-Bogoliubov model of superconductivity
[53-55] are characterized by the condensation phenomenon leading
to the symmetry-breaking ground state. It is particularly helpful to 
remember that in BCS-Bogoliubov theory the
importance of this phenomenon resides in the possibility suggested by 
Cooper [82] that in the case of an arbitrary weak interaction the pair, 
composed of two mutually interacting electrons above the quiescent Fermi 
sea, remains in a bound state. The electrons filling the Fermi sea do not 
interact with the pair and in the same time they block the levels below
the Fermi surface. The superconductive phase arises due to effective
attraction between electrons occurred by exchange of virtual phonons [83].
In BCS microscopic theory of superconductivity instead of bound states, with 
inception by Cooper, one has a state with strongly correlated electron pairs 
or condensed state in which the pairs form the condensate. The energy
of a system in the superconducting state is smaller than the energy in the
normal state described by the Bloch individual-particle model. The energy gap
arises is due to existence of the binding energy of a pair as a collective 
effect, the width of which is equal to twice the binding energy. According
to Pauli exclusion principle, only the electrons situated in the spherical 
thin shell near the Fermi surface can form bound pairs in which they have
opposite spin and momentum. The binding energy is maximum at absolute zero 
and decreases along the temperature increasing because of the disintegration 
of pairs.
Pursuing the analogy with these ideas in outlined here approach a serious 
problem is to find out the eligible mechanism 
leading to the formation of pairs, somewhat like Cooper mechanism, but
generalized for relativistic fermions, of course in absence of any lattice.
We suggest this mechanism in the framework of gauge invariance 
incorporated with the P-violation phenomenon in W-world.
To trace a maximum resemblance to the 
superconductivity theory,
within this section it will be convenient to describe our approach
in terms of four dimensional Minkowski space ${\M1_{W}}_{4}$ corresponding
to the internal W-world:
$\G1_{W}\rightarrow {\M1_{W}}_{4}$(subsec.2.1). 
Although we shall leave the suffix $(W)$  implicit, but it goes without 
saying that all results obtained within this section refer to the W-world.
According to previous section, we consider the 
isospinor-scalar $\varphi$-meson arisen in the W-world
$$
\varphi(x)=\gamma^{0}{\Psi_{L}}^{+}(x)\Psi_{R}(x),
$$
where $x \in M_{4}$ is a point of the W-world. The following notational
conventions will be employed throughout
$$
{\q1_{W}}{}_{L}\equiv {\ps1_{W}}{}_{L}(\x1_{W})\rightarrow \Psi_{L}(x),
\quad {\M1_{W}}_{4}\rightarrow M_{4},\quad
{\q1_{W}}{}_{R}\equiv {\ps1_{W}}{}_{R}(\x1_{W})\rightarrow \Psi_{R}(x),
$$
where 
$
\Psi_{R}(x)=\gamma(1+\vec{\sigma}\vec{\beta})\Psi_{L}(x), \quad 
\vec{\beta}=\FFr{\vec{v}}{c},\quad
\Psi_{L}(x)=\gamma(1-\vec{\sigma}\vec{\beta})\Psi_{R}(x), \quad
\gamma =\FFr{E}{m},
$
provided by the spin $\vec{\sigma}$, energy $E$ and velocity
$\vec{v}$ of particle. In terms of Fourier integrals
\begin{equation}
\label{eq: R19.1.4}
\Psi_{L}(x)=\FFr{1}{(2\pi)^{4}}\IIn\Psi_{L}(p_{L})
\,{\displaystyle e^{ip_{L}x}} \,d^{4}p_{L},
\quad
\Psi_{R}(x)=\FFr{1}{(2\pi)^{4}}\IIn\Psi_{R}(p_{R})
\,{\displaystyle e^{ip_{R}x}} \,d^{4}p_{R},
\end{equation}
it is readily to get
\begin{equation}
\label{eq: R19.1.5}
\varphi(k)=\IIn\varphi(x)\,
{\displaystyle e^{-ikx}}\,d^{4}x=
\gamma^{0}\IIn\FFr{d^{4}p_{L}}{(2\pi)^{4}}
\,{\Psi_{L}}^{+}(p_{L})\,\Psi_{R}(p_{L}+k)=
\gamma^{0}\IIn\FFr{d^{4}p_{R}}{(2\pi)^{4}}
\,{\Psi_{L}}^{+}(p_{R}-k)\,\Psi_{R}(p_{R})
\end{equation}
provided by the conservation law of fourmomentum
$
k = p_{R}-p_{L},
$
where $k=k(\omega, \vec{k})$, $\,p_{L,R}=p_{L,R}(E_{L,R}, \vec{p}_{L,R})$.
Our arguments on Bose-condensation are based on the local gauge invariance 
of the theory incorporated with the P-violation in weak interactions.
The rationale for this approach is readily forthcoming from the 
consideration of gauge transformations of the fields eq.(7.1.2) under
the P-violation in the W-world
$$
\Psi'_{L}(x)=U_{L}(x)\Psi_{L}(x), \quad
\Psi'_{R}(x)=U_{R}(x)\Psi_{R}(x),
$$
where the Fourier expansions carried out over corresponding {gauge quanta}
with wave fourvectors $q_{L}$ and $q_{R}$
\begin{equation}
\label{eq: R19.1.8}
U_{L}(x)=\IIn\FFr{d^{4}q_{L}}{(2\pi)^{4}}
\,{\displaystyle e^{iq_{L}x}} \,U_{L}(q_{L}), \quad
U_{R}(x)=\IIn\FFr{d^{4}q_{R}}{(2\pi)^{4}}
\,{\displaystyle e^{iq_{R}x}} \,U_{L}(q_{R}),
\end{equation}
and
$
U_{L}(x)\neq U_{R}(x).
$
They induce the gauge transformations implemented upon the $\varphi$-field
$
\varphi'(x) =U(x)\,\varphi(x).
$
The matrix of {\em induced gauge transformations} may be written down
in terms of {\em induced gauge quanta}
\begin{equation}
\label{eq: R19.1.11}
U(x)\equiv U^{+}_{L}(x)\,U_{R}(x)=
\IIn\FFr{d^{4}q}{(2\pi)^{4}}
\,{\displaystyle e^{iq x}} \,U(q), 
\end{equation}
where $q=-q_{L}+q_{R}, \quad q(q^{0},\vec{q})$. In momentum space
one gets
\begin{equation}
\label{eq: R19.1.12}
\begin{array}{l}
\varphi'(k')=\IIn\FFr{d^{4}q}{(2\pi)^{4}}
\,U(q) \,\varphi(k'-q)=
\IIn\FFr{d^{4}k}{(2\pi)^{4}}
\,U(k'-k) \,\varphi(k).
\end{array}
\end{equation}
Conservation of the fourmomentum requires that
$
k'=k + q.
$
According to eq.(7.1.2) and eq.(7.1.5), we have
$$
-p'_{L}+p'_{R}=-p_{L}+p_{R} + q =-p''_{L}+p_{R}=
-p_{L}+p''_{R},
$$
where 
$
p''_{L}=p_{L}-q,\quad
p''_{R}=p_{R}+q.
$
Whence the wave vectors of fermions imply the conservation law
$
\vec{p}_{L}+\vec{p}_{R}=\vec{p}''_{L}+\vec{p}''_{R},
$
characterizing the scattering process of two fermions with {\em effective
interaction caused by the mediating induced gauge quanta}.  We suggest the 
mechanism for the effective attraction between the fermions 
in the following manner: Among all induced gauge 
transformations with miscellaneous gauge quanta we distinguish 
a special subset with the induced gauge quanta of the frequencies
belonged to finite region characterized by the maximum frequency
$\FFr{\widetilde{q}}{\hbar}\quad (\widetilde{q}=max\{q^{0}\})$
greater than the frequency of inducing oscillations fermion force
$
\FFr{\bar{E_{L}}-\bar{E''_{L}}}{\hbar}< \FFr{\widetilde{q}}{\hbar}.
$
To the extent that this is a general phenomenon, we can expect under 
this condition the effective attraction 
(negative interaction) arisen between the fermions caused by exchange of
virtual induced gauge quanta if only the forced oscillations of
these quanta occur in the same phase with the oscillations of
inducing force (the oscillations of fermion density).
In view of this we may think of isospinor $\Psi_{L}$ and isoscalar
$\Psi_{R}$ fields as the fermion fields composing the iso-pairs with the 
same conserving net momentum $\vec{p}={\vec{p}}_{L}+{\vec{p}}_{R}$
and opposite spin, for which the maximum number of negative matrix
elements of operators composed by corresponding creation and annihilation
operators 
$
a^{+}_{\vec{p}''_{R}}\, \,a_{\vec{p}_{R}}\,\,
a^{+}_{\vec{p}''_{L}}\, \,a_{\vec{p}_{L}}
$
(designated by the pair wave vector $\vec{p}$)
may be obtained for coherent ground state with
$
\vec{p}=\vec{p}_{L}+\vec{p}_{R}=0.
$
In the mean time the interaction potential reads
\begin{equation}
\label{eq: R19.1.20}
V=\S_{\vec{p}''_{R},\, \,\vec{p}''_{L},\,\, \vec{p}_{R},\,\, \vec{p}_{L}}
\left( a^{+}_{\vec{p}_{L}} \right)^{+}a^{+}_{\vec{p}''_{R}}
\left( a^{+}_{\vec{p}''_{L}} \right)a_{\vec{p}_{R}}=
\S_{\vec{p}''_{R},\,\, \vec{p}''_{L},\,\, \vec{p}_{R},\,\,\vec{p}_{L}}
a^{+}_{\vec{p}''_{R}}\,\, a^{+}_{\vec{p}''_{L}}\,\,
a_{\vec{p}_{R}}\,\, a_{\vec{p}_{L}},
\end{equation}
implying the attraction between the fermions situated in the spherical
thin shall near the Fermi surface
\begin{equation}
\label{eq: R19.1.20}
V_{\vec{p}\vec{p}''}=\left\{ \begin{array}{ll}
-V\quad \mbox{at} \quad \mid E_{\vec{p}}-E_{F}\mid \leq \widetilde{q},
\quad \mid E_{\vec{p}''}-E_{F}\mid \leq \widetilde{q}, \\
0 \quad \mbox{otherwise}
\end{array}\right..
\end{equation}
The fermions filled up the Fermi sea block
the levels below Fermi surface. Hence, the fermions are in 
superconducting state if the condition eq.(7.1.7) holds. Otherwise, they are 
in normal state described
by Bloch individual particle model. Hence, the Bose-condensate arises
in the W-world as the collective mode of excitations of bound
quasi-particle iso-pairs described by the same wave function in the
superconducting phase
$
\Psi=<\Psi_{L}\Psi_{R}>,
$
where $<\cdots>$ is taken to denote the vacuum averaging. The vacuum of the 
W-world is filled up by such iso-pairs at absolute zero $T=0$.\\
We make a final observation that 
$\Psi_{R}\Psi_{R}^{+}=n_{R}$ is a scalar density number of
right-handed particles. It readily follows that:
\begin{equation}
\label{eq: R19.1.22}
(\Psi_{L}\Psi_{R})^{+}(\Psi_{L}\Psi_{R})=\Psi_{R}^{+}\Psi_{L}^{+}
\Psi_{L}\Psi_{R}=
\FFr{1}{n_{R}}
\Psi_{R}^{+}\gamma^{0}(\gamma^{0}\Psi_{L}^{+}\Psi_{R})
(\Psi_{R}^{+}\Psi_{L}\gamma^{0})
\gamma^{0}\Psi_{R}=\varphi\varphi^{+},
\end{equation}
where
$
\mid \Psi\mid^{2}=
<\varphi\,\varphi^{+}>=\mid <\varphi>\mid^{2}\equiv 
\mid \varphi \mid^{2}.
$
It is convenient to abbreviate the $<\varphi>$ by the 
symbol $\varphi $. The eq.(7.1.8) indeed shows that
the $\varphi$-meson actually arises as the collective mode of excitations of
bound quasi-particle iso-pairs.

\subsection{The Non-Relativistic Approximation}
\label{Appr}
In the approximation to non-relativistic limit
$(\beta \ll 1, \quad \Psi_{L}\simeq \Psi_{R}, \quad \gamma^{0}\rightarrow 1)$
by making use of Ginzburg-Landau's (GL) phenomenological theory [84] it is
straightforward to write down the free-energy functional for the order
parameter in equilibrium superconducting phase in presence of
magnetic field. The self-consistent coupled GL-equations are differential 
equations like Schr\"{o}dinger and Maxwell equations, which relate
the spatial variation of the order parameter $\Psi$ to the vector 
potential $\vec{A}$ and the current ${\vec{j}}_{s}$. In the papers
[58, 59], by means of thermodynamic Green's functions in well defined
limit Gor'kov was able to show that GL-equations are a consequence of
the BCS-Bogoliubov microscopic theory of superconductivity. The theoretical
significance of these works resides in the microscopic interpretation
of all physical parameters of GL-theory. Subsequently these ideas were 
extended to lower temperatures by others [85-87] using a requirement
that the order parameter and vector potential vary slowly over distances of 
the order of the coherence length and that the electrodynamics be local
(London limit). Namely, the validity of derived GLG-equations is restricted
to the temperature $T$, such $T_{c}-T\ll T_{c}$ and to the local
electrodynamics region $q\xi_{0}\ll 1$, where $T_{c}$ is transition 
temperature, $\xi_{0}$ is coherent length characterizing the spatial extent 
of the electron pair correlations, $q$ are the wave numbers of magnetic
field $\vec{A}$. The most important order parameter $\Psi$, 
the mass $m_{\Psi}$ and the coupling constant $\lambda_{\Psi}$ figured in 
GLG-equations read
\begin{equation}
\label{eq: R19.2.2}
\begin{array}{l}
\Psi(\vec{r})=
\FFr{(7\zeta(3)N)^{1/2}}
{4\pi k_{B}T_{c}}\Delta(\vec{r}),\quad
\Delta(T)=
\simeq 3.1k_{B}T_{c}\left( 1-\FFr{T}{T_{c}}\right)^{1/2},\quad
\xi_{0}=\simeq 0.18\FFr{\hbar v_{F}}{k_{B}T_{c}},\\\\
m^{2}_{\Psi}=
1.83\FFr{\hbar^{2}}{m}\FFr{1}{\xi^{2}_{0}}
\left( 1-\FFr{T}{T_{c}}\right),\quad
\lambda^{2}_{\Psi}=
1.4\FFr{1}{N(0)}\left( \FFr{\hbar^{2}}{2m\xi^{2}_{0}}\right)^{2}
\FFr{1}{(k_{B}T_{c})^{2}}.
\end{array}
\end{equation}
Reviewing the notation $\Delta(\vec{r})$ is the energy gap, 
$e^{*}=2e$ is the effective charge, $N(0)$ is the state density at Fermi
surface, $N$ is the number of particles per unit volume in normal mode,
$v_{F}$ is the Fermi velocity, $m\equiv \Sigma_{Q}=f_{Q}$ is the mass of
fermion field. The transition temperature
relates to gap at absolute zero $\Delta_{0}$ [53]. 
The estimate for the pair size at $v_{F}\sim 10^{8}cm/s,
\quad T_{c}\sim 1$ gives [88]
$
\xi_{0} \simeq 10^{-4}cm.
$

\subsection{The Relativistic Treatment}
\label{Rel}
We start with the Lagrangian eq.(6.7.7) of self-interacting
fermion field in W-world, which is arisen from the Lagrangian eq.(6.6.6)
of primary fundamental field after the rearrangement of the vacuum
of the Q-world
\begin{equation}
\label{eq: R19.3.1}
\begin{array}{l}
\L1_{W}(x)=
\FFr{i}{2} \{ 
\bp_{W}(x)\gamma^{\mu}{\pr_{W}}_{\mu}\ps1_{W}(x)-
\bp_{W}(x)\gamma^{\mu}{\lpr_{W}}_{\mu}\ps1_{W}(x)\}-
m\bp_{W}(x)\ps1_{W}(x)-\\\\
-\FFr{\lambda}{2}\bp_{W}(x)\left(\bp_{W}(x)\ps1_{W}(x)\right)\ps1_{W}(x).
\end{array}
\end{equation}
Here, $m =\Sigma_{Q}$ is the self-energy operator of the fermion field 
component in Q-world, the suffix $(W)$ just was put forth in 
illustration of a point at issue. For the sake of simplicity, we also admit 
${\bf\B1_{W}}(x)=0$, but, of course, one is free to restore the gauge field 
${\bf \B1_{W}}(x)$ whenever it will be needed. 
In lowest order the relation
$m\equiv m_{Q}\ll \lambda^{-1/2}$ holds.
The Lagrangian eq.(7.3.1) leads to the field equations
\begin{equation}
\label{eq: R19.3.2}
\begin{array}{l}
(\gamma p-m)\Psi(x)-\lambda \left(\bar{\Psi}(x)\Psi(x) \right)
\Psi(x)=0,\\ \\
\bar{\Psi}(x)(\gamma \overleftarrow{p}+m)+
\lambda \bar{\Psi}(x)\left(\bar{\Psi}(x)\Psi(x) \right)=0,
\end{array}
\end{equation}
where the indices have been suppressed as usual.
At non-relativistic limit the function $\Psi$ reads
$
\Psi \rightarrow e^{imc^{2}t}\Psi,
$
and Lagrangian eq.(7.3.1) leads to Hamiltonian used in [58]. 
In the following our discussion will be in close analogy with the latter. 
We make use of the Gor'kov's technique and evaluate the equations 
(7.3.2) in following manner: 
The spirit of the calculation will be to treat interaction between 
the particles as being absent everywhere except the thin spherical shell
$2\widetilde{q}$ near the Fermi surface. The Bose
condensate of bound particle iso-pairs occurred at zero momentum.
The scattering processes between the particles are absent. 
We consider the matrix elements 
$$
\begin{array}{l}
<T \left(\Psi_{\alpha}(x_{1})\Psi_{\beta}(x_{2})
\bar{\Psi}_{\gamma}(x_{3})\bar{\Psi}_{\delta}(x_{4})
\right)>=-<T \left(\Psi_{\alpha}(x_{1})\bar{\Psi}_{\gamma}(x_{3})\right)>
<T \left(\Psi_{\beta}(x_{2})\bar{\Psi}_{\delta}(x_{4}) \right)>\\\\
+<T \left( \Psi_{\alpha}(x_{1})\bar{\Psi}_{\delta}(x_{4})\right)>
<T \left(\Psi_{\beta}(x_{2})\bar{\Psi}_{\gamma}(x_{3})\right)>+ 
\\\\
<T \left( \Psi_{\alpha}(x_{1})\Psi_{\beta}(x_{2})\right)>
<T \left( \bar{\Psi}_{\gamma}(x_{3})\bar{\Psi}_{\delta}(x_{4})\right)>,
\end{array}
$$
where
$$
\begin{array}{l}
<T \left( \Psi_{\alpha}(x_{1})\Psi_{\beta}(x_{2})\right)>
<T \left( \bar{\Psi}_{\gamma}(x_{3})\bar{\Psi}_{\delta}(x_{4})\right)>
=\\\\
<T \left( \Psi^{+}_{\gamma}(x_{3})\gamma^{0}\Psi^{+}_{\delta}(x_{4})\right)>
<T \left( \gamma^{0} \Psi_{\alpha}(x_{1})\Psi_{\beta}(x_{2})\right)>,
\end{array}
$$
also introduce the functions
\begin{equation}
\label{eq: R19.3.11}
\begin{array}{l}
<N \mid
T \left( \gamma^{0} \Psi(x)\Psi(x')\right)
\mid N+2>=e^{-2i\mu't}F(x-x'), \\ \\
<N +2\mid
T \left( \Psi^{+}(x)\gamma^{0}\Psi^{+}(x')\right)
\mid N>=e^{2i\mu't}F^{+}(x-x').
\end{array}
\end{equation}
Here, $\mu'=\mu + m$, $\mu$ is a chemical potential. 
We omit a prime over $\mu$, but should
understand under it $\mu + m$. 
Let us now make use of Fourier integrals
$$
G_{\alpha\beta}(x-x')=
\IIn\FFr{d\omega\,d\vec{p}}{(2\pi)^{4}}
G_{\alpha\beta}(p)
{\displaystyle e^{i\vec{p}(\vec{x}-\vec{x'})-i\omega (t-t')}}
$$
etc, which render the equation (7.3.2) easier to handle in momentum space
\begin{equation}
\label{eq: R19.3.18}
\begin{array}{l}
(\gamma p-m)G(p)-i\lambda \gamma^{0}F(0+)\bar{F}(p)=
1, \\ \\
\bar{F}(p)(\gamma p+m-2\mu \gamma^{0})+
i\lambda \bar{F}(0+)G(p)=0,
\end{array}
\end{equation}
where
$
F_{\alpha\beta}(0+)=e^{2i\mu t}
<T \left( \gamma^{0} \Psi_{\alpha}(x)\Psi_{\beta}(x)\right)>=
\Lm_{x \rightarrow x'(t \rightarrow t')}F_{\alpha\beta}(x-x').
$
Next we substitute 
$$
(\gamma p-m)=(\omega' -\xi_{p})\gamma^{0},\quad
(\gamma p+m-2\mu \gamma^{0})=\gamma^{0}(\omega' +\xi_{p}^{+}),
$$
where
$$
\begin{array}{l}
\omega' =\omega -\mu'=\omega -m -\mu ,\quad
\xi_{p}=(\vec{\gamma}\vec{p}+m)\gamma^{0}-\mu'=
(\vec{\gamma}\vec{p}+m)\gamma^{0}-m-\mu, \\
\xi_{p}^{+}=\gamma^{0}({\vec{\gamma}}^{+}\vec{p}+m) -m-\mu,
\end{array}
$$
and omit a prime over $\omega'$ for the rest of this section.
We employ
$$
F(0+)=-JI, \quad 
I=\left( \matrix{
\hspace{0.3cm} 0 \quad 1\cr
-1 \quad 0 \cr
}\right),\quad
F^{+}(0+)F(0+)=-J^{2}I^{2}=J^{2}, 
$$
and
$\widehat{\omega} +\widehat{\xi}_{p}=
\gamma^{0}(\omega +\xi_{p})$.
The gap function $\Delta$ reads
$
\Delta^{2}=\lambda^{2} J^{2},
$
where
$
J=\IIn\FFr{d\omega\,d\vec{k}}{(2\pi)^{4}}F^{+}(p).
$
Making use of standard rules [81], one may pass over the poles. 
This method allows oneself to extend the study up to limit of
temperatures, such that $T_{c}-T\ll T_{c}$, by making use of thermodynamic 
Green's function.
Hence
\begin{equation}
\label{eq: R19.3.32}
\begin{array}{l}
F^{+}(p)=-i\lambda J (\omega-\xi_{p}+i\delta)^{-1}
(\omega+\xi_{p}-i\delta)^{-1} 
-\FFr{\pi\Delta}{\varepsilon_{p}}n(\varepsilon_{p})
\{ \delta(\omega-\varepsilon_{p})+
\delta(\omega+\varepsilon_{p})\}, 
\\\\
G(p)=\gamma^{0}\{u_{p}^{2}(\omega-\xi_{p}+i\delta)^{-1}+
v_{p}^{2}(\omega+\xi_{p}-i\delta)^{-1}+ 
2\pi i n(\varepsilon_{p})
[ u_{p}^{2}\delta(\omega-\varepsilon_{p})-\\\\
v_{p}^{2}\delta(\omega+\varepsilon_{p})]\},
\end{array}
\end{equation}
where 
$u_{p}^{2}=\FFr{1}{2}\left( 1+ \FFr{\xi_{p}}{\varepsilon_{p}}\right),
\quad
v_{p}^{2}=\FFr{1}{2}\left( 1- \FFr{\xi_{p}}{\varepsilon_{p}}\right),\quad
\varepsilon_{p}=(\xi_{p}^{2}+\Delta^{2}(T))^{1/2}.
$
and
$n(\varepsilon_{p})$ is the usual Fermi function
$
n(\varepsilon_{p})=\left( \exp\FFr{\varepsilon_{p}}{T} +1\right)^{-1}.
$
Then 
\begin{equation}
\label{eq: R19.3.35}
1=\FFr{\mid \lambda \mid}{2(2\pi)^{3}}
\IIn d\vec{k}\, \FFr{1-2n(\varepsilon_{k})}{\varepsilon_{k}(T)}
\quad \left( \mid \xi_{p}\mid
< \widetilde{q} \right),
\end{equation}
determining the energy gap $\Delta$ as a function of $T$. According to
eq.(17.3.6), the $\Delta(T)\rightarrow 0$ at $T\rightarrow T_{c}\sim
\Delta(0)$ [53].
\subsection{Self-Interacting Potential of Bose-Condensate}
\label{Pot}
To go any further in exploring the form and significance of obtained
results it is entirely feasible to include 
the generalization of the equations (6.1.3.4) in presence of spatially 
varying magnetic field with vector potential $\vec{A}(\vec{r})$, which is
straightforward $(t\rightarrow \tau=it)$
\begin{equation}
\label{eq: R19.4.1}
\begin{array}{l}
\left\{ 
-\gamma^{0}\FFr{\partial}{\partial \tau}-
i\vec{\gamma}\left(\FFr{\partial}{\partial \vec{r}}-ie\vec{A}(\vec{r}) 
\right)-m +\gamma^{0}\mu 
\right\}
G(x,x')+\gamma^{0}\Delta(\vec{r})\bar{F}(x,x')=
\delta (x-x'), \\ \\
\bar{F}(x,x')
\left\{ 
\gamma^{0}\FFr{\partial}{\partial \tau}+
i\vec{\gamma}\left(\FFr{\partial}{\partial \vec{r}}+ie\vec{A}(\vec{r}) 
\right)-m +\gamma^{0}\mu 
\right\}-
\Delta^{*}(\vec{r})\gamma^{0}G(x,x')=0,
\end{array}
\end{equation}
where the thermodynamic Green's function [89, 90] is used and
the energy gap function is in the form
$
\Delta^{*}(\vec{r})=\lambda F^{+}(\tau,\vec{r};\,\tau,\vec{r}).
$
This function is logarithmically divergent, but with a cutoff of
energy of interacting fermions at the spatial distances in order
of $\FFr{\hbar v}{\widetilde{\omega}}$ can be made finite, 
where $\widetilde{\omega} \equiv \FFr{\widetilde{q}}{\hbar}$.
If one uses the Fourier components of functions $G(x,x')$ and
$F(x,x')$
\begin{equation}
\label{eq: R19.4.3}
G(\vec{r},\vec{r}';u)=T\S_{n}e^{-i\omega u}
G_{\omega}(\vec{r},\vec{r}'), \quad
G_{\omega}(\vec{r},\vec{r}')=\FFr{1}{2}\IIn^{1/T}_{-1/T}e^{i\omega u}
G(\vec{r},\vec{r}';u)d\,u,
\end{equation}
where $u=\tau - \tau'$, $\omega$ is the discrete index $\omega =
\pi T(2n + 1), \quad n=0,\pm 1, \ldots$, then the eq.(7.4.1) reduces
to
\begin{equation}
\label{eq: R19.4.4}
\begin{array}{l}
\left\{ 
i\omega\gamma^{0}-
i\vec{\gamma}\left(\vec{\partial}_{\vec{r}}-ie\vec{A}(\vec{r}) 
\right)-m +\gamma^{0}\mu 
\right\}
G_{\omega}(\vec{r},\vec{r}')+
\gamma^{0}\Delta(\vec{r})
\bar{F_{\omega}}(\vec{r},\vec{r}')=
\delta (\vec{r}-\vec{r}'), \\ \\
\bar{F_{\omega}}(\vec{r},\vec{r}')
\left\{ 
-i\omega\gamma^{0}+
i\vec{\gamma}\left(\vec{\partial}_{\vec{r}}+ie\vec{A}(\vec{r}) 
\right)-m +\gamma^{0}\mu 
\right\}-
\Delta^{*}(\vec{r})\gamma^{0}G_{\omega}(\vec{r},\vec{r}')=0,
\end{array}
\end{equation}
where the gap function is defined by
$
\Delta^{*}(\vec{r})=\lambda T\S_{n}
F^{+}_{\omega}(\vec{r},\vec{r}').
$
The Bloch individual particle Green's function
$\widetilde{G}_{\omega}(\vec{r},\vec{r}')$ for the fermion in
normal mode is written
\begin{equation}
\label{eq: R19.4.6}
\left\{ 
i\omega\gamma^{0}-
i\vec{\gamma}\left(\vec{\partial}_{\vec{r}}-ie\vec{A}(\vec{r}) 
\right)-m +\gamma^{0}\mu 
\right\}
\widetilde{G}_{\omega}(\vec{r},\vec{r}')=
\delta (\vec{r}-\vec{r}'), 
\end{equation}
or the adjoint equation
\begin{equation}
\label{eq: R19.4.7}
\left\{ 
i\omega\gamma^{0}+
i\vec{\gamma}\left(\vec{\partial}_{\vec{r}'}+ie\vec{A}(\vec{r}') 
\right)-m +\gamma^{0}\mu 
\right\}
\widetilde{G}_{\omega}(\vec{r},\vec{r}')=
\delta (\vec{r}-\vec{r}'). 
\end{equation}
By means of eq.(7.4.5) the eq.(7.4.3) gives rise to
\begin{equation}
\label{eq: R19.4.8}
G_{\omega}(\vec{r},\vec{r}')=\widetilde{G}_{\omega}(\vec{r},\vec{r}')-
\IIn\widetilde{G}_{\omega}(\vec{r},\vec{s})
\gamma^{0}\Delta(\vec{s})
\bar{F_{\omega}}(\vec{s},\vec{r}')d^{3}s,
\end{equation}
and
\begin{equation}
\label{eq: R19.4.9}
\bar{F_{\omega}}(\vec{r},\vec{r}')=
\IIn\widetilde{G}_{\omega}(\vec{s},\vec{r}')
\Delta^{*}(\vec{s})\gamma^{0}
\widetilde{G}_{-\omega}(\vec{s},\vec{r})d^{3}s.
\end{equation}
The gap function $\Delta(\vec{r})$ as well as 
$\bar{F_{\omega}}(\vec{r},\vec{r}')$ are small ones at close neighbourhood
of transition temperature $T_{c}$ and varied slowly over a coherence
distance. This approximation, which went into the derivation of 
equations,  meets our interest in eq.(7.4.6), eq.(7.4.7).
Using standard procedure one may readily express them in power series of $\Delta$ and $\Delta^{*}$ by keeping
only the terms in $\bar{F_{\omega}}(\vec{r},\vec{r}')$ up to the cubic
and in $G_{\omega}(\vec{r},\vec{r}')$ - quadratic order in 
$\Delta$. After averaging over the polarization of particles the
following equation coupling $\Delta(\vec{r})$ and 
$\vec{A}(\vec{r})$ ensued:
\begin{equation}
\label{eq: R19.4.10}
\begin{array}{l}
\overline{\Delta^{*}(\vec{r})}=\lambda T\S_{n}
\IIn 
\overline{\widetilde{G}_{\omega}(\vec{r},\vec{r}')
\widetilde{G}_{-\omega}(\vec{r},\vec{r}')}
\Delta^{*}(\vec{r}')d^{3}r'
-\\\\ 
\lambda T\S_{n}
\IIn\IIn\IIn
\overline{\widetilde{G}_{\omega}(\vec{s},\vec{r})
\widetilde{G}_{-\omega}(\vec{s},\vec{l})
\widetilde{G}_{\omega}(\vec{m},\vec{l})
\widetilde{G}_{-\omega}(\vec{m},\vec{r})}
\Delta(\vec{s})\Delta^{*}(\vec{l})\Delta^{*}(\vec{m})
d^{3}s \, d^{3}l \, d^{3}m.
\end{array}
\end{equation}
It is worthwhile to determine the function
$\widetilde{G}_{\omega}(\vec{r},\vec{r}')$. 
In the absence of applied magnetic field, i.e. $\vec{A}(\vec{r})=0$
the eq.(7.4.4) reduces to
\begin{equation}
\label{eq: R19.4.11}
\left\{ 
(i\omega+\mu )\gamma^{0}-
\vec{\gamma}\vec{p} -m  
\right\}
\widetilde{G}^{0}_{\omega}(\vec{r},\vec{r}')=
\delta (\vec{r}-\vec{r}').
\end{equation}
It is well to study at this point certain properties of the solution
which we shall continually encounter
\begin{equation}
\label{eq: R19.4.12}
\widetilde{G}^{0}_{\omega}(\vec{r},\vec{r}')=
\FFr{1}{2m}\left\{ 
(i\omega+\mu )\gamma^{0}-
\vec{\gamma}\vec{p} +m  
\right\}
\widetilde{G}^{0}_{G\omega}(\vec{r}-\vec{r}'),
\end{equation}
where the function $\widetilde{G}^{0}_{G\omega}(\vec{r}-\vec{r}')$
satisfies the equation
\begin{equation}
\label{eq: R19.4.13}
\FFr{1}{2m}\left( q^{2}+\Delta
\right)
\widetilde{G}^{0}_{G\omega}(\vec{r}-\vec{r}')=
\delta (\vec{r}-\vec{r}'),
\end{equation}
provided by
$
q^{2}=(i\omega+\mu )^{2}-m^{2}=2im\Omega +p^{2}_{0}
$
and
$
\Omega =\omega\FFr{\mu}{m}, \quad
p^{2}_{0}=\mu^{2}-m^{2}.
$
At $\mu \gg \mid \omega\mid$ one has $\FFr{p^{2}_{0}}{2m}
\gg \mid \Omega\mid$, and
\begin{equation}
\label{eq: R19.4.16}
\FFr{1}{2m}
\left\{
2im\Omega +p^{2}_{0}+\Delta
\right\}
\widetilde{G}^{0}_{G\omega}(\vec{r}-\vec{r}')=
\delta (\vec{r}-\vec{r}'),
\end{equation}
the solution of which reads
$$
\widetilde{G}^{0}_{G\omega}(\vec{r}-\vec{r}')=
-\FFr{m}{2\pi R}\exp(iqR),
$$
where
$$
q=sgn\, \Omega \,p_{0}+i\FFr{\mid \Omega \mid}{v}, \quad
R=\mid \vec{r}-\vec{r}' \mid, \quad
sgn\, \Omega=\FFr{\Omega}{\mid \Omega \mid}.
$$
In approximation to non-relativistic limit $\vec{p}\rightarrow 0$
this Green's function reduces to\\
$\widetilde{G}^{0}_{G\omega}(\vec{r}-\vec{r}')$ used in [58].
Making use of Fourier integrals we readily get
$$
\widetilde{G}^{0}_{\omega}(\vec{p})=
\FFr{\mu \gamma^{0}+\vec{\gamma}\vec{p} +m }
{q^{2}-\vec{p}^{2}+i0}=
\widehat{\gamma}\,\widetilde{G}^{0}_{G\omega}(\vec{p}),
$$
where 
$
\widehat{\gamma}=\FFr{1}{2m}
(\mu \gamma^{0}+\vec{\gamma}\vec{p} +m ).
$
One has
$$
\widetilde{G}^{0}_{G\omega}(\vec{p})=
\widetilde{G}^{0}_{G\Omega}(\vec{p})=
\FFr{1}{i\Omega -\xi}=\widetilde{G}^{0}_{G\Omega}(-\vec{p}),
$$
where as usual $\xi=\FFr{\vec{p}}{2m}-\FFr{\vec{p}_{0}}{2m}$.
The Green's function $\widetilde{G}_{\omega}(\vec{r},\vec{r}')$ in presence
of magnetic field differs from 
$\widetilde{G}_{\omega}^{0}(\vec{r}-\vec{r}')$ 
only by phase multiplier [58]
$$
\widetilde{G}_{\omega}(\vec{r},\vec{r}')=
\exp\left\{
\FFr{ie}{c}\left(\vec{A}(\vec{r}), \vec{r}-\vec{r}' \right)
\right\}
\widetilde{G}_{\omega}^{0}(\vec{r}-\vec{r}').
$$
The technique now is to expand a second term in right-hand side of 
the eq.(7.4.8)
up to the terms quadratic in $(\vec{r}-\vec{r}')$.
After calculations it transforms 
\begin{equation}
\label{eq: R19.4.40}
\begin{array}{l}
\left\{
\left(i\hbar\vec{\nabla}+\FFr{e^{*}}{c}\vec{A} \right)^{2}+
\FFr{2m}{\nu}\left[ 
\FFr{2\pi^{2}}{\lambda m p_{0}}\left(\FFr{\mu}{m} \right)^{2}
\left( \FFr{\mu}{m}-1\right)+
\left(\FFr{\mu}{m} \right)^{2}
\left( \FFr{T}{T_{c\mu}}-1\right)+
\right. \right.\\\\ 
\left. \left.
\FFr{2}{N}\mid \Psi(\vec{r})\mid^{2}
\right]
\right\}\Psi(\vec{r})=0,
\end{array}
\end{equation}
where $\nu=\FFr{7\zeta(3)mv_{F}^{2}}{24(\pi k_{B}T_{c})^{2}}$ and 
$T_{c\mu}= \FFr{m}{\mu}T_{c}$.
Succinctly 
\begin{equation}
\label{eq: R19.4.41}
\left\{
\vec{p}_{A}^{2}-\FFr{1}{2}m_{\Psi}^{2}+\FFr{1}{4}\lambda_{\Psi}^{2}
\mid \Psi(\vec{r})\mid^{2}
\right\}\Psi(\vec{r})=0,
\end{equation}
provided by
\begin{equation}
\label{eq: R19.4.41}
\begin{array}{l}
m_{\Psi}^{2}(\lambda,T,T_{c\mu})=
\FFr{24}{7\zeta(3)}\left(\FFr{\hbar}{\xi_{0}} \right)^{2}
\left(\FFr{\mu}{m} \right)^{2}
\left[
1-\FFr{T}{T_{c\mu}}-\left( \FFr{\mu}{m}-1\right)
\ln\FFr{2\widetilde{\omega}}{\Delta_{0}}
\right],\\ \\
\lambda_{\Psi}^{2}(\lambda,T_{c})=
\FFr{96}{7\zeta(3)}\left(\FFr{\hbar}{\xi_{0}} \right)^{2}
\FFr{1}{N}, \quad 
\Psi(\vec{r})=\Delta(\vec{r})\FFr{\left(7\zeta(3)N\right)^{1/2}}
{4\pi k_{B}T_{c}}.
\end{array}
\end{equation}
According to the eq.(7.4.15), the 
magnitude of the relativistic effects, however, is found to be greater to
account for the large contribution to
the values $m_{\Psi}^{2}$ and $\lambda_{\Psi}^{2}$. While the transition
temperature decreases inversely by the relativistic factor
$\FFr{\mu}{m}$.
A spontaneous breakdown of symmetry of ground state occurs at
$
\eta_{\Psi}^{2}(\lambda,T < T_{c\mu}) > 0,
$
where
$
\eta_{\Psi}^{2}(\lambda,T,T_{c\mu})=\FFr{m_{\Psi}^{2}}{\lambda_{\Psi}^{2}}.
$
The eq.(7.4.14) splits into the couple of equations for
$\Psi_{L}$ and $\Psi_{R}$. Subsequently, a Lagrangian of the
$\varphi$
will be arisen  with
the corresponding values of mass 
$m_{\Psi}^{2}\equiv m_{\varphi}^{2}$
and coupling constant 
$\lambda_{\Psi}^{2}\equiv\lambda_{\varphi}^{2}$.

\subsection{The Four-Component Bose-Condensate in Magnetic Field}
\label{Four}
Now we are going to derive the equation of four-component 
bispinor field of Bose-condensate, which may sound strange, but due 
to self-interaction the spin part of it is vanished. We start with 
the nonsymmetric state $\Delta_{L}\neq \Delta_{R}$, 
where $\Psi_{L}$ and $\Psi_{R}$ are two eigenstates of chirality 
operator $\gamma_{5}$. In standard representation
\begin{equation}
\label{eq: R19.5.1}
\begin{array}{l}
\Psi=\FFr{1}{\sqrt{2}}
\left(
\matrix{
1 & 1\cr
1 & -1\cr
}
\right)
\left(
\matrix{
\Psi_{L}\cr
\Psi_{R}\cr
}
\right)
=\FFr{1}{\sqrt{2}}
\left(
\matrix{
\Psi_{L}+\Psi_{R}\cr
\Psi_{L}-\Psi_{R}\cr
}
\right)
\equiv
\left(
\matrix{
\Psi_{1}\cr
\Psi_{2}\cr
}
\right),\\ \\
\Delta
=
\left(
\matrix{
\Delta_{1}\cr
\Delta_{2}\cr
}
\right)=
\FFr{1}{\sqrt{2}}
\left(
\matrix{
\Delta_{L}+\Delta_{R}\cr
\Delta_{L}-\Delta_{R}\cr
}
\right), \quad
\Delta_{L}\neq \Delta_{R}.
\end{array}
\end{equation}
The eq.(7.4.14) enables to postulate the equation of four-component 
Bose-condensate in magnetic field and equilibrium state 
\begin{equation}
\label{eq: R19.5.2}
i\hbar \FFr{\partial \Psi}{\partial t}=
\left\{
c\vec{\alpha}\left(\vec{p}+ \FFr{e^{*}}{c}\vec{A} \right)+\beta mc^{2}+
M(F)+L(F)\mid \Psi\mid^{2}
\right\}\Psi=0,
\end{equation}
or succinctly
\begin{equation}
\label{eq: R19.5.3}
\left( \gamma p_{A} -m \right)\Psi=0.
\end{equation}
This is in standard notations
$$
\begin{array}{l}
\gamma^{0}=\beta=\left(
\matrix{
1 &0\cr
0 & -1\cr
}
\right), \quad \vec{\alpha}=\gamma^{0}\vec{\gamma}=
\left(
\matrix{
0 & \vec{\sigma}\cr
\vec{\sigma} & 0\cr
}
\right),\quad F=F_{\mu\nu}\sigma^{\mu\nu}=inv, 
\\ 
\sigma^{\mu\nu}=\FFr{1}{2}\left[\gamma^{\mu},\gamma^{\nu} \right],\quad
\FFr{i}{2}e^{*}F=-e^{*}\vec{\Sigma}\vec{H},\quad
\vec{H}=rot \vec{A},\quad 
F_{\mu\nu}=(0,\vec{H}), \quad \vec{\Sigma}=
\left(
\matrix{
\vec{\sigma} &0\cr
0 &\vec{\sigma}\cr
}
\right),\\
p_{A}=(p_{A0},\vec{p}_{A}), \quad 
\vec{p}_{A}=i\hbar\vec{\nabla}+
\FFr{e^{*}}{c}\vec{A},\quad p_{A0}=-\left(M(F)+L(F)\mid \Psi\mid^{2} 
\right).
\end{array}
$$
The $M(F)$ and $L(F)$ are some functions depending upon the invariant F,
which will be determined under the requirement that the second-order 
equations ensued from the eq.(7.5.3) must match onto eq.(7.4.14).
Defining the functions $M(F)$ and $L(F)$: 
$$
\begin{array}{l}
M(F)=\left(M^{2}_{0}+ \FFr{i}{2}e^{*}F \right)^{1/2}, \quad
M_{0}=\left( m^{2}+\FFr{1}{2}m^{2}_{\varphi} \right)^{1/2},\quad
L(F)=-\FFr{\lambda^{2}_{\varphi}}{8M(F)}, \\ \\
L_{0}=-\FFr{\lambda^{2}_{\varphi}}{8M_{0}},
M(F)L(F)=M_{0}L_{0}=-\FFr{1}{8}\lambda^{2}_{\varphi},
\end{array}
$$
and taking into account an approximation fitting our interest that the gap
function is small at close neighbourhood of transition temperature,
one gets
\begin{equation}
\label{eq: R19.5.8}
\left\{ \vec{p}_{A}^{2} +m^{2}-\left(M_{0}+L_{0}\mid \Psi\mid^{2} 
\right)^{2} \right\}\Psi(\vec{r})\equiv
\left\{ \vec{p}_{A}^{2}-\FFr{1}{2}m^{2}_{\varphi}+
\FFr{1}{4}\lambda^{2}_{\varphi}\mid \Psi\mid^{2} 
 \right\}\Psi(\vec{r})=0.
\end{equation}
This has yet another 
important consequence that at $\Delta_{L}\neq 0$ and
imposed constraint\\
$
\left( m+M(F) + L(F)\mid \Psi\mid^{2}\right)_{F\rightarrow 0}
\rightarrow 0
$
we have
\begin{equation}
\label{eq: R19.5.12}
\Delta_{2}=
\FFr{1}{\sqrt{2}}(\Delta_{L}-\Delta_{R})=0, \quad
\Psi_{2}=0,
\end{equation}
where $\mid \Psi_{0}\mid $ is the gap function symmetry-restoring 
value
$$
\Delta_{2}\left( \mid \Psi_{0}\mid^{2}=\FFr{m+M_{0}}{-L_{0}}\right)
=0,
\quad
\Delta_{L}\left( \mid \Psi_{0}\mid^{2}\right)=
\Delta_{R}\left( \mid \Psi_{0}\mid^{2}\right),
$$
where, according to eq.(7.5.4), one has
\begin{equation}
\label{eq: R19.5.15}
V\equiv \left[m^{2}-\left( M_{0}+L_{0}\mid \Psi\mid^{2}\right)^{2}\right]
\Psi^{2} =\left[-\FFr{1}{2}m^{2}_{\varphi}+
\FFr{1}{4}\lambda^{2}_{\varphi}\mid \Psi\mid^{2}\right]
\Psi^{2},
\end{equation}
and 
\begin{equation}
\label{eq: R19.5.16}
V\left( \mid \Psi_{0}\mid^{2}=\FFr{m+M_{0}}{-L_{0}}=
\FFr{1}{2}\eta^{2}_{\varphi}(\lambda,T,T_{c\mu})
\right)=0.
\end{equation}
It leads us to the conclusion that the field
of symmetry-breaking Higgs boson must be counted off from the
$\Delta_{L}=\Delta_{R}$ symmetry-restoring value of Bose-condensate
$
\mid \Psi_{0}\mid=\FFr{1}{\sqrt{2}}\eta_{\varphi}(\lambda,T,T_{c\mu})
$
as the point of origin describing the excitation in the neighbourhood
of stable vacuum eq.(7.5.7).
We may write down the Lagrangian corresponding to the eq.(7.5.3)
$$
L_{\Psi}=\FFr{1}{2}\left\{ 
\bar{\Psi} \gamma p_{A} \Psi -\bar{\Psi} \gamma \overleftarrow{p}_{A}\Psi
\right\}-m\bar{\Psi}\Psi.
$$
The gauge invariant Lagrangian eq.(6.7.7) takes the form
\begin{equation}
\label{eq: R19.5.19}
\L1_{W}(\D1_{W})=\FFr{i}{2}
\left\{ 
\bp_{W} \gamma \D1_{W} \ps1_{W} -
\bp_{W} \gamma \overleftarrow{\D1_{W}}\ps1_{W}
\right\}- 
\bp_{W}\left\{ 
m +\gamma^{0}\left[M(F) + L(F)\mid \Psi\mid^{2}\right] 
\right\}\ps1_{W}.
\end{equation}
At the symmetry-restoring point, this Lagrangian 
can be replaced 
$$
\L1_{W}(\D1_{W})\rightarrow {\L1_{W}}{}_{1}(\D1_{W})=
\FFr{1}{2}\left( \D1_{W}{\ps1_{W}}{}_{1} \right)^{2} -
\V_{W}\left(\mid {\ps1_{W}}{}_{1} \mid^{2} \right),
$$
provided by
$$
\V_{W}\left(\mid {\ps1_{W}}{}_{1} \mid^{2} \right)=
-\FFr{1}{2}m^{2}_{\varphi}{\ps1_{W}}{}_{1}^{2}+
\FFr{1}{4}\lambda^{2}_{\varphi}{\ps1_{W}}{}_{1}^{4}.
$$
Taking into account the eq.(7.1.8), in which
$
\mid {\ps1_{W}}{}_{1}\mid^{2}=\mid \varphi\mid^{2}=
\FFr{1}{2}\mid \eta_{\varphi} +\chi\mid^{2},
$
one gets
\begin{equation}
\label{eq: R19.5.23}
{\L1_{W}}{}_{\varphi}(\D1_{W})=
\FFr{1}{2}\left( \D1_{W}\varphi \right)^{2} -
\V_{W}\left(\mid \varphi \mid^{2} \right),
\quad
\V_{W}\left(\mid \varphi \mid^{2} \right)=
-\FFr{1}{2}m^{2}_{\varphi}\varphi^{2} +
\FFr{1}{4}\lambda^{2}_{\varphi}\varphi^{4}.
\end{equation}
The average value of well-defined current source expressed in terms of 
spinless field $\Psi$ eq.(7.5.3) is given by
$
\left.\vec{j}(\vec{r}) \right|_{\vec{\Sigma}=0}.
$
According to eq.(7.4.6) and eq.(7.4.7), one has
\begin{equation}
\label{eq: R19.6.6}
\begin{array}{l}
\left.\delta G(x,x')\right|_{x=x'}=
-T\S_{\omega}
\IIn\widetilde{G}_{G\Omega}(\vec{r},\vec{s})
\widetilde{G}_{G\Omega}(\vec{l},\vec{r})
\widetilde{G}_{G-\Omega}(\vec{l},\vec{s})
\Delta^{*}(\vec{l})\Delta(\vec{s})d^{3}s\,d^{3}l.
\end{array}
\end{equation}
Hence
\begin{equation}
\label{eq: R19.6.11}
\left.\vec{j}(\vec{r}) \right|_{\vec{\Sigma}=0}=
\simeq \FFr{m}{4(m-M_{0}-L_{0}\mid \Psi\mid^{2})}
\left( \gamma^{0}+\FFr{\mu}{m}\right)
\left\{ 
\left( \gamma^{0}+\FFr{\mu}{m} \right)^{2}+3\beta^{2}_{F}
\right\}
\vec{j}_{G}(\vec{r}),
\end{equation}
provided by
\begin{equation}
\label{eq: R19.6.10}
\vec{j}_{G}(\vec{r})=\left( \FFr{m}{\mu}\right)^{3}
\left\{ 
\FFr{-ie^{*}\hbar}{2m}
\left(\Psi\FFr{\partial \Psi^{*}}{\partial \vec{r}}- 
\Psi^{*}\FFr{\partial \Psi}{\partial \vec{r}}\right)-
\FFr{{e^{*}}^{2}}{mc}\vec{A} | \Psi |^{2}
\right\}.
\end{equation}
Below we write the eq.(7.5.3)
in the form on close analogy of the elementary excitations in 
superconductivity model described by coherent mixture of electrons and
holes near the Fermi surface [54, 55, 91]. In chirial representation
$
\Psi=
\left(
\matrix{
\Psi_{L}\cr
\Psi_{R}\cr
}
\right)
$
one has
$$
p_{A0}\Psi_{L}=\vec{\sigma}\vec{p}_{A}\Psi_{L}+m\Psi_{R}, \quad 
p_{A0}\Psi_{R}=-\vec{\sigma}\vec{p}_{A}\Psi_{R}+m\Psi_{L}, \quad 
p_{A0}=\pm \left(\vec{p}_{A}^{2}+m^{2} \right)^{1/2}.
$$
The two states of quasi-particle are separated in energy by 
$2\mid p_{A0} \mid$. In the ground state all quasi-particles should be
in lower (negative) energy states. It would take a finite 
energy $2\mid p_{A0} \mid\geq 2m$ to excite a particle to the upper
state (the case of Dirac particle). Thus, one may assume that the energy
gap parameter $m$ is also due to some interaction between massless
bare fermions [56, 57].
\subsection{Extension to Lower Temperatures}
\label{Ext}
It is worth briefly recording the question of whether or not it is 
possible to extend the ideas of former approach to lower temperatures as 
it was investigated in the case of Gor'kov's theory by others [85-87].
Here, as usual we admit that the order parameter and vector potential vary 
slowly over distances of the order of the coherence length. 
We restrict ourselves to the London limit and the derivation of equations
will be proceeded by iterating to a low order giving only the leading 
terms. Taking into account the eq.(7.4.1), eq.(7.4.3), eq.(7.4.6) 
and eq.(7.4.7), it is straightforward to derive the separate integral
equations for $G$ and $F^{+}$ in terms of $\Delta$, $\Delta^{*}$
and $\widetilde{G}$.
We introduce
\begin{equation}
\label{eq: R19.7.8}
K_{\Omega}(\vec{r},\vec{s})=\delta (\vec{r}-\vec{s})
\left\{
i\Omega + \FFr{1}{2m} \left(\FFr{\partial}{\partial \vec{s}} -
 i\FFr{e}{c}\vec{A}(\vec{s}) \right)^{2}+\mu_{0}
\right\}+
\Delta(\vec{r})\widehat{\gamma}_{A}^{2}(\vec{s})
\widetilde{G}_{-\Omega}(\vec{s},\vec{r})
\Delta^{*}(\vec{s}),
\end{equation}
provided by $\mu_{0}=\FFr{p_{0}^{2}}{2m}$,
and
$$
F^{+}_{\omega}(\vec{s},\vec{r}')=\widehat{\gamma}_{A}(\vec{s})
F^{+}_{\Omega}(\vec{s},\vec{r}'),\quad
\widehat{\gamma}_{A}(\vec{s})=\FFr{1}{2m}\{
\gamma^{0}(i\omega+\mu)-\vec{\gamma}{\vec{p}}_{A}(\vec{s})+m\}.
$$
We write down the coupled equations in the form
\begin{equation}
\label{eq: R19.7.10}
\IIn d^{3}s\, K_{\Omega}(\vec{r},\vec{s})G_{\Omega}(\vec{s},\vec{r}')=
\delta (\vec{r}-\vec{r}'),
\end{equation}
and
\begin{equation}
\label{eq: R19.7.11}
\IIn d^{3}s\, F_{\Omega}^{+}(\vec{s},\vec{r}')K_{-\Omega}(\vec{s},\vec{r})=
\Delta^{*}(\vec{r})\widehat{\gamma}_{A}(\vec{r}')
\widetilde{G}_{\Omega}(\vec{r},\vec{r}'),
\end{equation}
The mathematical structure of obtained equations is closely similar to
that studied by [85, 92, 93] in somewhat different context. So, adopting 
their technique we introduce sum and difference coordinates, and
Fourier transform with respect to the difference coordinates as follows:
\begin{equation}
\label{eq: R19.7.12}
K_{\Omega}(\vec{p},\vec{R})\equiv \IIn d^{3}(r-s)e^{-i\vec{p}
(\vec{r}-\vec{s})}K_{\Omega}(\vec{r},\vec{s})
\end{equation}
with $\vec{R}\equiv \FFr{1}{2}(\vec{r}+\vec{s})$. We also involve similar 
expansions for all other functions. Then, the eq.(7.6.2) and
eq.(7.6.3) reduce to following:
\begin{equation}
\label{eq: R19.7.13}
\begin{array}{l}
\Theta\left[K_{\Omega}(\vec{p},\vec{R}) 
G_{\Omega}(\vec{p}',\vec{R}')
\right]=1,\\\\ 
\Theta\left[
F^{+}_{\Omega}(\vec{p}',\vec{R}')
K_{-\Omega}(-\vec{p},\vec{R})
\right]=
\Theta\left[
\Delta^{*}(\vec{R})\widehat{\gamma}_{A}(\vec{p}',\vec{R}')
\widetilde{G}_{\Omega}(\vec{p}',\vec{R}')
\right]
\end{array}
\end{equation}
provided by the standard differential operator of finite order
defined under the requirement that it produces the Fourier transform of
the matrix product of two functions when it operates on the transforms
of the individual functions [92, 93]
\begin{equation}
\label{eq: R19.7.14}
\Theta\equiv \Lm_{\matrix{
\vec{R}'\rightarrow \vec{R}\cr
\vec{p}'\rightarrow \vec{p}\cr
}}\exp\left[ 
\FFr{i}{2}\left( 
\FFr{\partial}{\partial\vec{R}}\FFr{\partial}{\partial\vec{p}'}-
\FFr{\partial}{\partial\vec{p}}\FFr{\partial}{\partial\vec{R}'}
\right)
\right].
\end{equation}
One gets
\begin{equation}
\label{eq: R19.7.15}
\begin{array}{l}
K_{\Omega}(\vec{p},\vec{R})=i\Omega - \epsilon (\vec{p},\vec{R})+
\Lm_{\vec{R}',\vec{R}''\rightarrow \vec{R}}
\exp\left[ 
\FFr{i}{2}\FFr{\partial}{\partial\vec{p}}
\left( \FFr{\partial}{\partial\vec{R}'}-
\FFr{\partial}{\partial\vec{R}''}\right)
\right]\times \\\\ 
\times
\Delta^{*}(\vec{R}')\widehat{\gamma}_{A}(\vec{p},\vec{R}')
\widehat{\gamma}_{A}(-\vec{p},\vec{R})\Delta^{*}(\vec{R}''),
\end{array}
\end{equation}
where it is denoted
$
\epsilon (\vec{p},\vec{R})\equiv \FFr{1}{2m}\left( 
\vec{p}-\FFr{e}{c}\vec{A}(\vec{R})
\right)^{2}-\mu_{0}.
$
To obtain resulting expressions we shall proceed with further 
calculations, 
but shall forbear to write them out as they are so standard. 
There is only one thing to be noticed about the integration. That is,
due to the angular
integration in momentum space, as mentioned above, the terms linear in the
vector $\vec{p}$ will be vanished , as well as the integration over the 
energies removes the linear terms in $\epsilon (\vec{p})$. So, we may
expand the quantities in eq.(7.6.5) according to the degree of
inhomogenity somewhat like it we have done in equation (7.4.8) of
gap function $\Delta^{*}(\vec{r})$, which in mixed 
representation transforms to the following:
\begin{equation}
\label{eq: R19.7.17}
\Delta^{*}(\vec{R})=
T\S_{\omega}\IIn\FFr{d^{3}p}{(2\pi)^{3}}
F^{+}_{\omega}(\vec{p},\vec{R})
=T\S_{\omega}\IIn\FFr{d^{3}p}{(2\pi)^{3}}
\widehat{\gamma}_{A}(\vec{p},\vec{R})
F^{+}_{\Omega}(\vec{p},\vec{R}).
\end{equation}
The approximation was used to obtain the function $F^{+}_{\Omega}$
must be of one order higher 
$F^{+}_{\Omega}\simeq F^{(0)+}_{\Omega}+F^{(1)+}_{\Omega}+
F^{(2)+}_{\Omega}$ than that for function 
$\widetilde{G}_{\Omega}\simeq \widetilde{G}_{\Omega}^{(0)}+
\widetilde{G}_{\Omega}^{(1)}$. Employing an iteration method of solution
one replaces $K\rightarrow \widetilde{K},\quad 
G\rightarrow \widetilde{G}$ in eq.(7.6.5) and puts $\Theta^{(0)}=1,\quad
\widetilde{K}^{(1)}=0, \quad \widetilde{G}^{(1)}=0$. Hence 
$\widetilde{G}=\widetilde{G}^{(0)}$. 
The resulting equation
for gap function is similar to those occurring in [85], although 
not identical. The sole difference is that in the resulting equation we use
the expressions of $\Omega$ and $\xi$. With this replacement the  
equation reads 
\begin{equation}
\label{eq: R19.7.19}
\begin{array}{l}
\Delta^{*}=
T\S_{\omega}\IIn\FFr{d^{3}p}{(2\pi)^{3}}
\left\{ 
\FFr{\Delta^{*}}{\Omega^{2} +\xi^{2}}+
\left[ \left( \FFr{\partial}{\partial \vec{R}}+\FFr{2ie}{c}\vec{A} 
\right)^{2}\Delta^{*}\right.\right.
+\Delta\left( 
\left( \FFr{\partial}{\partial \vec{R}}+\FFr{2ie}{c}\vec{A} 
\right)\Delta^{*}\right)^{2}\FFr{\partial}{\partial|\Delta |^{2}}+\\\\ 
+\left.\left.\FFr{\Delta^{*}}{3}\FFr{\partial^{2}|\Delta |^{2}}
{\partial \vec{R}^{2}}
\FFr{\partial}{\partial|\Delta |^{2}}+
\FFr{\Delta^{*}}{6}\left(
\FFr{\partial |\Delta |^{2}}{\partial \vec{R}}\right)^{2}
\FFr{\partial^{2}}{\partial\left(|\Delta |^{2}\right)^{2}}
\right]\FFr{p^{2}/6m^{2}}{\left( \Omega^{2}+ \xi^{2}\right)^{2}}
\right\},
\end{array}
\end{equation}
where 
$
\xi(\vec{p},\vec{R})\equiv \left[\epsilon^{2}(\vec{p})+
|\Delta(\vec{R}) |^{2} \right]^{1/2}.
$
The average value of the operator of current
density at $T\sim T_{c}$ follows at once
\begin{equation}
\label{eq: R19.7.21}
\begin{array}{l}
\vec{j}_{G}(\vec{R})=\left(\overline{\gamma^{0}{\widehat{\gamma}}^{3}} 
\right)_{\Sigma=0}\FFr{2e}{m}
\left\{ -\FFr{i}{2}\left( \Delta^{*}(\vec{R})
\FFr{\partial \Delta (\vec{R}) }{\partial \vec{R}}-
\FFr{\partial \Delta^{*} (\vec{R}) }{\partial \vec{R}}\Delta(\vec{R})
\right)-\right.\\ \\
\left.-\FFr{2e}{c} |\Delta (\vec{R})|^{2} \vec{A}(\vec{R})\right\}
T\S_{\omega}\IIn\FFr{d^{3}p}{(2\pi)^{3}}
\FFr{p^{2}/3m}{\left( \Omega^{2}+ \xi^{2}(\vec{p},\vec{R})\right)^{2}},
\end{array}
\end{equation}
where 
$\left(\overline{\gamma^{0}{\widehat{\gamma}}^{3}} 
\right)_{\Sigma=0}=\FFr{1}{8}\left( \gamma^{0}+\FFr{\mu}{m}\right)
\left\{ \left( \gamma^{0}+\FFr{\mu}{m}\right)^{2}+3\beta_{F}^{2}
\right\}$. At $\Delta\ll \pi k_{B}T$ and $\vec{A}$ is independent of
position the eq.(7.6.9) and eq.(7.6.10) lead back to the equations
(7.4.13) and (7.5.11).
Actually, from such results it is then easy by ordinary manipulations
to investigate the pertinent
physical problem in several particular cases, but a separate calculation
for each case would be needed.

\renewcommand{\theequation}{\thesection.\arabic{equation}}

\section{Lagrangian of Electroweak Interactions;
The Transmission of the Electroweak Symmetry Breaking From 
the $W$-World to Spacetime Continuum}
\label{simul}
The results obtained within the previous subsections enable 
us to trace unambiguously rather general scheme of unified electroweak 
interactions, where the self-interacting isospinor 
scalar Higgs bosons have arisen as the collective modes of excitations of 
bound quasi-particle iso-pairs on the internal $W$-world. 
But, at the very first we remind some features
allowing  us to write down the final Lagrangian of electroweak 
interactions.\\
1. During the realization of the MW connections of weak interacting 
fermions under the action of the $Q$-world the P-violation compulsory 
occurred in the W-world incorporated with the symmetry reduction eq.(6.7.1)
characterized by the Weinberg mixing angle with
the fixed value at $30^{0}$. This gives rise to 
the local symmetry $SU(2)\otimes U(1)$, under which the
left-handed fermions transformed as six independent doublets, 
while the right-handed fermions transformed as twelve 
independent singlets.\\
2. Due to vacuum rearrangement in Q-world the Yukawa couplings arise 
between the fermion fields and corresponding isospinor-scalar $\varphi$-
meson in conventional form.\\
3. In the framework of suggested mechanism providing the effective 
attraction between the relativistic fermions caused by the exchange of 
the mediating induced gauge quanta in W-world, the 
self-interacting isospinor-scalar Higgs bosons arise 
as Bose-condensate, namely the $SU(2)$ multiplets of spinless $\varphi$-meson 
fields coupled to the gauge fields in a gauge invariant way. Thus,  
in the Lagrangian of $\varphi$-meson with the degenerate 
vacuum of the W-world the symmetry-breaking Higgs 
boson is counted off from the gap symmetry restoring
value as the point of origin.
In view of this the total Lagrangian ensues from the eq.(6.7.5)-
eq.(6.7.7), which is now invariant under the
local symmetry $SU(2)\otimes U(1)$, where a set of gauge fields are coupled 
to various multiplets of fields among which is also a multiplet of 
Higgs boson. Subsequently, we separate a piece of Lagrangian 
containing only the fields defined on four dimensional Minkowski 
flat spacetime continuum $M_{4}$.
To facilitate writing we shall forbear here to write out the piece of 
Lagrangian containing the terms of other fermion generations than one, 
as it is a somewhat lengthy and so standard. But, in the mean time, we shall 
retain the explicit terms of Higgs bosons arisen on the internal $W-$world 
to emphasize the specific mechanism of the electroweak symmetry breakdown 
discussed below.
The resulting Lagrangian reads 
\begin{equation}
\label{eq: R20.1}
\begin{array}{l}
L=-\FFr{1}{2}TrG_{\mu\nu}G^{\mu\nu}-\FFr{1}{4}F_{\mu\nu}F^{\mu\nu}+
i\bar{L}\hat{D}L +i\bar{e}_{R}\hat{D}e_{R}+ 
i\bar{\nu}_{R}\hat{D}\nu_{R}+|\D1_{W}{}_{\mu}\varphi |^{2}-\\\\
\FFr{1}{2}\lambda^{2}_{\varphi}\left( |\varphi |^{2}-
\FFr{1}{2}\eta^{2}_{\varphi}\right)^{2}- 
f_{e}\left(\bar{L} \varphi e_{R}+\bar{e}_{R} \varphi^{+}L \right)
-f_{\nu}\left(\bar{L} \varphi_{c} \nu_{R}+\bar{\nu}_{R} 
\varphi^{+}_{c}L \right)+\\ \\
\mbox{similar terms for other fermion generations}.
\end{array}
\end{equation}
The gauge fields ${\bf A}_{\mu}(x)$ and $B_{\mu}(x)$ associate 
respectively with the groups $SU(2)$ and $U(1)$,
where the gauge covariant curls are $F_{\mu\nu}, G_{\mu\nu}.$
The corresponding gauge covariant derivatives are in standard form.
One took into account corresponding values of the operators
${\bf T}$ and $Y$ for left- and right-handed fields, and for isospinor
$\varphi$-meson. The Yukawa coupling constants $f_{e}$ and
$f_{\nu}$ are inserted in subsec.6.10.
Since the electroweak symmetry is at any rate only approximate, the test 
of the theory will depend on its ability to account for its breaking as well. 
here the MSM creates a particular incentive for the study of such a breaking.
Then, just it remains to see how can such Higgs bosons arisen on the internal 
$W$-world break the gauge symmetry down in $M_{4}$ and lead to masses of the 
spacetime-components of the MW-fields? 
It is remarkable to see that the suggested MSM, in contrast to the 
SM, predicts the transmission of electroweak symmetry breaking from the 
$W-$world to the $M_{4}$ spacetime continuum. Actually,
in standard scenario for the simplest Higgs sector, a gauge invariance of 
the Lagrangian is broken when the $\varphi$-meson fields eq.(7.5.9) acquire a 
VEV $\quad\eta_{\varphi}\neq 0$ in the $W$-world.
While the mass $m_{\varphi}$ and coupling 
constant $\lambda_{\varphi}$ are in the form eq.(7.4.15). 
The spontaneous breakdown of symmetry is vanished at 
$\eta_{\varphi}^{2}(\lambda, T > T_{c\mu}) < 0$.
When this doublet obtains a VEV, three of the
gauge fields $\Z1_{W}{}^{0}_{\mu},\,\W1_{W}{}^{\pm}_{\mu}$ acquire masses.
These fields are the $W$-components of the mesons mediating the weak 
interactions. Certainly, the derivative 
$$\D1_{W}{}_{\mu}\,\varphi \equiv \left( \pr_{W}{}_{\mu}-
\FFr{i}{2} \,g\,
{\bf \tau  \W1_{W}}{}_{\mu}-
\FFr{i}{2} \,g'\,\X1_{W}{}_{\mu}\right)\,\varphi
$$
arisen in the eq.(8.1) leads to the masses  
$$
M_{W}=\FFr{g\,\eta_{\varphi}}{2}, \quad 
M^{2}_{Z}=\FFr{\left(g^{2}+g'{}^{2}\right)^{1/2}\,\eta_{\varphi}}{2}, \quad
\cos\,\theta_{W}=\FFr{M_{W}}{M_{Z}},
$$ 
respectively of the gauge field components
$$
\W1_{W}{}^{\pm}_{\mu}=\FFr{1}{\sqrt{2}}\left(\W1_{W}{}^{1}_{\mu} 
\pm \W1_{W}{}^{2}_{\mu}\right),
\quad
\Z1_{W}{}_{\mu}=\FFr{g\,\W1_{W}{}^{3}_{\mu}-
g'\,\X1_{W}{}_{\mu}}{\left(g^{2}+g'{}^{2}\right)^{1/2}}\equiv
\cos\,\theta_{W}\,\W1_{W}{}^{3}_{\mu}-\sin\,\theta_{W}\,\X1_{W}{}_{\mu}.
$$
Consequently, a massless gauge field 
$$
\A1_{W}{}_{\mu}=\FFr{g'\,\W1_{W}{}^{3}_{\mu}+
g\,\X1_{W}{}_{\mu}}{\left(g^{2}+g'{}^{2}\right)^{1/2}}\equiv
\sin\,\theta_{W}\,\W1_{W}{}^{3}_{\mu}+\cos\,\theta_{W}\,\X1_{W}{}_{\mu}.
$$
may be identified as the $W$-component of the photon field coupled to the 
electric current.
Therewith the $x$-components of the fields above simultaneously 
acquire corresponding masses too, since, according to the specific MW scheme 
(eq.(6.6.7)), all the components of corresponding frame fields are 
defined on the MW mass shells, i.e.,
$$
\Bx_{x}\,\W1_{x}{}_{\mu}=M^{2}_{W}\,\W1_{x}{}_{\mu},\quad
\Bx_{x}\,\Z1_{x}{}_{\mu}=M^{2}_{Z}\,\Z1_{x}{}_{\mu},\quad
\Bx_{x}\,\A1_{x}{}_{\mu}=M^{2}_{A}\,\A1_{x}{}_{\mu},
$$
provided by
$$
M^{2}_{W}\,\W1_{W}{}_{\mu}\equiv \Bx_{W}\,\W1_{W}{}_{\mu},\quad
M^{2}_{Z}\,\Z1_{W}{}_{\mu}\equiv \Bx_{W}\,\Z1_{W}{}_{\mu},\quad 
M^{2}_{A}\,\A1_{W}{}_{\mu}\equiv \Bx_{W}\,\A1_{W}{}_{\mu}=0. 
$$
The microscopic structure
of these fields reads
$$
\begin{array}{l}
W^{+}\equiv {\phi}{}_{W}(\eta)\, 
(q_{1}q_{2}q_{3})^{Q}(q\bar{q})^{W}, \quad
W^{-}\equiv {\phi}{}_{W}(\eta)\, 
(\overline{q_{1}q_{2}q_{3}})^{Q}(\bar{q}q)^{W}, \\\\
Z^{0}\equiv {\phi}{}_{Z}(\eta)\, 
(q\bar{q})^{Q}(q\bar{q})^{W},\quad
A \equiv {\phi}{}_{A}(\eta)\, 
(q\bar{q})^{Q}\A1_{W}(0).
\end{array}
$$
The values of the masses $M_{W}$ and $M_{Z}$ are changed if the Higgs
sector is built up more compoundly.
Due to Yukawa couplings the fermions acquire the masses after 
symmetry-breaking. The mass of electron reads
$
m_{e}=\FFr{\eta_{\varphi}}{\sqrt{2}}f_{e}
$
etc. One gets for the leptons
$
f_{e}:f_{\mu}:f_{\tau}=m_{e}:m_{\mu}:m_{\tau}.
$
This mechanism does not disturb the
renormalizability of the theory  [94, 95].
In approximation to lowest order
$
f=\Sigma_{Q}\simeq m_{Q}\ll \lambda^{-1/2}\quad
\left( \lambda^{-1}=\FFr{mp_{0}}{2\pi^{2}}
\ln\FFr{2\widetilde{\omega}}{\Delta_{0}}\right),
$
the Lagrangian eq.(8.1) produce the Lagrangian of phenomenological
SM, where at $f\sim 10^{-6}$ one gets $ \lambda\ll 10^{12}$.

\section{ The Two Solid Phenomenological Implications of the MSM}
\label{lagr}
Discussing now the relevance of our present approach to the physical 
realities we should attempt to provide some ground for checking the 
predictions of the MSM against experimental evidence. It is remarkable that
the resulting theory makes plausible following 
testable implications for the current experiments at LEP2, at the Tevatron and
LHC discussed below, which are drastically different from the predictions of 
conventional models:\\
1. Due to the specific mechanism 
of the electroweak symmetry breakdown given in the previous subsection,
the first important phenomenological implication of the MSM 
ensued at once that: 

$\bullet$ {\em the Higgs bosons never could emerge in spacetime 
continuum and, thus,  could not be discovered in experiments nor at any 
energy range}.\\\\
2. According to previous subsection, the lowest pole $m_{Q}$ of the 
self-energy operator $\Sigma_{Q}$ in eq.(8.1) has fixed the whole mass 
spectrum of the SM particles. But, in general, one must also take into 
account the mass spectrum of expected various collective excitations of 
bound quasi-particle pairs produced by higher-order interactions as a 
``superconductive'' solution obtained from a nonlinear spinor field 
Lagrangian of the $Q$-component possessed $\gamma_{5}$ invariance (subsec.6.7). 
These states must be considered as a direct effect of the same primary 
nonlinear fermion interaction which provides the mass of the $Q$-component of 
Fermi field, which itself is a collective effect. 
They would manifest themselves 
as stable or unstable states. The general features of mass spectrum of 
the collective excitations and their coupling with the fermions are 
discussed in [56, 57] through the use of the Bethe-Salpeter 
equation handled in the simplest ladder approximation incorporated with the 
self-consistency conditions, when one is 
still left with unresolved divergence problem. 
One can reasonably expect that these results for the bosons of small masses at 
low energy compared to the unbound fermion states are essentially correct in 
spite of the very simple approximations.
Therein, some bound states are predicted too the obtained mass values 
of which are rather high, and these states should decay very quickly. 
The high-energy poles may in turn determine the low-energy resonances.
All this prompt us to expect that the other poles different from those of 
lowest one in turn will produce the new heavy SM family partners.
Hence, one would expect a second important phenomenological implication of the 
MSM that: 

$\bullet$ {\em for each of the three SM families of quarks and 
leptons there are corresponding heavy family partners with the same 
$SU(3)_{c}\otimes SU(2)_{L}\otimes U(1)_{Y}$ quantum numbers at the energy 
scales related to next poles with respect to lowest one}.\\\\ 
To see its nature, now we may estimate the energy threshold of 
creation of such heavy family partners using the results far obtained in 
[56, 57]. It is therefore necessary under the simplifying assumption to 
consider in the $Q$-world a composite system of dressed fermion $(N_{*})$ 
made of the unbound fermion $(N)$ coupled with the different kind 
two-fermion bound states $(N\,\bar{N})$ at low energy, which all together 
represent the primary manifestation of the fundamental interaction. Such a 
dressed fermion would have a total mass $m_{*}\simeq m_{Q} + \mu$, where 
$m_{Q}$ and $\mu$ are the masses, respectively, of the unbound fermion and the 
bound state.
According to the general discussion of the mass spectrum of the collective 
excitations given in [56], here we are interested only in the following 
low-energy bound states written explicitly in spectroscopic notation 
$({}^{1}S_{0}){}_{N=0},\quad ({}^{1}S_{0}){}_{N=\pm 2},\quad 
({}^{3}P_{1}){}_{N=0}$ and $({}^{3}P_{0}){}_{N=0}$ with the expected masses 
$\mu=0,\,\,\,>\sqrt{2}m_{Q},\,\,
\sqrt{\FFr{8}{3}}m_{Q}$ and $2\,m_{Q}$, respectively, where the subscript $N$ 
indicates the nucleon number. One notes the peculiar symmetry existing 
between the pseudoscalar and the scalar states that the first has zero mass 
and binding energy $2m_{Q}$, while the opposite holds for the scalar state.
When the next pole $m_{*}$ to the lowest one $m_{Q}$ will be 
switched on, then due to the Yukawa couplings in the eq.(8.1) the all 
fermions will acquire the new masses with their common shift 
$\FFr{m_{*}}{m_{Q}}\equiv 1+k$ held upwards along the energy scale. To fix 
the energy threshold value all we have to do then is choose the heaviest 
member among the SM fermions, which is the top quark,  and to set up the 
quite obvious formula 
$$
E\geq E_{0}\equiv m_{t'}\,c^{2}=(1+k)\,m_{t}\,c^{2},
$$ 
where $m_{t}$ is the mass of the top quark. The top quark observed firstly in 
the two FNAL $p\,\bar{p}$ collider experiments in 1995, has the mass 
turned out to be startlingly large $m_{t}=(173.8 \pm 5.0)GeV/c^{2}$ 
compared to all the other SM fermion masses [96]. Thus, we get the most
important energy threshold scale estimate where the
heavy partners of the SM extra families of quarks and 
leptons likely would reside at:
$
E_{1} > (419.6 \pm 12.0)GeV,\quad E_{2}= (457.6 \pm 13.2)GeV$ and 
$
E_{3}=(521.4 \pm 15.0)GeV,
$
corresponded to the next nontrivial poles are written: 
$k_{1}>\sqrt{2}, \quad k_{2}=\sqrt{{8/3}}$ and $k_{3}= 2$, respectively.
We recognize well that the general results obtained in [56, 57], however, 
model-dependent and may be considerably altered, especially in the high 
energy range by using better approximation. In present state of the theory it 
seemed to be a bit premature to get exact high energy results, which will be 
important subject for the future investigations. But, in the same time we 
believe that the approximation used in [56, 57] has enough accuracy for the 
low-energy estimate made above . Anyhow, it is for one thing, the new scale 
where the family partners reside will be much higher than the 
electroweak scale and thus these heavy partners lie far above the electroweak 
scale.

\section{Quark Flavour Mixing and the Cabibbo Angles}
\label{Cabib}
An implication of quark generations into general scheme will be carried out
in the same way of the leptons.
But before proceeding further that it is profitable to enlarge it by the 
additional assumption without asking the reason behind it:

$\bullet$ The MW components imply
\begin{equation}
\label{eq: R16.1.3}
{}^{i}{\bar{\ps1_{u}}}^{A}(\cdots,\theta_{i_{1}},\cdots\theta_{i_{n}},\cdots)\,\,
{}^{j}{\ps1_{u}}^{B}(\cdots,\theta_{i_{1}},\cdots\theta_{i_{n}},\cdots)=
\delta_{ij}\S_{l=i_{1},\ldots,i_{n}}f^{AB}_{il}{}\,\,^{i}\left(
\bar{q}_{l}q_{l}\right),
\end{equation}
namely, the contribution of each individual subquark ${}^{i}q_{l}$, into the 
component of given world ($i$) is determined by the 
{\em partial formfactor} $f^{AB}_{il}$.
Under the group $SU(2)\otimes U(1)$ the 
left-handed quarks transform as three doublets, while the right-handed
quarks transform as independent singlets except of following 
differences:\\
1. The values of weak-hypercharge of quarks are changed due to their
fractional electric charges
$
q_{L}:Y^{w}=\FFr{1}{3},\quad u_{R}:Y^{w}=\FFr{4}{3},\quad
d_{R}:Y^{w}=-\FFr{2}{3},$
etc.\\
2. All Yukawa coupling constants have nonzero values.\\
3. An appearance of quark mixing and Cabibbo angle, which
is unknown in the scope of standard model.\\
4. An existence of CP-violating phase in unitary matrix of quark
mixing. We shall discuss it in the next 
section.\\
Below, we attempt to give an
explanation to quark mixing and Cabibbo angle.
Here for simplicity, we consider this problem on the example of four quarks 
$u,d,s,c$. The further implication of all quarks would complicate 
the problem only in algebraic sense.
Let consider four left-handed quarks forming a $SU(2)_{L}$ doublets mixed
with Cabibbo angle
$
\left(\matrix{
u'\cr
d\cr}\right)_{L}, \quad \mbox{and}\quad
\left(\matrix{
c'\cr
s\cr}\right)_{L},
$
where $u'=u\cos \theta + c\sin \theta, \quad
c'=-u\sin \theta + c\cos \theta$.
One must distinguish two kind of fermion states:
an eigenstate of gauge interactions, i.e. the fields of $u'$
and $c'$; an eigenstate of mass-matrices, i.e the fields of
$u$ and $c$. The qualitative properties
of Cabibbo flavour mixing could be understood in terms of Yukawa couplings.
Unlike the case of leptons, where the Yukawa couplings are
characterized  by two constants $f_{e}$ and $f_{\mu}$, the interaction
of Higgs boson with $u'$ and $c'$ is due to following three terms:
$$
\begin{array}{l}
\FFr{1}{\sqrt{2}}f_{u'}(\bar{u}'_{L}u'_{R}+\bar{u}'_{R}u'_{L})
(\eta +\chi)=
\FFr{1}{\sqrt{2}}f_{u'}(\bar{u}'u')(\eta +\chi), \\\\
\FFr{1}{\sqrt{2}}f_{c'}(\bar{c}'_{L}c'_{R}+\bar{c}'_{R}c'_{L})
(\eta +\chi)=
\FFr{1}{\sqrt{2}}f_{c'}(\bar{c}'c')(\eta +\chi), \\ \\
\FFr{1}{\sqrt{2}}f_{u'c'}(\bar{c}'_{L}u'_{R}+\bar{c}'_{R}u'_{L}+
\bar{u}'_{R}c'_{L}+\bar{u}'_{L}c'_{R})
(\eta +\chi)=
\FFr{1}{\sqrt{2}}f_{u'c'}(\bar{c}'u'+\bar{u}'c')(\eta +\chi).
\end{array}
$$
Our discussion here throughout a small part will be a standard one [97-101], 
except, instead of mixing of fields 
$d'$ and $s'$ we consider a quite equivalent mixing of $u'$ and $c'$.
The last expression may be diagonalized by means of rotation
right through Cabibbo angle. In the sequel one gets
$
m_{u}\bar{u} u+m_{c}\bar{c} c,
$
where $m_{u}$ and $m_{c}$ are masses of quarks $u$ and $c$. Straightforward
comparison of two states gives
$$
\begin{array}{l}
m_{u}=\FFr{1}{\sqrt{2}}(f_{u'}\cos^{2}\theta+f_{c'}\sin^{2}\theta-
2f_{u'c'}\cos\theta\sin\theta)\eta, \\ \\
m_{c}=\FFr{1}{\sqrt{2}}(f_{u'}\sin^{2}\theta+f_{c'}\cos^{2}\theta+
2f_{u'c'}\cos\theta\sin\theta)\eta, \\ \\
\tan 2\theta=\FFr{2f_{u'c'}}{f_{c'}-f_{u'}}\neq 0.
\end{array}
$$
Similar formulas can be worked out for the other mixing.
Hence, the nonzero value of Cabibbo angle arises due to nonzero coupling
constant $f_{u'c'}$. The problem is to calculate all coupling
constants $f_{u'c'}$,$\,\,f_{c't'}$, and $f_{t'u'}$ generating three Cabibbo 
angles
$$
\tan 2\theta_{3}=\FFr{2f_{u'c'}}{f_{c'}-f_{u'}},\quad
\tan 2\theta_{1}=\FFr{2f_{c't'}}{f_{t'}-f_{c'}},\quad
\tan 2\theta_{2}=\FFr{2f_{t'u'}}{f_{u'}-f_{t'}}.
$$
Taking into account the explicit form of Q-components of quark fields 
eq.(6.4.1)
$$
{\ps1_{Q}}_{u'}=(q_{2}q_{3})^{Q}, \quad
{\ps1_{Q}}_{c'}=(q_{3}q_{1})^{Q}, \quad
{\ps1_{Q}}_{t'}=(q_{1}q_{2})^{Q}, 
$$
also eq.(6.7.2) and eq.(6.10.1), we may write down
\begin{equation}
\label{eq: R21.8}
f_{u'}\rightarrow \FFr{1}{2}\left\{
{\bp_{u}}_{u'}\hat{p}_{u}{\ps1_{u}}_{u'}-
\left( {\bp_{u}}_{u'}\overleftarrow{\hat{p}}_{u} \right) {\ps1_{u}}_{u'}
\right\}
=\left( \bar{\Sigma}{}_{Q\,u}^{2}+\bar{\Sigma}{}_{Q\,u}^{3}\right)
{\bp_{u}}_{u'}{\ps1_{u}}_{u'}\rightarrow 
\left( \bar{\Sigma}{}_{Q\,u}^{2}+\bar{\Sigma}{}_{Q\,u}^{3}\right), 
\end{equation}
where according to the eq.(6.10.7) 
$\bar{\Sigma}{}_{Q\,u}^{i}=tr\left(\rho_{u}\,\Sigma^{i}_{Q}\right)$, 
$\rho_{u}$ is given in eq.(6.10.8) and 
$
\hat{p}_{Q}\,\,q^{Q}_{i}=\Sigma_{Q}^{i}\,\,q^{Q}_{i}.
$
In analogy, the $f_{c'}$ and $f_{u'c'}$ imply
\begin{equation}
\label{eq: R21.10}
f_{c'}\rightarrow \FFr{1}{2}\left\{
{\bp_{u}}_{c'}\hat{p}_{u}{\ps1_{u}}_{c'}-
\left( {\bp_{u}}_{c'}\overleftarrow{\hat{p}}_{u} \right) {\ps1_{u}}_{c'}
\right\}\rightarrow 
\left( \bar{\Sigma}{}_{Q\,c}^{3}+\bar{\Sigma}{}_{Q\,c}^{1}\right), 
\end{equation}
and 
\begin{equation}
\label{eq: R21.11}
\begin{array}{l}
f_{u'c'}\rightarrow \FFr{1}{4}\left\{
{\bp_{Q}}_{u'}\hat{p}_{Q}{\ps1_{Q}}_{c'}+
{\bp_{Q}}_{c'}\hat{p}_{Q}{\ps1_{Q}}_{u'}-
\left( {\bp_{Q}}_{u'}\overleftarrow{\hat{p}}_{Q} \right) {\ps1_{Q}}_{c'}-
\left( {\bp_{Q}}_{c'}\overleftarrow{\hat{p}}_{Q} \right) {\ps1_{Q}}_{u'}
\right\}=\\ \\
=\FFr{1}{2}\left\{
\left( \bar{\Sigma}{}_{Q\,u}^{2}+\bar{\Sigma}{}_{Q\,u}^{3}\right){\bp_{Q}}_{u'}{\ps1_{Q}}_{c'}+
\left( \bar{\Sigma}{}_{Q\,c}^{3}+\bar{\Sigma}{}_{Q\,c}^{1}\right)
{\bp_{Q}}_{c'}{\ps1_{Q}}_{u'}
\right\}.
\end{array}
\end{equation}
In accordance with eq.(10.1), one has
$$
\begin{array}{l}
{\bp_{Q}}_{u'}{\ps1_{Q}}_{c'}=
(\overline{q_{2}q_{3}})^{Q}(q_{3}q_{1})^{Q}=
f^{u'c'}_{Q3}(\bar{q}_{3}q_{3})^{Q},
\\\\
{\bp_{Q}}_{c'}{\ps1_{Q}}_{u'}=
(\overline{q_{3}q_{1}})^{Q}(q_{2}q_{3})^{Q}=
f^{c'u'}_{Q3}(\bar{q}_{3}q_{3})^{Q},
\end{array}
$$
where 
$$f^{u'c'}_{Q3}(\bar{q}_{3}q_{3})^{Q}=\left(
f^{u'c'}_{Q3}(\bar{q}_{3}q_{3})^{Q}
\right)^{*}=f^{c'u'}_{Q3}(\bar{q}_{3}q_{3})^{Q}\equiv
\bar{f}_{3}.$$
Hence
$
f_{u'c'}
=\FFr{\bar{f}_{2}}{3}
\left( \bar{\Sigma}{}_{Q\,u}^{2}+\bar{\Sigma}{}_{Q\,u}^{3}+
\bar{\Sigma}{}_{Q\,c}^{3}+\bar{\Sigma}{}_{Q\,c}^{1}\right)
$ and
\begin{equation}
\label{eq: R21.16}
\begin{array}{l}
\tan 2\theta_{3}=
\FFr{\bar{f}_{3}
\left( \bar{\Sigma}{}_{Q\,u}^{2}+\bar{\Sigma}{}_{Q\,u}^{3}+
\bar{\Sigma}{}_{Q\,c}^{3}+\bar{\Sigma}{}_{Q\,c}^{1}\right)}
{\left(\bar{\Sigma}{}_{Q\,c}^{3}+\bar{\Sigma}{}_{Q\,c}^{1}-
\bar{\Sigma}{}_{Q\,u}^{2}-\bar{\Sigma}{}_{Q\,u}^{3}\right)},\quad
\tan 2\theta_{1}=
\FFr{\bar{f}_{1}
\left( \bar{\Sigma}{}_{Q\,c}^{3}+\bar{\Sigma}{}_{Q\,c}^{1}+
\bar{\Sigma}{}_{Q\,t}^{1}+\bar{\Sigma}{}_{Q\,t}^{2}\right)}
{\left(\bar{\Sigma}{}_{Q\,t}^{1}+\bar{\Sigma}{}_{Q\,t}^{2}-
\bar{\Sigma}{}_{Q\,c}^{3}-\bar{\Sigma}{}_{Q\,c}^{1}\right)},\\\\
\tan 2\theta_{2}=
\FFr{\bar{f}_{2}
\left( \bar{\Sigma}{}_{Q\,t}^{1}+\bar{\Sigma}{}_{Q\,t}^{2}+
\bar{\Sigma}{}_{Q\,u}^{2}+\bar{\Sigma}{}_{Q\,u}^{3}\right)}
{\left(\bar{\Sigma}{}_{Q\,u}^{2}+\bar{\Sigma}{}_{Q\,u}^{3}-
\bar{\Sigma}{}_{Q\,t}^{1}-\bar{\Sigma}{}_{Q\,t}^{2}\right)},
\end{array}
\end{equation}
where the rest of $\bar{f}_{i}$ reads
$\bar{f}_{1}\equiv f^{c't'}_{Q1}=f^{t'c'}_{Q1}$ and
$\bar{f}_{2}\equiv f^{t'u'}_{Q2}=f^{u't'}_{Q2}$.
Thus, the Q-components of the quark fields $u',c'$ and $t'$
contain at least one identical subquark, due to which in
eq.(6.10.1) the partial formfactors 
$\bar{f}_{i}$ have nonzero values causing a
quark mixing with the Cabibbo angles eq.(10.5). Therefore, 
the unimodular orthogonal group of global rotations arises, and the quarks 
$u',c'$ and $t'$ come up in doublets 
$(u',c')$,$(c',t')$, and $(t',u')$. For the leptons
these formfactors equal zero 
$\bar{f}_{i}^{lept}\equiv 0$, because of
eq.(6.1.1), namely the lepton mixing is absent.
In conventional notation 
$
\left(\matrix{
u'\cr
d\cr}\right)_{L}, 
\left(\matrix{
c'\cr
s\cr}\right)_{L},
\left(\matrix{
t'\cr
b\cr}\right)_{L}\rightarrow
\left(\matrix{
u\cr
d'\cr}\right)_{L}, 
\left(\matrix{
c\cr
s'\cr}\right)_{L},
\left(\matrix{
t\cr
b'\cr}\right)_{L},
$
which gives rise to
$
f_{u'c'}\rightarrow f_{d's'},\quad
f_{c't'}\rightarrow f_{s'b'},\quad
f_{t'u'}\rightarrow f_{b'd'},\quad
f_{u'}\rightarrow f_{d'},\quad 
f_{c'}\rightarrow f_{s'},\quad 
f_{t'}\rightarrow f_{b'},\quad 
f_{d}\rightarrow f_{u},\quad
f_{s}\rightarrow f_{c},\quad
f_{b}\rightarrow f_{t}.
$
\section{The Appearance of the CP-Violating Phase}
\label{Phase}

The required magnitude of the CP-violating complex parameter $\varepsilon$
depends upon the specific choice of theoretical model for
explaining the $ K^{0}_{2}\rightarrow 2\pi $ decay [102].
From the experimental data it is somewhere
$
| \varepsilon |\simeq 2.3\times 10^{-3}.
$
In the framework of Kobayashi-Maskawa (KM) parametrization of unitary matrix
of quark mixing [103], this parameter may be expressed in terms of 
three Eulerian angles of global rotations in the three dimensional quark 
space and one phase parameter. Below we attempt to derive the KM-matrix with
an explanation given to an appearance of the CP-violating 
phase.
We recall that during the realization of multiworld structure 
the P-violation 
compulsory occurred in the W-world provided by the spanning
eq.(6.8.1). The three dimensional effective space 
$W^{loc}_{v}(3)$ arises as follows:
\begin{equation}
\label{eq: R22.3}
\begin{array}{l}
W^{loc}_{v}(3)\ni q^{(3)}_{v}=
\left( \matrix{
q^{w}_{R}(\vec{T}=0)\cr
\cr
q^{w}_{L}(\vec{T}=\FFr{1}{2})\cr
}\right)\equiv\\ \\
\equiv
\left( \matrix{
u_{R},d_{R}\cr
\cr
\left( \matrix{
u'\cr
d\cr
}\right)_{L}
\cr
}\right),
\left( \matrix{
c_{R},s_{R}\cr
\cr
\left( \matrix{
c'\cr
s\cr
}\right)_{L}
\cr
}\right),
\left( \matrix{
t_{R},b_{R}\cr
\cr
\left( \matrix{
t'\cr
b\cr
}\right)_{L}
\cr
}\right)
\equiv
\left( \matrix{
q^{w}_{3}\cr
\cr
\left( \matrix{
q^{w}_{1}\cr
q^{w}_{2}\cr
}\right)
\cr
}\right),
\left( \matrix{
q^{w}_{1}\cr
\cr
\left( \matrix{
q^{w}_{2}\cr
q^{w}_{3}\cr
}\right)
\cr
}\right),
\left( \matrix{
q^{w}_{2}\cr
\cr
\left( \matrix{
q^{w}_{3}\cr
q^{w}_{1}\cr
}\right)
\cr
}\right),
\end{array}
\end{equation}
where the subscript $(v)$ formally specifies a vertical direction of
multiplet, the subquarks $q^{w}_{\alpha} (\alpha=1,2,3)$ associate with 
the local rotations around corresponding axes of three dimensional
effective space $W^{loc}_{v}(3)$. The local gauge transformations
$f^{v}_{exp}$ are implemented upon the multiplet 
${q'}^{(3)}_{v}=f^{v}_{exp} q^{(3)}_{v}$, where 
$f^{v}_{exp}\in SU^{loc}(2)\otimes U^{loc}(1)$. 
If for the moment we leave it intact and make a closer examination of the 
content of the middle row in eq.(11.1), then we distinguish the other 
symmetry arisen along the horizontal line $(h)$. 
Hence, we may expect a situation similar to those of subsec.6.8 will be held
in present case. The procedure just explained therein can be followed again.
We have to realize that due to the specific structure of 
W-world implying the condition of realization of the MW connections 
eq.(6.1.5) with $\vec{T}\neq 0, \quad Y^{w}\neq 0$, the subquarks 
$q^{w}_{\alpha}$ tend to be compulsory involved into triplet. They form
one ``doublet''  $\vec{T}\neq 0$ and one singlet $Y^{w}\neq 0$. Then the
quarks $u'_{L},c'_{L}$ and $t'_{L}$ form
a $SO^{gl}(2)$ ``doublet'' and a $U^{gl}(1)$ singlet 
\begin{equation}
\label{eq: R22.4}
\begin{array}{l}
\left( \left( u'_{L},c'_{L}\right)t'_{L}\right)\equiv
\left( \left( q^{w}_{1},q^{w}_{2}\right)q^{w}_{3}\right)\equiv
q^{(3)}_{h}\in W^{gl}_{h}(3),\\\\ 
\left( u'_{L},\left( c'_{L},t'_{L}\right)\right)\equiv
\left( q^{w}_{1},\left( q^{w}_{2},q^{w}_{3}\right)\right),\quad
\left( \left( t'_{L},u'_{L}\right)c'_{L}\right)\equiv
\left( \left( q^{w}_{3},q^{w}_{1}\right),q^{w}_{2}\right).
\end{array}
\end{equation}
Here $W^{gl}_{h}(3)$ is the three dimensional effective space in
which the global rotations occur. They are implemented upon the
triplets through the transformation matrix $f^{h}_{exp}$:\\ 
$
{q'}^{(3)}_{h}=f^{h}_{exp} \,q^{(3)}_{h},
$
which reads (eq.(11.2))
$$
f^{h}_{exp}=\left( \matrix{
f_{33} &0 &0\cr
0 & c & s\cr
0 & -s & c\cr
}\right)
$$
in the notation $c=\cos \theta, \quad s=\sin \theta$. This 
implies the incompatibility relation eq.(4.4.5), namely
\begin{equation}
\label{eq: R22.7}
\|f^{h}_{exp}\|=f_{33}(f_{11}f_{22}-f_{12}f_{21})=
f_{33}\varepsilon_{123}\varepsilon_{123}\|f^{h}_{exp}\|f^{*}_{33}.
\end{equation}
That is
$
f_{33}f^{*}_{33}=1,
$
or
$
f_{33}=e^{i\delta}
$
and $\|f^{h}_{exp}\|=1$.
The general rotation in $W^{gl}_{h}(3)$ is described by Eulerian three angles
$\theta_{1},\theta_{2},\theta_{3}$. If we put the arisen phase only in the 
physical sector then a final KM-matrix of quark flavour mixing would result
\begin{equation}
\label{eq: R22.10}
\begin{array}{l}
\left( \bar{u}_{L},\bar{c}_{L},\bar{t}_{L}\right)V_{K-M}
\left( \matrix{
d\cr
s\cr
b\cr
}\right)\equiv
\left( \bar{u}'_{L},\bar{c}'_{L},\bar{t}'_{L}\right)
\left( \matrix{
d\cr
s\cr
b\cr
}\right)\equiv
\left( \bar{u}_{L},\bar{c}_{L},\bar{t}_{L}\right)
\left( \matrix{
d'\cr
s'\cr
b'\cr
}\right)=\\\\ 
=
\left( \bar{u}_{L},\bar{c}_{L},\bar{t}_{L}\right)
\left( \matrix{
1 &0 &0\cr
0 &c_{2} &s_{2}\cr
0 &-s_{2} &c_{2}\cr
}\right)
\left( \matrix{
c_{1} &s_{1} &0\cr
-s_{1} &c_{1} &0\cr
0 &0 &e^{i\delta}\cr
}\right)
\left( \matrix{
1 &0 &0\cr
0 &c_{3} &s_{3}\cr
0 &-s_{3} &c_{3}\cr
}\right)
\left( \matrix{
d\cr
s\cr
b\cr
}\right),
\end{array}
\end{equation}
where
$$
\left( \bar{u}'_{L},\bar{c}'_{L},\bar{t}'_{L}\right)\equiv
\left( \bar{u}_{L},\bar{c}_{L},\bar{t}_{L}\right)V_{K-M}, \quad
\left( \matrix{
d'\cr
s'\cr
b'\cr
}\right)\equiv
V_{K-M}
\left( \matrix{
d\cr
s\cr
b\cr
}\right).
$$
The CP-violating parameter $\varepsilon$ 
approximately is written [97, 101]
$
\varepsilon\sim s_{1}s_{2}s_{3}\sin \delta\neq 0.
$
While the spanning
$W^{loc}_{v}(2)\rightarrow W^{loc}_{v}(3)$ eq.(11.1) underlies 
the P-violation and the expanded symmetry
$G^{loc}_{v}(3)=SU^{loc}(2)\otimes U^{loc}(1)$, the 
CP-violation stems from the similar
spanning $W^{gl}_{h}(2)\rightarrow W^{gl}_{h}(3)$ eq.(11.2) 
with the expanded global symmetry group.

\section{The Mass-Spectrum of Leptons and Quarks}
\label{Mass}
The mass-spectrum of leptons and quarks
stems from their internal MW-structure eq.(6.3.1) and eq.(6.4.1)
incorporated with the quark mixing eq.(10.5). We start a discussion 
with the leptons. It might be worthwhile 
to adopt a simple viewpoint on Higgs sector.
Following the sec.8, the explicit expressions of the lepton masses read
$m_{i}=\FFr{\eta}{\sqrt{2}}f_{i}$ and 
$m_{i}^{\nu}=\FFr{\eta}{\sqrt{2}}f_{i}^{\nu}$, that 
$m_{e}:m_{\mu}:m_{\tau}=f_{e}:f_{\mu}:f_{\tau}=
L_{1}^{2}:L_{2}^{2}:L_{3}^{2}$ provided by 
$L_{1}^{2}=\FFr{m_{i}}{M}$ and $\sqrt{M}=\S_{i}\sqrt{m_{i}}$. Thus, 
$L_{1}=(8.9;7.8)\times 10^{-3},\quad L_{2}= 0.13; 0.11,\quad L_{3}=
0.9; 0.88$. \\
Taking into account the 
eq.(6.1.6) and eq.(6.10.7) the coupling constants of the quarks $d, s$ and 
$b$ can be written
\begin{equation}
\label{eq: R23.8}
\begin{array}{l}
f_{d}=L_{1}\,tr(\rho_{d}\,\Sigma_{Q})\equiv L_{1}\,\widetilde{f}_{d}, \quad
f_{s}=L_{2}\,tr(\rho_{s}\,\Sigma_{Q})\equiv L_{2}\,\widetilde{f}_{s}, \quad
f_{b}=L_{3}\,tr(\rho_{b}\,\Sigma_{Q})\equiv L_{3}\,\widetilde{f}_{b}, \\\\
\rho_{d}=\rho^{Q}\rho_{d}^{B},\quad
\rho_{s}=\rho^{Q}\rho_{s}^{B}\rho^{s},\quad
\rho_{b}=\rho^{Q}\rho_{b}^{B}\rho^{b}.
\end{array}
\end{equation}
Hence
$m_{d}=\FFr{\eta}{\sqrt{2}}f_{d}, \quad
m_{s}=\FFr{\eta}{\sqrt{2}}f_{s}, \quad m_{b}=\FFr{\eta}{\sqrt{2}}f_{b}, \quad$
and 
$m_{d}:m_{s}:m_{b}=(L_{1}\,\widetilde{f}_{d}):(L_{2}\,\widetilde{f}_{s}):
(L_{3}\,\widetilde{f}_{b}).$
According to the subsec.6.10, we derive the masses of the $u,c$ and $t$ quarks
\begin{equation}
\label{eq: R23.8}
\begin{array}{l}
m_{u}=\FFr{\eta}{\sqrt{2}}\left\{ 
\left( \bar{\Sigma}{}^{2}_{Q\,u}+ \bar{\Sigma}{}^{3}_{Q\,u} \right)\cos^{2} 
\theta_{3}
+\left( \bar{\Sigma}{}^{3}_{Q\,c}+ \bar{\Sigma}{}^{1}_{Q\,c}\right)\sin^{2} 
\theta_{3} -\FFr{\bar{f}_{3}}{2}
\left(  \bar{\Sigma}{}^{2}_{Q\,u}+ \bar{\Sigma}{}^{3}_{Q\,u}+ 
\right.\right.\\ \\
\quad\quad\left.\left.
\bar{\Sigma}{}^{3}_{Q\,c}+ \bar{\Sigma}{}^{1}_{Q\,c}\right)
\sin 2\theta_{3}
\right\}=\FFr{\eta}{\sqrt{2}}\left\{ 
\left( \bar{\Sigma}{}^{2}_{Q\,u}+ \bar{\Sigma}{}^{3}_{Q\,u} \right)\cos^{2} 
\theta_{2}
+\left( \bar{\Sigma}{}^{1}_{Q\,t}+ \bar{\Sigma}{}^{2}_{Q\,t}\right)\sin^{2} 
\theta_{2} +
\right.\\ \\
\quad\quad\left.
\FFr{\bar{f}_{2}}{2}
\left( \bar{\Sigma}{}^{1}_{Q\,t}+ \bar{\Sigma}{}^{2}_{Q\,t} +
\bar{\Sigma}{}^{2}_{Q\,u}+ \bar{\Sigma}{}^{3}_{Q\,u}+ \right)
\sin 2\theta_{2}
\right\},
\end{array}.
\end{equation}
\begin{equation}
\label{eq: R23.8}
\begin{array}{l}
m_{c}=\FFr{\eta}{\sqrt{2}}\left\{ 
\left( \bar{\Sigma}{}^{2}_{Q\,u}+ \bar{\Sigma}{}^{3}_{Q\,u} \right)\sin^{2} 
\theta_{3}
+\left( \bar{\Sigma}{}^{3}_{Q\,c}+ \bar{\Sigma}{}^{1}_{Q\,c}\right)\cos^{2} 
\theta_{3} +\FFr{\bar{f}_{3}}{2}
\left(  \bar{\Sigma}{}^{2}_{Q\,u}+ \bar{\Sigma}{}^{3}_{Q\,u}+ 
\right.\right.\\ \\
\quad\quad\left.\left.
\bar{\Sigma}{}^{3}_{Q\,c}+ \bar{\Sigma}{}^{1}_{Q\,c}\right)
\sin 2\theta_{3}
\right\}=\FFr{\eta}{\sqrt{2}}\left\{ 
\left( \bar{\Sigma}{}^{3}_{Q\,c}+ \bar{\Sigma}{}^{1}_{Q\,c} \right)\cos^{2} 
\theta_{1}
+\left( \bar{\Sigma}{}^{1}_{Q\,t}+ \bar{\Sigma}{}^{2}_{Q\,t}\right)\sin^{2} 
\theta_{1} -
\right.\\ \\
\quad\quad\left.
\FFr{\bar{f}_{1}}{2}
\left( \bar{\Sigma}{}^{3}_{Q\,c}+ \bar{\Sigma}{}^{1}_{Q\,c} +
\bar{\Sigma}{}^{1}_{Q\,t}+ \bar{\Sigma}{}^{2}_{Q\,t}+ \right)
\sin 2\theta_{1}
\right\},
\end{array}.
\end{equation}
\begin{equation}
\label{eq: R23.8}
\begin{array}{l}
m_{t}=\FFr{\eta}{\sqrt{2}}\left\{ 
\left( \bar{\Sigma}{}^{1}_{Q\,t}+ \bar{\Sigma}{}^{2}_{Q\,t} \right)\cos^{2} 
\theta_{1}
+\left( \bar{\Sigma}{}^{3}_{Q\,c}+ \bar{\Sigma}{}^{1}_{Q\,c}\right)\sin^{2} 
\theta_{1} +\FFr{\bar{f}_{1}}{2}
\left(  \bar{\Sigma}{}^{1}_{Q\,t}+ \bar{\Sigma}{}^{2}_{Q\,t}+ 
\right.\right.\\ \\
\quad\quad\left.\left.
\bar{\Sigma}{}^{3}_{Q\,c}+ \bar{\Sigma}{}^{1}_{Q\,c}\right)
\sin 2\theta_{1}
\right\}=\FFr{\eta}{\sqrt{2}}\left\{ 
\left( \bar{\Sigma}{}^{1}_{Q\,t}+ \bar{\Sigma}{}^{2}_{Q\,t} \right)\cos^{2} 
\theta_{2}
+\left( \bar{\Sigma}{}^{2}_{Q\,u}+ \bar{\Sigma}{}^{3}_{Q\,u}\right)\sin^{2} 
\theta_{2} -
\right.\\ \\
\quad\quad\left.
\FFr{\bar{f}_{2}}{2}
\left( \bar{\Sigma}{}^{1}_{Q\,t}+ \bar{\Sigma}{}^{2}_{Q\,t} +
\bar{\Sigma}{}^{2}_{Q\,u}+ \bar{\Sigma}{}^{3}_{Q\,u}+ \right)
\sin 2\theta_{2}
\right\}.
\end{array}.
\end{equation}

\section{The Physical Outlook and Concluding Remarks}
\label{Conc}
In this section we briefly expose the main features 
of our physical outlook and draw a number of conclusions.\\\\
Our purpose in the part I is to develop the OM  formalism, which is
the mathematical framework for our physical outlook embodied in the idea 
that the geometry and fields, with the internal symmetries and all 
interactions, as well the four major principles of relativity (special and 
general), quantum, gauge and colour confinement, are derivative. They come 
into being simultaneously in the stable system of the underlying 
``primordial structures'' involved in the ``linkage'' establishing 
processes (sec.4). The OM formalism is the generalization of secondary 
quantization of the field theory with appropriate expansion over the 
geometric objects leading to the quantization of geometry different from all 
existing schemes. 
Below, once again, we resume its relevant steps and major points:

$\bullet$ 1. An extension of the four-dimensional Minkowski space $M_{4}$ by 
the additional one sample of the internal world (a simplified 
case of the one $u$-channel) in order to introduce the mass operator of the 
fields defined on the internal world (sec.2). 2. A two-step passage for each 
sample of the $M_{4}\rightarrow G_{6}$ (sec.2), which restores the complete 
equivalence between the three spacial and the three time components with the 
subsequent rotation of the basis vectors on the $45^{0}$ angle providing an 
adequate algebra for geometry quantization. \\3. We deal in terms of first 
degree of the line element, which entails an additional phase multiplier for 
vectors (the origin of the fields). All this leads to the 
simplified scheme of smooth 12-dimensional manifold $G$ (sec.2-4). The 
passage back to the $M_{4}$ may be performed whenever it will be necessary 
(subsec.2.1).

$\bullet$ Building up the OM formalism we proceed at once to the  
quantization of geometry by substituting the basis vectors for the creation 
and annihilation operators acting in the configuration space of occupation 
numbers (subsec.2.1). They include also the Pauli's matrices, due to 
which all the states defined on the OM are degenerate at the very outset with 
degeneracy degree equal 2. It implies the half-spin quantum number, which 
subsequently gives rise to the spins of the particles. This rule for spin 
quantum number is not without an important reason. The argument for this 
conclusion is compulsory suggested by the properties of Pauli's operators 
(subsec.2.1).

$\bullet$ In this framework we derive the matrix element of the line element 
eq.(2.2.2), which gives the most important relation for the realization of 
the $G$.

$\bullet$ Based on configuration space mechanics with antisymmetric state 
functions, we discuss in detail the quantum field and differential geometric 
aspects of the OM (subsec.2.1, App.A). 

$\bullet$ We have chosen a simple setting and considered the primordial  
structures, which are designed to posses certain physical properties 
satisfying the stated in subsec.4.1 general rules and have involved in the 
linkage establishing processes. The processes of their creation and 
annihilation in the lowest state (the regular structures) just are described 
by the OM formalism. In all the higher states the primordial structures are 
distorted ones, namely they have undergone the distortion transformations 
(subsec.4.2). These transformations yield the ``quark'' and ``antiquark'' 
fields defined on the simplified geometry (one $u$-channel) given in the 
subsec.4.3, and skeletonized for illustrative purposes. Due to geometry 
realization conditions held in the stable systems of primordial structures 
they emerge in confined phase (subsec.4.2). This scheme still should be 
considered as the preliminary one, which is further elaborated in the 
subsec.5.3 to get the physically more realistic picture.

$\bullet$
The distortion transformation 
functions are the operators acting in the space of the internal degrees of 
freedom (colours) and imply the incompatibility relations eq.(4.4.5), which 
hold for both the local and the global distortion rotations. They underly the 
most important symmetries such as the internal symmetries $U(1), SU(2), 
SU(3)$, the $SU(2)\otimes U(1)$ symmetry of electroweak interactions, etc., 
(see part II).

$\bullet$
We generalize the OM formalism via the concept of the OMM yielding the
MW geometry involving the spacetime continuum and the internal worlds of the 
given number (sec.5). In an enlarged framework of the OMM we define and 
clarify the conceptual basis of subquarks and their
characteristics stemming from the various symmetries of the internal worlds
(subsec.5.3). They imply subcolour confinement and gauge principle.

By this we have arrived at an entirely satisfactory answer to the question of 
the physical origin of the geometry and fields, the internal symmetries and interactions, as well the principles of relativity, quantum, gauge and subcolour confinement.
\\\\
The value of the present version of hypothesis of 
existence of the MW structures resides in solving in part II some key 
problems of the SM, wherein we attempt to suggest a microscopic approach to 
the properties of particles and interactions. 

$\bullet$
Within this approach the 
fields have composite nontrivial internal structure (sec.6). 
The condition of
realization of the MW connections is arisen due to the symmetry of 
Q-world of electric charge and embodied in the Gell-Mann-Nishijima relation 
(subsec.6.1). 
During the realization of the MW-structure the symmetries of 
corresponding internal worlds are unified into more higher symmetry 
including also the operators of isospin and hypercharge.
Such approach enables to conclude that only
possible at low-energy the three lepton generations consist of six lepton 
fields with integer electric and leptonic charges and being free of 
confinement (subsec.6.3). Also the three quark 
generations exist composed of six possible quark fields. They
carry fractional electric and baryonic charges and obey confinement 
condition (subsec.6.4). The global group unifying all global symmetries of the internal
worlds of quarks is the flavour group $SU_{f}(6)$ (subsec.6.5).
The whole complexity of leptons, quarks and other composite particles, 
and their interactions arises from the primary
field, which has nontrivial MW internal structure
and involves nonlinear fermion self-interaction of the components 
(subsec.6.6). This Lagrangian
contains only two free parameters, which are the coupling constants of
nonlinear fermion and gauge interactions.
Due to specific structure of the W-world of weak interactions implying the 
condition of realization of the MW connections, the spanning eq.(6.8.1) 
takes place, which underlies the P-violation in W-world. 
It is expressed in the reduction of initial symmetry of the right-handed 
subquarks. Such reduction is characterized by the Weinberg mixing angle 
with the value fixed at $30^{0}$ (subsec.6.9). It gives rise to 
the expanded local symmetry $SU(2)\otimes U(1)$, under which the 
left-handed fermions transform as six independent $SU(2)$ doublets, while
the right-handed fermions transform as twelve independent singlets 
(subsec.6.8).
Due to vacuum rearrangement in Q-world the Yukawa couplings arise between 
the fermion fields and corresponding isospinor-scalar $\varphi$-meson in
conventional form (subsec.6.10). 

$\bullet$
We suggest the microscopic approach to Higgs 
bosons with self-interaction and Yukawa couplings (sec.7). 
It involves the Higgs 
bosons as the collective excitations of bound quasi-particle iso-pairs.
In the framework of local gauge invariance of the theory incorporated with 
the P-violation in weak interactions we propose a mechanism providing the
Bose-condensation of iso-pairs, which is due to effective attraction
between the relativistic fermions caused by the exchange of the mediating
induced gauge quanta in the W-world. 
We consider the four-component Bose-condensate, where
due to self-interaction its spin part is vanished.
Based on it we show that the field of 
symmetry-breaking Higgs boson always must be counted 
off from the gap symmetry restoring value as
the point of origin. Then the Higgs boson describes the excitations in the
neighbourhood of stable vacuum of the W-world. 

$\bullet$
In contrast to the SM, the suggested approach predicts the electroweak 
symmetry breakdown in the $W$-world by the VEV of spin zero Higgs bosons and 
the transmission of electroweak symmetry breaking from the 
$W-$world to the $M_{4}$ spacetime continuum (sec.8). 
The resulting Lagrangian of unified electroweak 
interactions of leptons and quarks ensues, which 
in lowest order approximation
leads to the Lagrangian of phenomenological SM. In general,
the self-energy operator underlies the Yukawa coupling constant, which takes 
into account a mass-spectrum of all expected collective excitations of bound 
quasi-particle pairs. 

$\bullet$
If the MSM proves viable it becomes an crucial 
issue to hold in experiments the two testable solid implications given in 
sec.9, which are drastically different from those of conventional models.

$\bullet$
The implication of quarks into this scheme is carried out in the same manner
except that of appearance of quark mixing with Cabibbo angles
and the existence of CP-violating complex phase in unitary matrix of
quark mixing. The
Q-components of the quarks $u',c'$ and $t'$ contain at least 
one identical subquark, due to which the partial formfactors 
gain nonzero values. This underlies the quark mixing with 
Cabibbo angles (sec.10). In the case of the leptons these formfactors
are vanished  and the mixing is absent.
The CP-violation stems from the spanning
eq.(11.2) incorporated with the expanded group of global rotations.
With a simple viewpoint on Higgs sector the masses of leptons and quarks 
are given in sec.12.
\vskip 0.5truecm
\centerline {\large \bf Acknowledgements}
\vskip 0.5\baselineskip
\noindent
I am pleased to mention the most valuable discussions with the late 
V.H.Ambartsumian on the various issues treated in this paper.
I wish to thank S.P.Novikov for useful discussions of some points of 
mathematical framework, especially, of the reflection formalism (App. B)
I express my gratitude to G.Jona-Lasinio for fruitful comments and 
suggestions. This work was supported in part by the program of Jumelage. It 
is pleasure to thank for their hospitality the ``Centre de 
Physique Theorique'' (CNRS Division 7061, Marseille) and its director 
P.Chiappetta, and 'D.A.R.C.(Observatoir de Paris-CNRS). I am grateful to 
R.Triay for comments and suggestions, and acknowledge useful discussions 
with H.Fliche, G.Sigle and M.Lemoine as well as all the participants of the 
seminars.
I'm indebted to A.M.Vardanian and K.L.Yerknapetian
for support.

\vskip 0.5truecm
\section * {Appendix A}
\label {app}
\setcounter{section}{0}
\renewcommand {\theequation}{A.\thesection.\arabic {equation}}
\section{Mathematical Background}
$\bullet${\bf Field Aspect of the OM}\\\\
The free state of $i$-type fermion with definite values of momentum
$p_{i}$ and spin projection $s$ is described by means of plane waves
($\hbar=1, c=1$)[1]:
$
{\ps1_{\eta}}{}_{p_{\eta}}(\eta)={\left({\FFr{m}{E_{\eta}}}
\right)} ^{1/2}
\U1_{\eta}(p_{\eta},s)\,e^{-ip_{\eta}\eta} 
$,
etc,  where $E_{i}\equiv {\p1_{i}}{}_{0}=\mid \vec{\p1_{i}}{}_{0}\mid ,
\quad
{\p1_{i}}{}_{0\alpha}=\FFr{1}{\sqrt{2}}({\p1_{i}}{}_{(+\alpha)}+
{\p1_{i}}{}_{(-\alpha)}),\quad
\vec{\p1_{i}}=\FFr{1}{\sqrt{2}}(\vec{\p1_{i}}{}_{+}-
\vec{\p1_{i}}{}_{-}),
\quad
p^{2}_{\eta}=E^{2}_{\eta}-{\vec{p}}^{2}_{\eta}=p^{2}_{u}=
E^{2}_{u}-{\vec{p}}^{2}_{u}=m^{2}.$\\
Suppose that the $i$-th fermion is found in the state $r_{i}$ 
with the vector function $\Phi_{r_{i}}^{(\lambda_{i},\mu_{i},\alpha_{i})}=
\zeta_{r_{i}}^{(\lambda_{i},\mu_{i},\alpha_{i})}
\Phi_{r_{i}}^{\lambda_{i},\mu_{i}}(\zeta_{r_{i}})$ and 
\begin{center}
$\zeta_{r_{i}^{\lambda_{i},\mu_{i}}}=
\S^{3}_{\alpha_{i}=1}
e^{r_{i}}_{(\lambda_{i},\mu_{i},\alpha_{i})}
\zeta_{r_{i}}^{(\lambda_{i},\mu_{i},\alpha_{i})}$, \quad
$\zeta_{r_{i}}=\S^{2}_{\lambda_{i},\mu_{i}=1}
\zeta_{r_{i}^{\lambda_{i},\mu_{i}}}\in \widetilde{\cal U}_{r_{i}}$, 
\end{center}
the $\widetilde{\cal U}
_{r_{i}}$
is the open neighbourhood of the point $\zeta_{r_{i}};$ the $r_{i}$ 
implies a set $\left(r_{i}^{11},r_{i}^{12},r_{i}^{21},r_{i}^{22}\right)$.
Let the ${\cal H}^{(1)}$ is a Hilbert space used for quantum mechanical 
description of one particle, namely ${\cal H}^{(1)}$ is a finite or infinite
dimensional complex space provided with scalar product $(\Phi,\Psi),$
which is linear with respect to $\Psi$ and antilinear to $\Phi$.  
The ${\cal H}^{(1)}$ is complete in norm 
$|\Phi|=(\Phi,\Phi)^{1/2}$,
i.e. each fundamental sequence $\{\Phi_{n}\}$ of vectors of the 
${\cal H}^{(1)}$ has converged by norm on ${\cal H}^{(1)}$. 
One particle state function is written 
$\Phi^{(1)}_{r_{i}}=\displaystyle \prod^{2}_{\lambda_{i},\mu_{i}=1}
\Phi^{(1)}_{r_{i}^{\lambda_{i},\mu_{i}}}\in {\cal H}^{(1)}_{r_{i}}$, where
${\cal H}^{(1)}_{r_{i}}=
\displaystyle \prod^{2}_{\lambda_{i},\mu_{i}=1}\otimes
{\cal H}^{(1)}_{r_{i}^{\lambda_{i},\mu_{i}}}$. Define
\begin{equation}
\label{R39}
\widetilde{\Phi}^{(1)}=\zeta_{i}\Phi^{(1)}_{r_{i}}\in 
\widetilde{G}^{(1)}_{r_{i}}=\widetilde{\cal U}^{(1)}_{r_{i}}\otimes
{\cal H}^{(1)}_{r_{i}}.
\end{equation}
For description of n particle system we introduce Hilbert space
\begin{equation}
\label{R319}
{\bar{\cal H}}^{(n)}_{(r_{1},\ldots,r_{n})}=
{\cal H}^{(1)}_{r_{1}}\otimes\cdots\otimes
{\cal H}^{(1)}_{r_{n}}
\end{equation}
by considering all sequences
\begin{equation}
\label{R310}
\Phi^{(n)}_{(r_{1},\ldots,r_{n})}=
\{\Phi^{(1)}_{r_{1}},\ldots,\Phi^{(1)}_{r_{n}}\}=
\Phi^{(1)}_{r_{1}}\otimes \cdots \otimes \Phi^{(1)}_{r_{n}},
\end{equation}
where $\Phi^{(1)}_{r_{i}}\in {\cal H}^{(1)}_{r_{i}}$ provided, 
as usual, with the scalar product
\begin{equation}
\label{R312}
(\Phi^{(n)}_{(r_{1},\ldots,r_{n})},\Psi^{(n)}_{(r_{1},\ldots,r_{n})})=
\prod^{n}_{i=1}(\Phi^{(1)}_{r_{i}},\Psi^{(1)}_{r_{i}}).
\end{equation}
We consider the space ${\cal H}^{(n)}_{(r_{1},\ldots,r_{n})}$ of all the 
limited linear combinations of eq.(A.1.2) and continue by linearity the 
scalar product eq.(A.1..4) on the ${\cal H}^{(n)}_{(r_{1},\ldots,r_{n})}$. 
The wave function
$\Phi^{(n)}_{(r_{1},\ldots,r_{n})} 
\in {\cal H}^{(n)}_{(r_{1},\ldots,r_{n})}$ 
must be antisymmetrized over its arguments. We distinguish the
antisymmetric part ${}^{A}{\bar{\cal H}}^{(n)}$ of Hilbert space
${\bar{\cal H}}^{(n)}$ by considering the functions
\begin{equation}
\label{R313}
{}^{A}\Phi^{(n)}_{(r_{1},\ldots,r_{n})}=\FFr{1}{\sqrt{n!}}\S_{\sigma\in S(n)}
sgn(\sigma)\Phi^{(n)}_{\sigma(r_{1},\ldots,r_{n})}.
\end{equation}
The summation is extended over all permutations of indices
$(r_{1}^{\lambda\mu},\ldots, r_{n}^{\lambda\mu})$ of the integers
$1,2,\ldots,n,$ where the antisymmetrical eigenfunctions are sums
of the same terms with alternating signs in dependence of a parity
$sgn(\sigma)$ of transposition. One continues the reflection 
$\Phi^{(n)}\rightarrow {}^{A}\Phi^{(n)}$ by linearity on the 
${\cal H}^{(n)}$, which is limited and 
enables the expansion by linearity on the ${}^{A}{\bar{\cal H}}^{(n)}$.
The region of values of this
reflection is the ${}^{A}{\bar{\cal H}}^{(n)}$, namely an antisymmetrized
tensor product of $n$ identical samples of ${}{\cal H}^{(1)}$.
We introduce
\begin{equation}
\label{R314}
\begin{array}{l}
{}^{A}{\widetilde{\Phi}}^{(n)}_{(r_{1},\ldots,r_{n})}=
\FFr{1}{\sqrt{n!}}\S_{\sigma\in S(n)}
sgn(\sigma){\widetilde {\Phi}}^{(n)}_{\sigma(r_{1},\ldots,r_{n})}=\\ 
=\FFr{1}{\sqrt{n!}}\S_{\sigma\in S(n)}
sgn(\sigma){\widetilde{\Phi}}^{(1)}_{r_{1}}\otimes \cdots \otimes 
{\widetilde{\Phi}}^{(1)}_{r_{n}} \in 
{}^{A}{\widetilde{G}}^{(n)}_{(r_{1},\ldots,r_{n})}
=\widetilde{\cal U}^{(n)}_{(r_{1},\ldots,r_{n})}\otimes
{}^{A}{\hat{\cal H}}^{(n)}_{(r_{1},\ldots,r_{n})}.
\end{array}
\end{equation}
and consider a set ${}^{A}\widetilde{\cal F}$ of all sequences
$
{}^{A}{\widetilde{\Phi}}=\{ {}^{A}{\widetilde{\Phi}}^{(0)},
{}^{A}{\widetilde{\Phi}}^{(1)}\ldots ,
{}^{A}{\widetilde{\Phi}}^{(n)}\ldots \},
$
with a finite number of nonzero elements. Therewith, the set 
${}^{A}{\cal F}q:\,\, {}^{A}\Phi=\{ {}^{A}\Phi^{(0)},
{}^{A}\Phi^{(1)}\ldots ,
{}^{A}\Phi^{(n)}\ldots \}
$
is provided by the structure of the Hilbert subspace implying
the composition rules
\begin{equation}
\label{R317}
\begin{array}{l}
{}^{A}(\lambda\Phi +\mu\Psi)^{(n)}=
\lambda {}^{A}\Phi^{(n)} + \mu {}^{A}\Psi^{(n)}, \quad 
\forall \lambda, \mu \in C,\\ 
({}^{A}\Phi,{}^{A}\Psi)=\S^{\infty}_{n=0}({}^{A}\Phi^{(n)},{}^{A}\Psi^{(n)}).
\end{array}
\end{equation}
The wave manifold ${\cal G}$ stems from the
${}^{A}\widetilde{\cal F}$ because of the expansion by metric 
induced as a scalar product on ${}^{A}\cal F$
\begin{equation}
\label{R318}
{\cal G}=\S^{\infty}_{n=0}{\cal G}^{(n)}=
\S^{\infty}_{n=0}\left(\widetilde{\cal U}^{(n)}\otimes {}^{A}
\bar{\cal H}^{(n)}\right).
\end{equation}
The creation ${\hat{\gamma}}_{r}$ and
annihilation ${\hat{\gamma}}{}^{r}$ operators for each
${}{\cal H}^{(1)}$ can be defined as follows:
one must modify the basis operators in order to provide
an anticommutation in arbitrary states
\begin{equation}
\label{R319}
{\hat{\gamma}}{}_{(\lambda,\mu,\alpha)}^{r}\rightarrow
{\hat{\gamma}}{}_{(\lambda,\mu,\alpha)}^{r}\,\eta_{r}^{\lambda\mu},
\quad
{(\eta_{r}^{\lambda\mu})}^{+}=\eta_{r}^{\lambda\mu},
\end{equation}
for given $\lambda,\mu,\alpha$,
where $\eta_{r}$ is a diagonal operator in the space 
of occupation numbers,
while, at $r_{i}<r_{j}$ one gets
$
{\hat{\gamma}}{}^{r_{i}}\,\eta_{r_{j}}=
-\eta_{r_{j}}\,{\hat{\gamma}}{}^{r_{i}},
\quad
{\hat{\gamma}}{}^{r_{j}}\,\eta_{r_{i}}=
\eta_{r_{i}}\,{\hat{\gamma}}{}^{r_{j}}.
$
The operators of corresponding occupation numbers (for given 
$\lambda,\mu,\alpha)$ are
$
{\hat{N}}_{r}={\hat{\gamma}}^{r}
{\hat{\gamma}}_{r}$.
Since the diagonal operators $(1-2{\hat{N}}_{r})$ 
anticommute with the
${\hat{\gamma}}^{r}$, then
$
\eta_{r_{i}}=\prod_{r =1}^{r_{i}-1}(1-2{\hat{N}}_{r}),
$
where
\begin{equation}
\label{R325}
\eta_{r_{i}}^{11}\Phi(n_{1},\ldots,n_{N};0;0;0)=\pd_{r =1}^{r_{i}-1}
(-1)^{n_{r}}\Phi(n_{1},\ldots,n_{N};0;0;0),
\end{equation}
etc.
Here the occupation numbers $n_{r}(m_{r},q_{r},t_{r})$ are introduced, 
which refer to the $r$-th states corresponding to operators
${\hat{\gamma}}_{(1,1,\alpha)}^{r}$, etc.,
either empty ($n_{r},\ldots,t_{r}=0$) or occupied
($n_{r},\ldots,t_{r}=1$).
To save writing we abbreviate the modified operators by the
same symbols. For example,
acting on free state $\mid 0>_{r_{i}}$ 
the creation operator ${\hat{\gamma}}_{r_{i}}$
yields the one occupied state
$\mid 1>_{r_{i}}$ with the phase $+$ or $-$ depending of parity of the number
of quanta in the states $ r < r_{i}$.
Modified operators satisfy the same anticommutation relations
of the basis operators (subsec.2.2). 
It is convenient to make use of notation
${\hat{\gamma}}{}^{(\lambda,\mu,\alpha)}_{r}\equiv
{e}^{(\lambda,\mu,\alpha)}_{r}\,{\hat{b}}{}^{\lambda\mu}_{(r\alpha)},$
and abbreviate the pair of indices $(r\alpha)$ by the single symbol $r$. 
Then for each $\Phi \in {}^{A}{\cal H}^{(n)}$
and any vector $f \in {\cal H}^{(1)}$ the operators
$\hat{b}(f)$ and
$\hat{b}^{*}(f)$ imply
\begin{equation}
\label{R327}
\begin{array}{l}
\hat{b}(f)\Phi=\FFr{1}{\sqrt{(n-1)!}}\S_{\sigma\in S(n)}
sgn(\sigma)\left( f\,\Phi^{(1)}_{\sigma{(1)}} \right)
\Phi^{(1)}_{\sigma{(2)}}\otimes\cdots\otimes \Phi^{(1)}_{\sigma{(n)}},\\
\hat{b}^{*}(f)\Phi=\FFr{1}{\sqrt{(n+1)!}}\S_{\sigma\in S(n+1)}
sgn(\sigma)\Phi^{(1)}_{\sigma{(0)}}\otimes
\Phi^{(1)}_{\sigma{(1)}}\otimes\cdots\otimes \Phi^{(1)}_{\sigma{(n)}},
\end{array}
\end{equation}
where $\Phi^{(1)}_{(0)}\equiv f$. One continues the 
$\hat{b}(f)$ and $\hat{b}^{*}(f)$ by linearity to linear reflections, 
which are denoted by the same symbols acting  respectively from 
${}^{A}{\cal H}^{(n)}$ onto ${}^{A}{\cal H}^{(n-1)}$ or 
${}^{A}{\cal H}^{(n+1)}$.
They are limited over the values $\sqrt{n}|f|$ and $\sqrt{(n+1)}|f|$
and can be expanded by continuation up to the reflections acting 
from ${}^{A}{\bar{\cal H}}^{(n)}$
onto ${}^{A}{\bar{\cal H}}^{(n-1)}$ or ${}^{A}{\bar{\cal H}}^{(n+1)}$.
Finally, they must be continued by linearity up to the linear
operators acting from ${}^{A}{\cal F}$ onto ${}^{A}{\cal F}$
defined on the same closed region in
${}^{A}{\bar{\cal H}}^{(n)}$, namely in ${}^{A}{\cal F}$, which is 
invariant with respect to reflections
$\hat{b}(f)$ and $\hat{b}^{*}(f)$. Hence, at $f_{i}, g_{i}\in 
{\cal H}^{(1)}$ $(i=1,\ldots,n; j=1,\ldots,m)$ all polynomials over 
$\{\hat{b}^{*}(f_{i}) \}$ and $\{\hat{b}(g_{j}) \}$ are completely defined 
on ${}^{A}{\cal F}$. While, for given $\lambda,\mu$, one has
\begin{equation}
\label{R328}
\begin{array}{l}
<\lambda,\mu\mid\{ {\hat{b}}^{\lambda\mu}_{r}(f),{\hat{b}}_{\lambda\mu}
^{r'}(g)\}\mid \lambda,\mu>=
\delta^{r'}_{r}.
\end{array}
\end{equation}
The mean values $<\varphi; \,{\hat{b}}_{r}^{\lambda\mu}(f)\,
{\hat{b}}^{r}_{\lambda\mu}(f)>$ calculated at fixed $\lambda,\mu$ for
any element $\Phi \in {}^{A}{\cal F}$ equal to mean values of the symmetric 
operator of occupation number in terms of
${\hat{N}}^{r}={\hat{b}}_{r}(f)\,
{\hat{b}}^{r}(f)$, with a wave function $f$ in the 
state described by $\Phi$. 
Here, as usual, it is denoted $<\varphi; A\Phi> = Tr \,
P_{\varphi}A = (\Phi,A\Phi)$ for each vector $\Phi \in {\cal H}$ with
$|\Phi |=1$, while the $P_{\varphi}$ is projecting operator onto one
dimensional space $\{\lambda \Phi\left.\right| \lambda\in C\}$ generated by
$\Phi$. Therewith, the probability of transition $\varphi \rightarrow \psi$  
is given $Pr\{\varphi\left.\right|\psi\} =\left| (\psi,\varphi)\right|^{2}$.
The linear operator $A$ defined on the elements 
of linear manifold ${\cal D}(A)$ of ${\cal H}$ takes the values in
${\cal H}$. The ${\cal D}(A)$ is an overall closed region of definition
of $A$, namely the closure of ${\cal D}(A)$ by the norm given in
${\cal H}$ coincides with ${\cal H}$. Meanwhile, the ${\cal D}(A)$  
included in ${\cal D}(A^{*})$ and  $A$ coincides with the reduction of 
$A^{*}$ on ${\cal D}(A)$, because ${\cal D}(A)$ is the symmetric operator
such that the linear operator $A^{*}$ is the maximal conjugated to $A$. 
That is,
any operator $A'$ conjugated to $A$ - $(\Psi, A'\Phi)=(A'\Psi, \Phi)$ 
for all $\Phi \in {\cal D}(A)$ and $\Psi \in {\cal D}(A')$ coincides 
with the reduction of $A^{*}$ on some linear manifold ${\cal D}(A')$
included in ${\cal D}(A^{*})$. Thus, the operator $A^{**}$ is closed 
symmetric expansion of operator $A$, namely it is a closure of $A$.
Self conjugated operator $A$, the closure of which is self conjugated as 
well, allows only the one self conjugated expansion $A^{**}$. Hence, 
self conjugated closure $\hat{N}$ of operator 
$\S_{i=1}^{\infty}
\hat{b}^{*}(f_{i})\,\hat{b}(f_{i})$, where $\{f_{i}\left.\right| i=1,\ldots, 
n\}$ is an arbitrary orthogonal basis on ${\cal H}^{(1)}$, can be regarded
as the operator of occupation number. For the vector $\chi^{0} \in 
{}^{A}{\cal F}$ and $\chi^{0(n)}=\delta_{0n}$ one gets $<\chi^{0(n)},
\hat{N}(f)>=0$ for all $f \in {\cal H}^{(1)}$.
Thus, $\chi^{0}$ is the vector of vacuum state:
$\hat{b}(f)\chi^{0}=0$ for all $f \in {\cal H}^{(1)}$. If
$f =\{ f_{i}\left.\right| i=1,2,\ldots\}$ is an arbitrary orthogonal basis 
on ${\cal H}^{(1)}$, then due to irreducibility of operators 
$\hat{b}^{*}(f_{i})\left.\right| f_{i} \in f$, the ${}^{A}{\cal H}$
includes the $0$ and whole space ${}^{A}{\cal H}$ as invariant subspaces 
with respect to all $\hat{b}^{*}(f)$. 
To define the 12 dimensional operator manifold 
$\hat{G}$ we consider a set $\hat{{\cal F}}$ of all the sequences
$\hat{\Phi}=\{\hat{\Phi}^{(0)},\hat{\Phi}^{(1)},\ldots,\hat{\Phi}^{(n)},
\ldots \}$ with a finite number of nonzero elements provided by
\begin{equation}
\label{R329}
\begin{array}{l}
\hat{\Phi}^{(n)}_{(r_{1},\ldots,r_{n})}=\hat{\Phi}^{(1)}_{r_{1}}
\otimes \cdots \otimes \hat{\Phi}^{(1)}_{r_{n}}\in \hat{G}^{(n)},\quad
\hat{\Phi}^{(1)}_{r_{i}}=\hat{\zeta}_{r_{i}}\Phi^{(1)}_{r_{i}}\in
\hat{G}^{(1)}_{i}=\hat{\cal U}^{(1)}_{i}\otimes {\cal H}^{(1)}_{i},\\
\hat{\zeta}_{r_{i}}\equiv \S^{3}_{\alpha_{i}=1}
{\hat{\gamma}}^{r_{i}}_{(\lambda_{i},\mu_{i},\alpha_{i})}
\zeta_{r_{i}}^{(\lambda_{i},\mu_{i},\alpha_{i})}\in
\hat{\cal U}^{(1)}_{r_{i}}, \quad \hat{G}^{(n)}=\hat{\cal U}^{(n)}
\otimes \bar{\cal H}^{(n)}, \quad
\hat{\cal U}^{(n)}_{(r_{1},\ldots,r_{n})}=\hat{\cal U}^{(1)}_{r_{1}},
\otimes\cdots\otimes \hat{\cal U}^{(1)}_{r_{n}}.
\end{array}
\end{equation}
Then, on the analogy of eq.(A.1.8) the operator manifold $\hat{G}$
ensues
\begin{equation}
\label{R330}
\hat{G}=\S^{\infty}_{n=0}\hat{G}^{(n)}=
\S^{\infty}_{n=0}\left(\hat{\cal U}^{(n)}\otimes {\bar{\cal H}}^{(n)}\right).
\end{equation}
To define the 
secondary quantized form of one particle observable $A$ on $\cal H$, 
following [45] let consider a set of identical samples $\hat{\cal H}_{i}$ of
one particle space ${\cal H}^{(1)}$ and operators $A_{i}$ acting on them.
To each closed linear operator $A^{(1)}$ in ${\cal H}^{(1)}$ with 
overall closed region of definition ${\cal D}(A^{(1)})$  
following operators are corresponded:
\begin{equation}
\label{R331}
\begin{array}{l}
A^{(n)}_{1}=A^{(1)}\otimes I \otimes\cdots \otimes I,\\
\ldots\ldots\ldots\ldots\ldots\ldots\ldots\ldots \\
A^{(n)}_{n}=I\otimes I \otimes\cdots \otimes A^{(1)}.
\end{array}
\end{equation}
The sum $\S^{n}_{j=1}A^{(n)}_{j}$ is given on the intersection of regions of
definition of operator terms including a linear manifold
${\cal D}(A^{(1)})\otimes\cdots \otimes {\cal D}(A^{(n)})$ closed
in $\hat{\cal H}^{(n)}$. While, the $A^{(n)}$ is a minimal closed expansion 
of this sum with ${\cal D}(A^{(n)})$. One considers a linear manifold
${\cal D}(\Omega(A))$ in ${\cal H}=\S^{\infty}_{n=0}\hat{\cal H}^{(n)}$
defined as a set of all the vectors $\Psi \in {\cal H}$ such as 
$\Psi^{(n)} \in {\cal D}(A^{(n)})$ and $\S^{\infty}_{n=0}\left|
A^{(n)}\Psi^{(n)}\right|^{2} < \infty$ . The manifold
${\cal D}(\Omega(A))$ is closed in ${\cal H}$. On this manifold one defines 
a closed linear operator $\Omega(A)$ acting as $\Omega(A)^{(n)}=
A^{(n)}\Psi^{(n)}$, namely $\Omega(A)\Phi = \S^{\infty}_{n=0}
A^{(n)}\Psi^{(n)}$, while the $\Omega(A)$ is self conjugated  operator
with overall closed region of definition. We suppose that the vector
$\Phi^{(n)} \in {\cal H}^{(n)}$ is in the form eq.(A.1.3),
where $\Phi_{i}\in {\cal D}(A)$. Then, $<\varphi^{(n)};A^{(n)}>=
\S^{n}_{i=1}<\varphi_{i};A>$, which enables the expansion by 
continuing onto ${\cal D}(A)$. Thus, $A^{(n)}$ is the $n$ particle 
observable corresponding to one particle observable $A$. So
$<\varphi;\Omega(A)>=
\S^{\infty}_{n=0}<\varphi^{(n)};A^{(n)}>$ for any $\Phi_{i}\in 
{\cal D}\left(\Omega(A)\right)$. While, the $\Omega(A)$ reflects
${}^{A}{\cal D}={\cal D}\left(\Omega(A)\right)\frown {}^{A}{\cal H}$
into ${}^{A}{\cal H}$.
The reduction of $\Omega(A)$ on ${}^{A}{\cal H}$ is the self conjugated
in the region ${}^{A}{\cal D}$, because ${}^{A}{\cal H}$
is the closed subspace of the ${\cal H}$. Hence, the $\Omega(A)$ is the 
secondary quantized form of one particle observable $A$ on the ${\cal H}$.\\
The vacuum state reads 
\begin{equation}
\label{R216}
\begin{array}{lll}
\chi^{0}(\nu_{1},\nu_{2},\nu_{3},\nu_{4})=
\mid 1,1>^{\nu_{1}}\cdot\mid 1,2>^{\nu_{2}}\cdot
\mid 2,1>^{\nu_{3}}\cdot\mid 2,2>^{\nu_{4}},\\
\nu_{i}= \left\{ \begin{array}{ll}
                   1   & \mbox{if $\nu=\nu_{i}$}\quad  \mbox{for some $i$,} \\
                   0   & \mbox{otherwise},
                   \end{array}
\right. ,
\end{array}
\end{equation}
where
$\mid\chi_{-}(1)>\equiv\chi^{0}(1,0,0,0),\quad   
\mid\chi_{+}(1)>\equiv\chi^{0}(0,0,0,1),\quad
\mid\chi_{-}(2)>  \equiv\chi^{0}(0,0,1,0),\quad   
\mid\chi_{+}(2)>\equiv\chi^{0}(0,1,0,0), \quad
<\chi_{\pm}(\lambda)\mid\chi_{\pm}(\mu)>=\delta_{\lambda\mu},$ and  
$<\chi_{\pm}(\lambda)\mid\chi_{\mp}(\mu)>=0,$
provided by
$
<\chi_{\pm}\mid A\mid \chi_{\pm}>\equiv
\S_{\lambda}<\chi_{\pm}(\lambda)\mid A\mid \chi_{\pm}(\lambda)>
$
and the normalization condition
$$
<\chi^{0}(\nu'_{1},\nu'_{2},\nu'_{3},\nu'_{4})\mid
\chi^{0}(\nu_{1},\nu_{2},\nu_{3},\nu_{4})>=\prod_{i=1}^{4}
\delta_{\nu_{i}\nu'_{i}}.
$$
The state vectors are introduced
\begin{equation}
\label{R333}
\begin{array}{l}
\chi({\{n_{r}\}}^{N}_{1};{\{m_{r}\}}^{M}_{1};
{\{q_{r}\}}^{Q}_{1};{\{t_{r}\}}^{T}_{1};
{\{\nu_{r}\}}^{4}_{1})=
{(\hat{b}_{N}^{11})}^{n_{\scriptscriptstyle N}}\cdots
{(\hat{b}_{1}^{11})}^{n_{1}}\cdot\\ 
\cdot {(\hat{b}_{M}^{12})}^{m_{\scriptscriptstyle M}}\cdots
{(\hat{b}_{1}^{12})}^{m_{1}}\cdot
{(\hat{b}_{Q}^{21})}^{q_{\scriptscriptstyle Q}}\cdots
{(\hat{b}_{1}^{21})}^{q_{1}}\cdot
\cdot{(\hat{b}_{T}^{22})}^{t_{\scriptscriptstyle T}}\cdots
{(\hat{b}_{1}^{22})}^{t_{1}}
\chi^{0}(\nu_{1},\nu_{2},\nu_{3},\nu_{4}),
\end{array}
\end{equation}
where ${\{n_{r}\}}^{N}_{1}=n_{1},\ldots,n_{N}$, etc., which are the 
eigenfunctions
of modified operators. They form a whole set of orthogonal vectors
\begin{equation}
\label{R334}
\begin{array}{l}
<\chi,\chi'>
=\displaystyle \prod_{r=1}^{N}\delta_{n_{r}n'_{r}}\cdot
\displaystyle \prod_{r=1}^{M}\delta_{m_{r}m'_{r}}\cdot
\displaystyle \prod_{r=1}^{Q}\delta_{q_{r}q'_{r}}\cdot
\displaystyle \prod_{r=1}^{T}\delta_{t_{r}t'_{r}}\cdot
\displaystyle \prod_{r=1}^{4}\delta_{\nu_{r}\nu'_{r}}.
\end{array}
\end{equation}
Considering an arbitrary superposition
\begin{equation}
\label{R335}
\chi=
\S^{1}_{
a={\{n_{r}\}}^{N}_{1},{\{m_{r}\}}^{M}_{1},
{\{q_{r}\}}^{Q}_{1},{\{t_{r}\}}^{T}_{1}=0
}
c'(a)\, \chi(a),
\end{equation}
the coefficients $c'$ of expansion are the corresponding amplitudes
of probabilities.
Taking into account eq.(A.1.12), the nonvanishing matrix elements of
operators $\hat{b}^{11}_{r_{k}}$ and $\hat{b}_{11}^{r_{k}}$ read
\begin{equation}
\label{R337}
\begin{array}{l}
<\chi({\{n'_{r}\}}^{N}_{1};0;0;0;1,0,0,0)\left|   \right.
\hat{b}^{11}_{r_{k}}
\chi({\{n_{r}\}}^{N}_{1};0;0;0;1,0,0,0)>=\\
\\
=<1,1\mid \hat{b}_{11}^{r'_{1}}\cdots\hat{b}_{11}^{r'_{n}}\cdot
\hat{b}^{11}_{r_{k}}\cdot
\hat{b}^{11}_{r_{n}}\cdots\hat{b}^{11}_{r_{1}}\mid 1,1>=\\
\\
= \left\{ \begin{array}{ll}
(-1)^{n'-k'}  & \mbox{if $n_{r}=n'_{r}$ for $r\neq r_{k}$ and $n_{r_{k}}=0;n'_{r_{k}}=1$}, \\
0           & \mbox{otherwise},
\end{array}  \right.   \\

<\chi({\{n'_{r}\}}^{N}_{1};0;0;0;1,0,0,0)\left|   \right.
\hat{b}_{11}^{r_{k}}
\chi({\{n_{r}\}}^{N}_{1};0;0;0;1,0,0,0)>=\\
\\
=<1,1\mid \hat{b}_{11}^{r'_{1}}\cdots\hat{b}_{11}^{r'_{n}}\cdot
\hat{b}_{11}^{r_{k}}\cdot
\hat{b}^{11}_{r_{n}}\cdots\hat{b}^{11}_{r_{1}}\mid 1,1>=\\
\\
= \left\{ \begin{array}{ll}
(-1)^{n-k}  & \mbox{if $n_{r}=n'_{r}$ for $r\neq r_{k}$ and $n'_{r_{k}}=0;n_{r_{k}}=1$}, \\
0           & \mbox{otherwise},
\end{array} \right.
\end{array}
\end{equation}
where one denotes
$
n=\S_{r=1}^{N}n_{r}, \quad n'=\S_{r=1}^{N}n'_{r},
$
the $r_{k}$ and $r'_{k}$ are $k$-th and $k'$-th terms of regulated sets of
$\{r_{1},\ldots ,r_{n}\} \quad (r_{1}<r_{2}<\cdots <r_{n})$ and
$\{r'_{1},\ldots ,r'_{n}\} \quad (r'_{1}<r'_{2}<\cdots <r'_{n})$,
respectively.
Continuing along this line we get a whole set of explicit forms of
matrix elements of the rest of operators
$\hat{b}_{r_{k}}$ and $\hat{b}^{r_{k}}$.
Hence
\begin{equation}
\label{R339}
\S_{\{\nu_{r}\}=0}^{1}<\chi^{0}\mid\hat{\Phi}(\zeta)\mid\chi>= 
\S_{r=1}^{N}c'_{n_{r}}e_{(1,1,\alpha)}^{n_{r}}
{\Phi}^{(1,1,\alpha)}_{n_{r}}
+\cdots,
\end{equation}
provided by
\begin{equation}
\label{R340}
c'_{n_{r}}\equiv\delta_{1n_{r}}c'(0,\ldots, n_{r},\ldots,0;0;0;0),
\cdots
\end{equation}
Hereinafter we change a notation of the coefficients
\begin{equation}
\label{R341}
\begin{array}{l}
\bar{c}(r^{11})=c'_{n_{r}},\quad \bar{c}(r^{21})=c'_{q_{r}},
\quad N_{11}=N, \quad N_{21}=Q, \\ 
\bar{c}(r^{12})=c'_{m_{r}}, \quad \bar{c}(r^{22})=c'_{t_{r}},
\quad N_{12}=M, \quad N_{22}=T,
\end{array}
\end{equation}
and make use of convention
\begin{equation}
\label{R342}
F_{r^{\lambda\mu}}= \S_{\alpha}
e_{(\lambda,\mu,\alpha)}^{r^{\lambda\mu}}
{\Phi}^{(\lambda,\mu,\alpha)}_{r^{\lambda\mu}},
\quad
\S_{\{\nu_{r}\}=0}^{1}<\chi^{0}\mid\hat{A}\mid\chi>
\equiv<\chi^{0}\parallel\hat{A}\parallel\chi>,
\end{equation}
The matrix elements of operator vector and covector fields
take the final forms
\begin{equation}
\label{R343}
\begin{array}{l}
<\chi^{0}\parallel\hat{\Phi}(\zeta)\parallel\chi>=
\S_{\lambda\mu=1}^{2}\S_{r^{\lambda\mu}=1}^{N_{\lambda\mu}}
\bar{c}(r^{\lambda\mu})\,F_{r^{\lambda\mu}}(\zeta),\\
<\chi\parallel\bar{\hat{\Phi}}(\zeta)\parallel\chi^{0}>=
\S_{\lambda\mu=1}^{2}\S_{r^{\lambda\mu}=1}^{N_{\lambda\mu}}
{\bar{c}}^{*}(r^{\lambda\mu})\,F^{r^{\lambda\mu}}(\zeta).
\end{array}
\end{equation}
In the following we shall use a convention:
$\left\{\S_{\lambda\mu}^{2} \right\}_{1}^{n}
\equiv \S_{\lambda_{1}\mu_{1}}^{2}\ldots\S_{\lambda_{n}\mu_{n}}^{2},\quad$
$r^{\lambda\mu}_{i}\equiv r^{\lambda_{i}\mu_{i}}$ and
$ 
\bar{c}(r^{11}_{1},\ldots,r^{11}_{n})=c'(n_{1},\ldots,n_{N};0;0;0),
$ etc.
The anticommutation relations ensue
\begin{equation}
\label{R346}
<\chi_{-}\mid \{ {\hb_{i}}{}^{+}_{r},\,{\hb_{i}}{}_{+}^{r'} \}\mid\chi_{-}>=
<\chi_{+}\mid \{ {\hb_{i}}{}^{-}_{r},\,{\hb_{i}}{}_{-}^{r'} \}\mid\chi_{+}>=
{\delta}^{r'}_{r},
\end{equation}
provided by 
${\hgam_{i}}{}^{(\lambda\alpha)}_{r}=
{\he_{i}}{}^{(\lambda\alpha)}_{r}\,\,{\hb_{i}}{}^{\lambda}_{(r\alpha)}, \quad
(r\alpha)\rightarrow r.$
The state functions
\begin{equation}
\label{R347}
\chi=
{({\hb_{\eta}}_{N}^{+})}^{n_{\scriptscriptstyle N}}\cdots
{({\hb_{\eta}}_{1}^{+})}^{n_{1}}
\cdot
{({\hb_{\eta}}_{M}^{-})}^{m_{\scriptscriptstyle M}}\cdots
{({\hb_{\eta}}_{1}^{-})}^{m_{1}}\cdot
{({\hb_{u}}_{Q}^{+})}^{q_{\scriptscriptstyle Q}}\cdots
{({\hb_{u}}_{1}^{+})}^{q_{1}}
\cdot{({\hb_{u}}_{T}^{-})}^{t_{\scriptscriptstyle T}}\cdots
{({\hb_{u}}_{1}^{-})}^{t_{1}}\cdot
\chi_{-}(\lambda)\chi_{+}(\mu),
\end{equation}
form a whole set of orthogonal
eigenfunctions of corresponding operators of occupation numbers
${\hN_{i}}{}^{\lambda}_{r}={\hb_{i}}{}_{r}^{\lambda}\,\,
{\hb_{i}}{}^{r}_{\lambda}$
with the expectation values 0,1.\\\\

$\bullet${\bf Differential Geometric Aspect of the OM}\\\\
An explicit form of matrix element of 
operator tensor reads
\begin{equation}
\label{R46}
\begin{array}{l}
\FFr{1}{\sqrt{n!}}<\chi^{0}\parallel\hat{\Phi}({\zeta}_{1})\otimes
\cdots \otimes\hat{\Phi}({\zeta}_{n})\parallel\chi>=\\
=\left\{\S_{\lambda\mu=1}^{2}\right\}_{1}^{n}\,\,
\S_{r_{1}^{\lambda\mu},\ldots ,r_{n}^{\lambda\mu}=1}^{N_{\lambda\mu}}
\bar{c}(r_{1}^{\lambda\mu},\ldots ,r_{n}^{\lambda\mu})\,
F_{r_{1}^{\lambda\mu}}(\zeta_{1})\wedge\cdots\wedge
F_{r_{n}^{\lambda\mu}}(\zeta_{n}),
\end{array}
\end{equation}
where $\wedge$ stands for exterior product. \\
The linear operator form of $1$ degree $\hat{\bf \omega}^{1}$ is a linear
operator valued function on $\hat{\bf T}_{\Phi_{p}}$, namely 
$\hat{\bf \omega}^{1}(\hat{\bf A}_{p}):\hat{\bf T}_{\Phi_{p}} 
\rightarrow \hat{R}$,
where $\hat{\bf A}_{p} \in \hat{\bf T}_{\Phi_{p}}$, and the operator
$\hat{\bf \omega}^{1}(\hat{\bf A})=<\hat{\bf \omega}^{1},{\bf A}> 
\in \hat{R}$ corresponds to $\hat{\bf A}_{p}$ at the point 
${\bf \Phi}_{p}$, provided, according to eq.(A.1.25), with
\begin{equation}
\label{R47}
<\chi \| \hat{\bf \omega }^{1} \|
\chi^{0} >=
\S_{\lambda,\mu=1}^{2}
\S_{r^{\lambda\mu}=1}^{N_{\lambda\mu}} \hat{c}^{*}(r^{\lambda\mu})\,
{\bf \omega }^{1}_{r^{\lambda\mu}},
\end{equation}
where ${\bf \omega}^{1}_{r^{\lambda\mu}}=
e_{r^{\lambda\mu}}^{(\lambda,\mu,\alpha)}
\omega^{r^{\lambda\mu}}_{(\lambda,\mu,\alpha)}$, the 
$<{\bf \omega}^{1}_{r^{\lambda\mu}},{\bf A}>=
\omega ^{1}_{r^{\lambda\mu}}({\bf A})$ is a linear form on 
${\bf T}_{p}$, and 
\begin{equation}
\label{R48}
\begin{array}{l}
\hat{\bf \omega}^{1}(\lambda_{1}\hat{\bf A}_{1} + 
\lambda_{2}\hat{\bf A}_{2})=\lambda_{1}\hat{\bf \omega}^{1}(\hat{\bf A}_{1})+
\lambda_{2}\hat{\bf \omega}^{1}(\hat{\bf A}_{2}),\\ 
\forall \lambda_{1},\lambda_{2} \in R,\quad
\hat{\bf A}_{1}, \hat{\bf A}_{2} \in \hat{\bf T}_{\Phi_{p}}.
\end{array}
\end{equation}
The set of all linear operator forms defined at the point ${\bf \Phi}_{p}$
fill up the operator vector space $\hat{\bf T}_{\Phi_{p}}^{*}$ 
dual to $\hat{\bf T}_{\Phi_{p}}$. While, the
$\{ \hat{\gamma}_{r} \}$ serves as a basis for them.
The operator $n$ form is defined as the exterior product of operator 
1 forms
\begin{equation}
\label{R49}
\hat{\bf \omega}^{n}(\hat{\bf A}_{1},\ldots,\hat{\bf A}_{n})=
\left(\hat{\bf \omega}^{1}_{1}\wedge \cdots \wedge 
\hat{\bf \omega}^{1}_{n}\right)
\left(\hat{\bf A}_{1},\ldots,\hat{\bf A}_{n}\right) =\\ 
\left\|
\begin{array}{lll}
\hat{\bf \omega}^{1}_{1}(\hat{\bf A}_{1}) 
\cdots\cdots 
&\hat{\bf \omega}^{1}_{n}(\hat{\bf A}_{1}\\
\vdots  &\vdots\\
\hat{\bf \omega}^{1}_{1}(\hat{\bf A}_{n}) 
\cdots\cdots 
&\hat{\bf \omega}^{1}_{n}(\hat{\bf A}_{n})
\end{array}
\right \| .
\end{equation}
Here as well as for the rest of this section 
we abbreviate the set of indices $(\lambda_{i},\mu_{i},\alpha_{i})$
by the single symbol $i$. 
If $\{\hat{\gamma}{}_{i}^{r_{i}}\}$ and $\{\hat{\gamma}{}^{i}_{r_{i}}\}$
are dual basis respectively in $\hat{\bf T}_{\Phi_{p}}$ and
$\hat{\bf T}_{\Phi_{p}}^{*}$, then the
$\{\hat{\gamma}{}_{1}^{r_{1}}\otimes\cdots\otimes\hat{\gamma}{}_{p}^{r_{p}}
\otimes\hat{\gamma}{}^{1}_{s_{1}}\otimes\cdots\otimes\hat{\gamma}
{}^{q}_{s_{q}}\}$
will be the basis in operator space
$
\hat{\bf T}^{p}_{q}=\underbrace{\hat{\bf T}_{\Phi_{p}}\otimes\cdots \otimes 
\hat{\bf T}_{\Phi_{p}}}_{p}
\otimes\underbrace{\hat{\bf T}^{*}_{\Phi_{p}}\otimes\cdots \otimes 
\hat{\bf T}^{*}_{\Phi_{p}}}_{q}.
$
Any operator tensor $\hat{\bf T} \in \hat{\bf T}^{p}_{q}({\bf \Phi}_{p})$
can be written
$$
\hat{\bf T}=T^{i_{1}\cdots i_{p}}_{j_{1}\cdots j_{q}}
\left(r_{1},\ldots,r_{p},s_{1},\ldots,s_{q}\right)
\hat{\gamma}{}_{i_{1}}^{r_{1}}\otimes\cdots\otimes\hat{\gamma}
{}_{i_{p}}^{r_{p}}
\otimes\hat{\gamma}{}^{j_{1}}_{s_{1}}\otimes\cdots\otimes
\hat{\gamma}{}^{j_{q}}_{s_{q}},
$$
where 
$
T^{i_{1}\cdots i_{p}}_{j_{1}\cdots j_{q}} 
\left(r_{1},\ldots,r_{p},s_{1},\ldots,s_{q}\right)=
T\left( 
\hat{\gamma}{}^{i_{1}}_{r_{1}}\otimes\cdots\otimes\hat{\gamma}
{}^{i_{p}}_{r_{p}}
\otimes\hat{\gamma}{}_{j_{1}}^{s_{1}}\otimes\cdots\otimes
\hat{\gamma}{}_{j_{q}}^{s_{q}}
\right)
$
are the components of $\hat{\bf T}$ in 
$\{\hat{\gamma}{}_{i}^{r_{i}}\}$ and $\{\hat{\gamma}{}^{i}_{r_{i}}\}$.
Any antisymmetric operator tensor of $\widehat{(0,n)}$ type
reads 
\begin{equation}
\label{R421}
\hat{\bf T}^{*}=T_{i_{1} \cdots i_{n}}\gamma^{\widehat{\imath}_{1}}
\otimes\cdots\otimes \gamma^{\widehat{\imath}_{n}}
=\S_{i_{1}<\cdots<i_{n}}
T_{i_{1} \cdots i_{n}} d\,\Phi^{\widehat{\imath_{1}}}\wedge\cdots\wedge
d\,\Phi^{\widehat{\imath_{n}}}.
\end{equation}
Let the $\hat{\cal D}_{1}$ and $\hat{\cal D}_{2}$ are two compact 
convex parallelepipeds in oriented $n$ dimensional operator space
$\hat{\bf R}^{n}$ and the $f:\hat{\cal D}_{1}\rightarrow \hat{\cal D}_{2}$
is differentiable reflection of interior of $\hat{\cal D}_{1}$ into
$\hat{\cal D}_{2}$ retaining an orientation, namely for any function
$\varphi \in C^{\infty}$ defined on $\hat{\cal D}_{2}$ it holds
$\varphi\circ f\in C^{\infty}$ and $f^{*}\varphi\left({\bf \Phi}_{p}\right)=
\varphi\left( f \left({\bf \Phi}_{p} \right)\right)$, where $f^{*}$
is an image of function $\varphi\left( f \left({\bf \Phi}_{p} \right)\right)$
on $\hat{\cal D}_{1}$ at the point ${\bf \Phi}_{p}$. Hence, the function
$f$ induces a linear reflection $\hat{d}\, f:\hat{\bf T}\left(
\hat{\cal D}_{1}\right)\rightarrow \hat{\bf T}\left(
\hat{\cal D}_{2}\right)$ as an operator differential of $f$ implying
$\hat{d}\,f\left(\hat{\bf A}_{p} \right)\varphi=\hat{\bf A}_{p}
(\varphi\circ f)$ for any operator vector $\hat{\bf A}_{p} \in
\hat{\bf T}_{\Phi_{p}}$ and for any function $\varphi \in C^{\infty}$ defined
in the neighbourhood of ${\bf \Phi'}_{p}=f\left({\bf \Phi}_{p}\right)$.
If the function $f$ is given in the form ${\Phi'}^{i}={\Phi'}^{i}\left(
\Phi_{p}\right)$ and
$\hat{\bf A}_{p}=\left(A^{i}\left.\hat{\partial}\right/ \partial\Phi^{i}
\right)_{p}$, then in terms of local coordinates one gets
$
\left(\hat{d}\,f\right)\hat{\bf A}_{p}=A^{i}\left(\FFr{\partial{\Phi'}^{j}}
{\partial\Phi^{i}}\right)_{p}
\left(\FFr{\hat{\partial}}
{\partial{\Phi'}^{j}}\right)_{p'}.
$
So, if $f_{1}:\hat{\cal D}_{1}\rightarrow \hat{\cal D}_{2}$ and
$f_{2}:\hat{\cal D}_{2}\rightarrow \hat{\cal D}_{3}$ then
$\hat{d}\,\left(f_{2}\circ f_{1}\right)=\hat{d}\,f_{2}\circ \hat{d}\,
f_{1}$. The differentiable reflection 
$f:\hat{\cal D}_{1}\rightarrow \hat{\cal D}_{2}$ induces the reflection
$\hat{f}^{*}:\hat{\bf T}^{*}\left(\hat{\cal D}_{2}\right)\rightarrow 
\hat{\bf T}^{*}\left(\hat{\cal D}_{1}\right)$ conjugated to $\hat{f}_{*}$.
The latter is the operator differential of $f$, while
\begin{equation}
\label{R423}
<\hat{f}^{*}\hat{\omega'}^{1},\hat{\bf A}>_{\Phi_{p}}=
\left.<\hat{\omega'}^{1},\hat{f}_{*}\hat{\bf A}>\right|_{f\left(
\Phi_{p}\right)},
\end{equation}
where $\left.\hat{\bf A}\right|_{f\left(\Phi_{p}\right)}=
\left(\hat{d}\,f\right)\hat{\bf A}_{p}$ and $\hat{\omega'}^{1}\in 
\left.\hat{\bf T}^{*}\right|_{f\left(\Phi_{p}\right)}$. Hence
$
\hat{f}^{*}\left(\hat{d}\,\varphi\right)=\hat{d}\,
\left(\hat{f}^{*}\varphi\right)
$
and
\begin{equation}
\label{R426}
\begin{array}{l}
\hat{f}^{*}\left.T\left(\hat{\bf A}_{1},\ldots,
\hat{\bf A}_{n}\right)\right|_{\Phi_{p}}=
\left.T\left(\hat{f}_{*}\hat{\bf A}_{1},\ldots,\hat{f}_{*}
\hat{\bf A}_{n}\right)\right|_{f\left(\Phi_{p}\right)},\\
\left.T\left(\hat{f}^{*}\hat{\bf \omega}^{1}_{1},\ldots,
\hat{f}^{*}\hat{\bf \omega}^{1}_{n}\right)\right|_{\Phi_{p}}=
\hat{f}_{*}\left.T\left(\hat{\bf \omega}^{1}_{1},\ldots,\hat{f}^{*}
\hat{\bf \omega}^{1}_{n}\right)\right|
_{f\left(\Phi_{p}\right)}.
\end{array}
\end{equation}
For any differential operator $n$ form $\hat{\bf \omega}^{n}$ on
$\hat{\cal D}_{2}$ the reflection $f$ induces the operator $n$ form
$\hat{f}^{*}\hat{\bf \omega}^{n}$ on $\hat{\cal D}_{1}$
\begin{equation}
\label{R427}
\left(\hat{f}^{*}\hat{\bf \omega}^{n}\right)
\left.\left(\hat{\bf A}_{1},\ldots,
\hat{\bf A}_{n}\right)\right|_{\Phi_{p}}=\hat{f}_{*}\hat{\bf \omega}^{n}
\left.\left(\hat{f}_{*}\hat{\bf A}_{1},\ldots,\hat{f}_{*}
\hat{\bf A}_{n}\right)\right|_{f\left(\Phi_{p}\right)}.
\end{equation}
If $\hat{\bf \omega'}^{1}=\alpha'_{i}d\,\,{\Phi'}^{\widehat{\imath}}$
then
$
\hat{f}^{*}\left(\alpha'_{i}d\,\,{\Phi'}^{\widehat{\imath}}\right)=
\alpha'_{i}\,\,\FFr{\partial {\Phi'}^{i}}{\partial {\Phi}^{j}}
\,d\,\Phi^{\widehat{\jmath}}.
$
This can be extended up to
$\hat{\bf \omega'}^{n} \rightarrow \hat{\bf \omega}^{n}$
\begin{equation}
\label{R429}
\hat{f}^{*}\left(
\S_{i_{1}<\cdots<i_{n}}
{T'}_{i_{1} \cdots i_{n}} d\,{\Phi'}^{\widehat{\imath_{1}}}\wedge\cdots\wedge
d\,{\Phi'}^{\widehat{\imath_{n}}}\right)
=\S_{\begin{array}{l}
{\scriptstyle i_{1}<\cdots<i_{n}} \\
{\scriptstyle j_{1}<\cdots<j_{n}}
\end{array}}
{T'}_{i_{1} \cdots i_{n}} 
\FFr{\partial {\Phi'}^{i_{1}}}{\partial {\Phi}^{j^{1}}}\cdots
\FFr{\partial {\Phi'}^{i_{n}}}{\partial {\Phi}^{j^{n}}}\,
d\,{\Phi}^{\widehat{\imath_{1}}}\wedge\cdots\wedge
d\,{\Phi}^{\widehat{\imath_{n}}},
\end{equation}
namely
$
\hat{f}^{*}\hat{\bf \omega'}^{n} = J_{\Phi}\,\hat{\bf \omega}^{n}=
\left( det\,\,d f\right)\hat{\bf \omega}^{n},
$
where $J_{\Phi}$ is the Jacobian of reflection $J_{\Phi}=\left\|
\FFr{\partial {\Phi'}^{i}}{\partial {\Phi}^{j}}\right\|$.
While
$$
\left(\hat{f}_{1}\circ\hat{f}_{2}\right)^{*}=
\hat{f}^{*}_{1}\circ\hat{f}^{*}_{2},\quad
\hat{f}^{*}\left(\hat{\bf \omega}_{1}\wedge
\hat{\bf \omega}_{2}\right)=
\hat{f}^{*}\left(\hat{\bf \omega}_{1}\right)\wedge
\hat{f}^{*}\left(\hat{\bf \omega}_{2}\right).
$$
We may consider the integration of operator $n$ form implying
$
\IIn_{\hat{\cal D}_{1}}\hat{f}^{*}\hat{\bf \omega}^{n} = 
\IIn_{\hat{\cal D}_{2}}\hat{\bf \omega}^{n}.
$
In general, let $\hat{\cal D}_{1}$ is the limited convex
$n$ dimensional parallelepiped in the $n$ dimensional operator space
$\hat{\bf R}^{n}$. One defines the $n$ dimensional $i$-th piece of 
integration path $\hat{\sigma}^{i}$ in $\hat{G}$ as $\hat{\sigma}^{i}=
\left(\hat{\cal D}_{i}, f_{i}, Or_{i}\right)$, where $\hat{\cal D}_{i}
\in \hat{\bf R}^{n}, \quad f_{i}:\hat{\cal D}_{i}\rightarrow \hat{G}$
and the $Or_{i}$ is an orientation of $\hat{\bf R}^{n}$. Then, the integral
over the operator $n$ form $\hat{\bf \omega}^{n}$ along the operator
$n$ dimensional chain $\hat{c}_{n}=\S m_{i}\hat{\sigma}^{i}$ may be
written
$$
\IIn_{\hat{c}_{n}}\hat{\bf \omega}^{n}=\S m_{i}
\IIn_{\hat{\sigma}^{i}}\hat{\bf \omega}^{n}=\S m_{i}
\IIn_{\hat{\cal D}_{i}}\hat{f}^{*}\hat{\bf \omega}^{n},
$$
where the $m_{i}$ is a multiple number. Taking into account the eq.(A.1.29),
the matrix element yields
\begin{equation}
\label{R434}
<\chi \| \IIn_{\hat{c}_{n}}\hat{\bf \omega}^{n}\|
\chi^{0} >
\rightarrow 
\left\{\S_{\lambda\mu=1}^{2}\right\}_{1}^{n}
\S_{r_{1}^{\lambda\mu},\ldots ,r_{n}^{\lambda\mu}=1}^{N_{\lambda\mu}}
\S m_{i}\bar{c}(r_{1}^{\lambda\mu},\ldots ,r_{n}^{\lambda\mu})
\IIn_{\hat{\cal D}_{i}}\hat{f}^{*}\hat{\bf \omega}^{n}
(r_{1}^{\lambda\mu},\ldots ,r_{n}^{\lambda\mu}).
\end{equation}
Next, we may apply the analog of exterior differentiation. We define 
the operator $(n+1)$ form $\hat{d}\,\hat{\bf \omega}^{n}$ on $(n+1)$
operator vectors $\hat{\bf A}_{1},\ldots,
\hat{\bf A}_{n+1}\in \hat{\bf T}_{\Phi_{p}}$ by considering 
diffeomorphic reflection $f$ of the neighbourhood of the point
$0$ in $\hat{\bf R}^{n}$ into neighbourhood of the point ${\bf \Phi}_{p}$
in $\hat{G}$. The prototypes of operator vectors
$\hat{\bf A}_{1},\ldots,
\hat{\bf A}_{n+1}\in \hat{\bf T}_{\Phi_{p}}\left(\hat{G}\right)$
at the operator differential of $f$ belong to tangent operator space
$\hat{\bf R}^{n}$ in $0$. Namely, the prototypes are the operator vectors
$\hat{\bf \xi}_{1},\ldots,\hat{\bf \xi}_{n+1} \in \hat{\bf R}^{n}$. Let
$f$ reflects the parallelepiped $\hat{\bf \Pi}^{*},$ stretched over 
the $\hat{\bf \xi}_{1},\ldots,\hat{\bf \xi}_{n+1}$, onto the $(n+1)$
dimensional piece $\hat{\bf \Pi}$ on the $\hat{G}$.
While the border of the $n$ dimensional chain $\partial \hat{\bf \Pi}$ 
in $\hat{\bf R}^{n+1}$ defined as follows: the pieces $\hat{\sigma}^{i}$
of the chain $\partial \hat{\bf \Pi}$ are $n$ dimensional facets
$\partial \hat{\bf \Pi}_{i}$ of parallelepiped $\partial \hat{\bf \Pi}$
with the reflections embedding the facets into $\hat{\bf R}^{n+1}$:
$\quad f_{i}:\hat{\bf \Pi}_{i} \rightarrow
\hat{\bf R}^{n+1}$, and the
orientations $Or_{i}$ has defined as $\partial \hat{\bf \Pi}=\S\hat{\sigma}^{i},
\quad \hat{\sigma}^{i}=\left(\hat{\bf \Pi}_{i},f_{i},Or_{i}\right)$
Considering the curvilinear parallelepiped
$$
F\left(\hat{\bf A}_{1},\ldots,\hat{\bf A}_{n}\right)=
\IIn_{\partial \hat{\bf \Pi}}\hat{\bf \omega}^{n},
$$
one may state that the unique operator of the $(n+1)$-form $\hat{\Omega}$
exists on $\hat{\bf T}_{\Phi_{p}}$, which is the principle $(n+1)$ linear
part in $0$ of integral over the border of 
$F\left(\hat{\bf A}_{1},\ldots,\hat{\bf A}_{n}\right)$, namely
\begin{equation}
\label{R436}
F\left(\varepsilon\hat{\bf A}_{1},\ldots,
\varepsilon\hat{\bf A}_{n}\right)=
\varepsilon^{n+1}\hat{\Omega}
\left(\hat{\bf A}_{1},\ldots,\hat{\bf A}_{n+1}\right)+
O\left(\varepsilon^{n+1}\right),
\end{equation}
where $\hat{\Omega}$ is independent of choice of the coordinates
used in definition of $F$. 
The prove of it is the same to those of similar one given in the 
differential geometry [46].
If in local coordinates 
$
\hat{\bf \omega}^{n}=\S_{i_{1}<\cdots<i_{n}}
T_{i_{1} \cdots i_{n}} d\,\Phi^{\widehat{\imath_{1}}}\wedge\cdots\wedge
d\,\Phi^{\widehat{\imath_{n}}},
$
then
\begin{equation}
\label{R438}
\hat{\Omega}=\hat{d}\,\hat{\bf \omega}^{n}=
\S_{i_{1}<\cdots<i_{n}}
\hat{d}\,T_{i_{1} \cdots i_{n}} d\,\Phi^{\widehat{\imath_{1}}}
\wedge\cdots\wedge d\,\Phi^{\widehat{\imath_{n}}}.
\end{equation}
The operator of exterior differential $\hat{d}$ commutes with the
reflection 
$
f:\hat{G}\rightarrow \hat{G}
$
$$
\hat{d}\,\left(\hat{f}^{*}\hat{\bf \omega}^{n}\right)=
\hat{f}^{*}\left(\hat{d}\,\hat{\bf \omega}^{n}\right).
$$
Define the exterior differential by operator (n+1) form
\begin{equation}
\label{R440}
\hat{d}\hat{\omega}^{n}=\S_{\begin{array}{l}
{\scriptstyle i_{0}} \\
{\scriptstyle i_{1}<\ldots<i_{n}}
\end{array}}
\FFr{\partial T_{i_{1}\ldots i_{n}}}{\partial\Phi^{i_{0}}}
d\Phi^{\widehat{i_{0}}}\wedge
d\Phi^{\widehat{i_{1}}}\wedge\cdots\wedge d\Phi^{\widehat{i_{n}}}
=\S_{i_{1}<\ldots<i_{n}}(\hat{d}T_{i_{1}\ldots i_{n}})\wedge
d\Phi^{\widehat{i_{1}}}\wedge\cdots\wedge d\Phi^{\widehat{i_{n}}},
\end{equation}
one gets
\begin{equation}
\label{R441}
\begin{array}{l}
<\chi\parallel \hat{d}\hat{\omega}^{n} \parallel\chi^{0}>\rightarrow \\
\S_{i_{1}<\ldots<i_{n}}
\left\{\S_{\lambda,\mu=1}^{2}\right\}_{1}^{n}
\S_{r_{1}^{\lambda\mu},\ldots ,r_{n}^{\lambda\mu}=1}^{N_{\lambda\mu}}
\bar{c}(r_{1}^{\lambda\mu},\ldots r_{n}^{\lambda\mu})
{(dT(r_{1}^{\lambda\mu},\ldots ,r_{n}^{\lambda\mu})}_{i_{1}\ldots i_{n}})
\wedge
d\Phi^{i_{1}}_{r_{1}^{\lambda\mu}}\wedge\cdots\wedge 
{d\Phi}^{i_{n}}_{r_{n}^{\lambda\mu}}.
\end{array}
\end{equation}
Reflected upon the results far obtained within this section
we may draw a conclusion that the matrix element 
of any geometric object of operator 
manifold $\hat{G}$ yields corresponding geometric object of wave 
manifold $\cal {G}$.

\section * {Appendix B}
\label {appe}
\setcounter{section}{0}
\renewcommand {\theequation}{B.\thesection.\arabic {equation}}
\section{Reflection of the Fermi Fields}
The rotation angles (subsec.4.2) are determined from the constraint imposed 
upon distortion transformations that a sum of distorted parts
of corresponding basis vectors $O_{\lambda}$ and 
$\sigma_{\beta}$ should be zero for given $\lambda$
\begin{equation}
\label{R52}
<O_{(\lambda\alpha)},O_{\tau}>_{\tau \neq \lambda}+\frac{1}{2}
\varepsilon_{\alpha\beta\gamma}\frac{<\sigma_{(\lambda\beta)},\sigma_{\gamma}>}
{<\sigma_{(\lambda\beta)},\sigma_{\beta}>}=0,
\end{equation}
where $\varepsilon_{\alpha\beta\gamma}$ is an antisymmetric unit tensor.
Thereupon, $\tan\theta_{(\lambda\alpha)}=-\kappa a_{(\lambda\alpha)}$,
where $\theta_{(\lambda\alpha)}$ is the particular rotation angle around the 
axis $\sigma_{\alpha}$. Since the $R$
should be independent of the sequence of rotation axes, then  it implies
the mean value $R=\displaystyle \frac{1}{6} \sum_{i \neq j \neq k}
R^{(ijk)}$, where $R^{(ijk)}$ the matrix of rotations occurred
in the given sequence $(ijk)$ $(i,j,k=1,2,3)$.
The field $a_{f}$ is due to the distortion of basis
pseudovector $O_{\lambda}$, while the distortion of $\sigma_{\alpha}$
follows from eq.(B.1.1).
We consider the reflection
of the Fermi fields and their dynamics from the flat manifold $\G1_{u}$ 
to
distorted manifold $\widetilde{\G1_{u}}$, and vice versa. 
While we construct a diffeomorphism
$u(u_{f}):\G1_{u}\rightarrow \widetilde{\G1_{u}}$, where
the holonomic functions 
$u(u_{f})$ satisfy defining
relation
\begin{equation}
\label {eq: RB.4}
e\psi=e^{f} + \chi^{f}({\bf B}_{f}).
\end{equation}
Here $e^{f}$ and $e$ are the basis vectors
on $\G1_{u}$ and $\widetilde{\G1_{u}}$.
The $\psi$ is taken to denote
$
\psi\equiv 
\FFr{\partial\,u}{\partial\,u_{f}}.
$
The covector
\begin{equation}
\label {eq: RB.7}
\chi^{f}_{(\tau\beta)}({\bf B}_{f})=
e_{(\lambda\alpha)}\chi^{(\lambda\alpha)}_{(\tau\beta)}=
-\FFr{1}{2}e_{(\lambda\alpha)}\IIn_{0}^{u_{f}}
({\pr_{u}}{}^{f}_{(\rho\gamma)}D^{(\lambda\alpha)}_{(\tau\beta)}-
{\pr_{u}}{}^{f}_{(\tau\beta)}D^{(\lambda\alpha)}_{(\rho\gamma)})
du^{(\rho\gamma)}_{f}
\end{equation}
realizes the coordinates $u$ 
by providing a criteria of integration and undegeneration [47,48]\footnote 
{I wish to thank S.P.Novikov for valuable discussion of this question}.
The Lagrangian $L(x)$ of fields $\Psi(u)$
may be obtained under the reflection from the Lagrangian $L_{f}(u_{f})$ of 
corresponding {\em shadow fields} $\Psi_{f}(u_{f})$ and vice versa.
The $\Psi_{f}(u_{f})$ is defined as the section of vector bundle 
associated with the primary gauge group G by reflection 
$\Psi_{f}:  \G1_{u}\rightarrow E$ such that
$p\,\Psi_{f}(u_{f})=u_{f}$, where 
$u_{f} \in  \G1_{u}$ is the point of flat manifold  $\G1_{u}$ 
(specified by index $({}_{f})$). The $\Psi_{f}$ takes value in standard 
fiber $F_{u_{f}}$
upon $u_{f} : p^{-1}(U^{(f)})=U^{(f)}\times F_{u_{f}}$, where 
$U^{(f)}$ is the region of base of principle bundle 
upon which an expansion into direct product $p^{-1}(U^{(f)})=
U^{(f)}\times G$ is defined.  The fiber is Hilbert vector space 
on which a linear representation $U_{f}$ of the group G is given. 
Respectively
$\Psi(u)\subset F_{u}$, where $F_{u}$
is the fiber upon $u:p^{-1}(U)=U\times F_{u}$, $U$ is the region of base 
 $\widetilde{\G1_{u}}$. Thus, the reflection of bispinor fields may be written down
\begin{equation}
\label {eq: RB.1}
\begin{array}{l}
\Psi(u)=R({\bf B}_{f})\Psi_{f}(u_{f}), \quad
\bar{\Psi}(u)=\bar{\Psi}_{f}(u_{f})\widetilde{R}^{+}({\bf B}_{f}),\quad
g(u)\nabla\Psi(u)=\\
S(B_{f})R({\bf B}_{f})\gamma_{f}D\Psi_{f}(u_{f}),\quad
\left( \nabla\bar{\Psi}(u)\right)g(u)
=S(B_{f})\left( D\bar{\Psi}_{f}(u_{f})\right)
\gamma_{f}\widetilde{R}^{+}({\bf B}_{f}).
\end{array}
\end{equation}
Reviewing the notation
${\bf B}(u_{f})=T^{a}{\bf B}^{a}(u_{f})$
is the gauge field of distortion
with the values in Lie algebra of group G, $R({\bf B}_{f})$ is the reflection 
matrix (see eq.(B.1.7)),
$\widetilde{R}=\gamma^{0}R\gamma^{0},$
$D=\partial^{f}- ig {\bf B},
\quad g^{(\lambda\alpha)}(\theta)=V^{(\lambda\alpha)}_{(i,l)}
(\theta){\gamma}^{(i,l)}_{f}$, 
$V^{(\lambda\alpha)}_{(i,l)}(\theta)$ are congruence parameters of curves
(Latin indices refer to tetrad components).
The matrices
${\gamma}^{(\pm\alpha)}_{f}=\FFr{1}{\sqrt{2}}(\gamma^{0}\sigma^{\alpha}
\pm \gamma^{\alpha})$, 
$\gamma^{0},\gamma^{\alpha}$ are Dirac matrices. 
$\nabla$ is 
covariant derivative defined on  $\widetilde{\G1_{u}}$:
$\nabla={\pr_{u}}+
\Gamma$, where
the connection
$\Gamma(\theta)$ in terms of Ricci 
rotation coefficients reads
$
\Gamma_{(\lambda\alpha)}(\theta)=\FFr{1}{4}
\Delta_{(\lambda\alpha)(i,l)(m,p)}\,\gamma^{(i,l)}_{f}\,\gamma^{(m,p)}_{f},
\quad
\bar{\Gamma}_{(\lambda\alpha)}(\theta)=\FFr{1}{4}
\Delta_{(\lambda\alpha)(i,l)(m,p)}\,\gamma^{(m,p)}_{f}\,\gamma^{(i,l)}_{f}.
$\\
According to the general gauge principle [42,44], the physical system of
the fields $\Psi(u)$ is required to be invariant under the finite 
local gauge transformations 
\begin{equation}
\label {eq: RB.8}
\begin{array}{l}
\Psi'(u)=U_{R}\Psi(u),\quad
\left(
g(u) \nabla \Psi(u)\right)'=
U_{R}\left(g(u)\nabla \Psi(u)
\right), \quad U_{R}=R({\bf B}'_{f})U_{f}R^{-1}({\bf B}_{f}),
\end{array}
\end{equation}
of the Lie group of gravitation $G_{R} (\ni  U_{R})$ generated by G, 
where the gauge field
${\bf B}_{f}(u_{f})$ is transformed under 
G in standard form.
The physical meaning of the general principle is as follows: one has 
conventional G-gauge theory on flat manifold in terms of curviliniear 
coordinates if curvature tensor is zero, to which the zero vector 
eq.(B.1.3) is corresponded.
Otherwise it yields the gravitation interaction. \\
Out of a set of arbitrary curvilinear coordinates in $\widetilde{\G1_{u}}$ 
the
real curvilinear coordinates may be distinguished, which satisfy
eq.(B.1.2) under all the possible Lorentz and gauge transformations. There is
a single~-valued conformity between corresponding tensors with various
suffixes on $\widetilde{\G1_{u}}$ and $ \G1_{u}$. While, each 
index transformation is incorporated with the function $\psi$. 
The transformation of real curvilinear coordinates
$u \rightarrow u'$ is due to some Lorentz $(\Lambda)$ and gauge
$(B_{f}\rightarrow B'_{f})$ transformations
\begin{equation}
\label {R213}
\frac{\partial u'}{\partial u}=\psi(B'_{f})
\psi(B_{f})\Lambda.
\end{equation}
There would then exist preferred systems and group of transformations of
real curvilinear coordinates in  the $\widetilde{\G1_{u}}$. The wider group of 
transformations of arbitrary curvilinear coordinates in the 
$\widetilde{\G1_{u}}$ 
would then be of no consequence for the field dynamics.
A straightforward calculation gives
the reflection matrix 
\begin{equation}
\label{eq: R2.8}
R(u,u_{f})=R_{f}(u_{f})R_{g}(u)=\exp\left[ 
-i\Theta_{f}(u_{f})-
\Theta_{g}(u)\right],
\end{equation}
where
\begin{equation}
\label{eq: R2.9}
\Theta_{f}(u_{f})=
g\IIn_{0}^{u_{f}}{\bf B}(u_{f})du_{f},
\quad 
\Theta_{g}(u)=\frac{1}{2}\IIn_{0}^{u} R^{+}_{f}\left\{ 
g\Gamma R_{f}, g \,du\right\}.
\end{equation}
Then
\begin{equation}
\label {eq: RB.9}
S(B_{f})=\frac{1}{8K}\psi\left\{\widetilde{R}^{+}g\,R,
\gamma^{0}\right\}=inv,
\end{equation}
where
$
K =\widetilde{R}^{+}R=\widetilde{R}^{+}_{g}R_{g}
= 1.
$
and
$
\widetilde{U}_{R}^{+}U_{R}=\gamma^{0}U_{R}^{+}\gamma^{0}U_{R}=1.
$
The Lagrangian of shadow Fermi field may be written
\begin{equation}
\label {eq: RB.15}
\begin{array}{l}
L_{f}(u_{f})=J_{\psi} L(u)=\\
J_{\psi}\left\{S(B_{f})
\displaystyle {\frac{i}{2}}\left[ 
\bar{\Psi}_{f}(u_{f})\gamma_{f}D
\Psi_{f}(u_{f})-\right.\right. 
\left.\left.
\left(D\bar{\Psi}_{f}(u_{f})\right)\gamma_{f}
\Psi_{f}(u_{f})\right]-
m\bar{\Psi}_{f}(u_{f})\Psi_{f}(u_{f})\right\}.
\end{array}
\end{equation}
provided by
$
J_{\psi}= \| \psi\|\,\sqrt{-g} \equiv\left(1+2\|
<e^{f},\chi^{f}>\|
+ \| <\chi^{f},\chi^{f}>\|\right)^{1/2}.
$
The Lagrangian $L(u)$ of the field $\Psi(u)$ reads
\begin{equation}
\label {eq: RB.16}
\sqrt{-g}L(u)=\FFr{\sqrt{-g}}{2} \left\{-i \bar{\Psi}(u)
g({\pr_{u}}-\Gamma)\Psi(u)+ i\bar{\Psi}(u)
({\lpr_{u}}-\bar{\Gamma})g\bar{\Psi}(u)+
2m\bar{\Psi}(u)\Psi(u)\right\},
\end{equation}
yielding the field equations
\begin{equation}
\label {eq: RB.18}
\left[ i g ({\pr_{u}}-\Gamma )- m \right]\Psi(u) =0,\quad
\bar{\Psi}(u)\left[
i({\lpr_{u}}-\bar{\Gamma}) g - m \right] =0.
\end{equation}

\section * {Appendix C}
\label {sol}
\setcounter{section}{0}
\renewcommand {\theequation}{C.\thesection.\arabic {equation}}
\section{The Solution of Wave Equation of Distorted Structure}
To solve the equation (B.1.12)
\begin{equation}
\label {eq: RB.18}
\left[ i g ({\pr_{u}}-\Gamma )- m \right]\Psi(u) =0,
\end{equation}
we transform it into
\begin{equation}
\label{R144}
\{-\partial^{2}-m^{2}-{(g\Gamma)}^{2}+2(\Gamma g)+
(g \partial)(g\Gamma) \}\Psi=0,
\end{equation}
where we abbreviate the
indices $(\lambda\alpha)$ by the single symbol $\mu$, and Latin indices
$(im)\quad (i=\pm, m=1,2,3)$ by $i$, also denote 
${\hat{\p1}_{u}}\equiv 
\hat{p}$ and
\begin{equation}
\label{R145}
\begin{array}{l}
\FFr{1}{2}\sigma^{\mu\nu}F_{\mu\nu}=(g\partial)(g\Gamma)-
(\partial\Gamma),\quad (g\partial)(g\Gamma)=
g^{\mu}g^{\nu}\partial_{\mu}\Gamma_{\nu}, \\
\FFr{1}{2}\sigma^{\mu\nu}[\Gamma_{\mu},\Gamma_{\nu} ]=
{(g\Gamma)}^{2}- \Gamma^{2}, \quad
\partial^{2}=\partial^{\mu}\partial_{\mu}, \quad
\Gamma^{2}=\Gamma^{\mu}\Gamma_{\mu}, \\
2g^{\mu\nu}=\{ g^{\mu},g^{\nu} \},\quad
2\sigma^{\mu\nu}= [g^{\mu},g^{\nu} ],\quad
\F_{\mu\nu}=\partial_{\mu}\Gamma_{\nu}-
\partial_{\nu}\Gamma_{\mu}.
\end{array}
\end{equation}
We are looking for a solution given in the form
$\Psi=e^{-ipu}F(\varphi)$,
where $p_{\mu}$ is a constant sixvector $pu=p_{\mu}u_{\mu}$,
and admit that the field of distortion may be switched on
at $u_{0}=-\infty $ smoothly. Then the function $\Psi$ must match
onto the wave function of ordinary regular structure. 
Smoothness requires that the numbers $p_{\mu}$ become the components of 
link momentum of regular structure and satisfy the boundary condition
$p_{\mu}p_{\mu}=m^{2}=p^{2}_{\eta}$.
Due to it we cancel unwanted solutions
and clarify the normalization of wave functions 
\begin{equation}
\label{R146}
\int\Psi^{*}_{p'}\Psi_{p}d^{3}u=\int{\bar{\Psi}}_{p'}\gamma^{0}\Psi_{p}
d^{3}u={(2\pi)}^{3}\delta(\vec{p'}-\vec{p}).
\end{equation}
We suppose that at $\sqrt{-g}\neq1 $ the gradient of the function 
$\varphi $ reads
$$\partial_{\mu}\varphi=V^{i}_{\mu}k_{i}, \quad
\partial^{\mu}\varphi=V_{i}^{\mu}k^{i},$$
where $k_{i}$ are arbitrary constant numbers satisfying the condition
$k_{i}k_{i}=0$. Thus
$\partial^{\mu}\varphi\,\partial_{\mu}\varphi=
(V_{i}^{\mu}\,V^{j}_{\mu})\,k^{i}\,k_{j}=0$.
The eq.(C.1.2) gives rise to $F'=A(\theta)F$,
where $(\cdots)'$ stands for the derivative with respect to $\varphi$, and
\begin{equation}
\label{R147}
A(\theta)=\FFr{2i(p\Gamma) + m^{2}-p^{2}+{(g\Gamma)}^{2}-
(g\partial)(g\Gamma)}{2i(kVp)-(kDV)};
\end{equation}
where
\begin{equation}
\label{R148}
\begin{array}{ll}
(kVp)=k^{i}V_{i}^{\mu}p_{\mu},\quad
(kDV)=k^{i}D_{\mu}V_{i}^{\mu}, 
\quad
D_{\mu}=\partial_{\mu}-2\Gamma_{\mu}, \\
p^{2}=p^{\mu}p_{\mu}=g^{\mu\nu}(\theta)p_{\mu}p_{\nu}\quad
(kVdu)=k_{i}V^{i}_{\mu}du^{\mu}.
\end{array}
\end{equation}
We are interested in the right-handed eigenvectors $F_{r}\quad
(r=1,2,3,4)$ corresponding to eigenvalues $\mu_{r}$ of matrix $A:
AF_{r}=\mu_{r}F_{r}$, which are the roots of polynomial characteristic
equation 
$$
c(\mu)=\left\|(\mu I-A)\right\|=0.
$$
Thus, one gets $F'_{r}=\mu_{r}F_{r}$ and
$F=\displaystyle \prod^{4}_{r=1}F_{r}$.
Hence
$(\ln F)'=\S^{4}_{r=1}\mu_{r}=trA$
and $(\ln F)'=X_{R}(\theta)-iX_{J}(\theta)$,
provided
\begin{equation}
\label{R149}
\begin{array}{l}
X_{R}(\theta)=
trA_{R}(\theta)=tr \left\{ \FFr{-(kDV)\left[ m^{2}-p^{2}+
{(g\Gamma)}^{2}-(g\partial)(g\Gamma)\right]+4(kVp)
(p\Gamma)}{(kDV)^{2}+4(kVp)^{2}} \right\},\\
X_{J}(\theta)=
trA_{J}(\theta)=2tr \left\{ \FFr{(kVp)\left[ m^{2}-p^{2}+
{(g\Gamma)}^{2}-
(g\partial)(g\Gamma)\right]+(kDV)(p\Gamma)}{(kDV)^{2}+
4(kVp)^{2}} \right\}.
\end{array}
\end{equation}
The solution of eq.(C.1.1) reads
\begin{equation}
\label{R1410}
F(\theta)=C{\left( \frac{m}{E_{u}} \right)}^{1/2} U\exp
\{\chi_{R}(\theta)-i\chi_{J}(\theta)\}, 
\end{equation}
where $C=1$ is the normalization constant, $U$ is the constant bispinor, and
\begin{equation}
\label{R1410}
\chi_{R}(\theta)=\int_{0}^{u^{\mu}}(kVdu)X_{R}(\theta),\quad
\chi_{J}(\theta)=\int_{0}^{u^{\mu}}(kVdu)X_{J}(\theta).
\end{equation}

\section * {Appendix D}
\label {fieldomm}
\setcounter{section}{0}
\renewcommand {\theequation}{D.\thesection.\arabic {equation}}
\section{Field Aspect of the OMM}
The quantum field and differential geometric aspects of the $\hat{G}_{N}$
may be discussed on the analogy
of $\hat{G}_{N=1}$. Here we turn only to some points of the field aspect. 
A Lagrangian of free field reads
\begin{equation}
\label{eq: RC.22.1}
\widetilde{L}_{0}(D)=
\FFr{i}{2} \{ \bar{\Psi}_{e}(\zeta)
{}^{i}\gamma^{(\lambda,\mu,\alpha)}
{\pr_{i}}{}_{(\lambda,\mu,\alpha)}\Psi_{e}(\zeta)-
{\pr_{i}}{}_{(\lambda,\mu,\alpha)}\bar{\Psi}_{e}(\zeta)
{}^{i}\gamma^{(\lambda,\mu,\alpha)}
\Psi_{e}(\zeta) \},
\end{equation}
where we have adopted the following convention:
\begin{equation}
\label{eq: RC.22.2}
\begin{array}{l}
\Psi_{e}(\zeta)=e\otimes\Psi(\zeta)=
\left( \matrix{
1 &1\cr
1 &1\cr
} 
\right)  
\otimes\Psi(\zeta), 
\quad
\bar{\Psi}_{e}(\zeta)=e\otimes \bar{\Psi}(\zeta),\quad
\bar{\Psi}(\zeta)=\Psi^{+}(\zeta)\gamma^{0}, \\ 
{}^{i}\gamma^{(\lambda,\mu,\alpha)}=
{}^{i}{\widetilde{O}}^{\lambda,\mu}
\otimes{\widetilde{\sigma}}^{\alpha}, \quad
{}^{i}{\widetilde{O}}^{\lambda,\mu}=\FFr{1}{\sqrt{2}}
\left( \nu_{i}\xi_{0}\otimes
{\widetilde{O}}^{\mu}+\varepsilon_{\lambda}\xi \otimes{}^{i}
{\widetilde{O}}^{\mu}\right),\\ 
\varepsilon_{\lambda}=\left\{ \matrix{
1 &\lambda=1\cr
-1 &\lambda=2\cr
}\right., \quad
<\nu_{i},\nu_{j}>=\delta_{ij},\quad
\left\{ {}^{i} {\widetilde{O}}^{\lambda},{}^{j}{\widetilde{O}}^{\mu}
\right\}=\delta_{ij}{}^{*}\delta^{\lambda\mu},\\ 
{\widetilde{O}}^{\mu}=\FFr{1}{\sqrt{2}}
\left( \xi_{0}+
\varepsilon_{\mu}\xi \right),\quad
{\widetilde{O}}^{\lambda}=
{}^{*}\delta^{\lambda\mu}{\widetilde{O}}_{\mu}=
{({\widetilde{O}}_{\lambda})}^{+},\quad 
{}^{i}{\widetilde{O}}^{\mu}=\FFr{1}{\sqrt{2}}
\left( \xi_{0i}+
\varepsilon_{\mu}\xi_{i} \right),
\\
{\pr_{i}}{}_{(\lambda,\mu,\alpha)}=\partial/\partial\,{}^{i}\zeta^
{(\lambda,\mu,\alpha)}, \quad 
\xi_{0}=\left( \matrix{
1 &0 \cr
0 &-1\cr
}\right)
\quad
\xi=\left( \matrix{
0 &1 \cr
-1 &0\cr
}\right),\\ 
{\xi_{0}}^{2}=-\xi^{2}=-{\xi_{0i}}^{2}=\xi^{2}_{i}=1,\quad
\{\xi_{0},\xi\}=\{\xi_{0},\xi_{0i}\}=\{\xi_{0},\xi_{i}\}=\\ 
=\{\xi,\xi_{0i}\}=\{\xi,\xi_{i}\}=\{\xi_{0i},\xi_{j}\}_{i\neq j}=
\{\xi_{0i},\xi_{0j}\}_{i\neq j}=\{\xi_{i},\xi_{j}\}_{i\neq j}=0.
\end{array}
\end{equation}
Field equations are written
\begin{equation}
\label{eq: RC.22.3}
\begin{array}{l}
(\hp1_{\eta}-m)\ps1_{\eta}(\eta)=0,\quad       
\bar{\ps1_{\eta}}(\eta)(\hp1_{\eta}-m)=0,\\ 
(\hp1_{u}-m)\ps1_{u}(u)=0, \quad \bar{\ps1_{u}}(u)(\hp1_{u}-m)=0,
\end{array}
\end{equation}
where
\begin{equation}
\label{eq: RC.22.4}
\begin{array}{ll}
\hp1_{\eta}=i\hpr_{\eta}, \quad \hp1_{u}=i\hpr_{u},\quad 
\hpr_{u}={}^{i}\gamma^{(\lambda\alpha)}{\pr_{u_{i}}}_{(\lambda\alpha)},\quad
{\pr_{\eta}}{}_{(\lambda\alpha)}=\partial/\partial\eta^{(\lambda\alpha)},
\quad{\pr_{u_{i}}}{}_{(\lambda\alpha)}=\partial /
\partial u^{(\lambda\alpha)}_{i},\\ 
{}^{i}{\gam_{\eta}}^{(\lambda\alpha)}=
{}^{i}{\widetilde{\O1_{\eta}}}^{\lambda}\otimes
{\widetilde{\sigma}}^{\alpha}=
\nu_{i}\xi_{0}\otimes\gamma^{(\lambda\alpha)}=
\nu_{i}\xi_{0}\otimes{\widetilde{O}}^{\lambda}\otimes
{\widetilde{\sigma}}^{\alpha},\\ 
{}^{i}{\gam_{u}}^{(\lambda\alpha)}=
{}^{i}{\widetilde{\O1_{u}}}^{\lambda}\otimes
{\widetilde{\sigma}}^{\alpha}=
\xi\otimes{}^{i}\gamma^{(\lambda\alpha)}=
\xi\otimes{}^{i}{\widetilde{O}}^{\lambda}\otimes
{\widetilde{\sigma}}^{\alpha},\\ 
\left(\gamma^{(\lambda\alpha)}\right)^{+}=
{}^{*}\delta^{\lambda\tau}\delta^{\alpha\beta}\gamma^{(\tau\beta)}=
\gamma_{(\lambda\alpha)},\quad 
\left({}^{i}{\gam_{u}}^{(\lambda\alpha)}\right)^{+}=
-{}^{i}{\gam_{u}}_{(\lambda\alpha)}.
\end{array}
\end{equation}
The state of free ordinary structure of ${}^{i}u$-type with the given values 
of link momentum $\p1_{u_{i}}$ and spin projection $s_{i}$ is described 
by means of plane wave.
It is necessary to consider also the solution of negative
frequencies with the normalized bispinor amplitude.

\end{document}